\documentclass{emulateapj}

\usepackage{natbib}
\usepackage{color}

\newcommand{\simgt}{\lower.5ex\hbox{$\; \buildrel > \over \sim \;$}}
\newcommand{\simlt}{\lower.5ex\hbox{$\; \buildrel < \over \sim \;$}}

\def\btheta{\mbox{\boldmath $\theta$}}

\def\hkpc{\mathrel{h^{-1}{\rm kpc}}}
\def\hMpc{\mathrel{h^{-1}{\rm Mpc}}}
\def\Mvir{\mathrel{M_{\rm vir}}}
\def\cvir{\mathrel{c_{\rm vir}}}
\def\rvir{\mathrel{r_{\rm vir}}}

\def\hMsol{\mathrel{h^{-1}M_\odot}}

\slugcomment{}

\shorttitle{Subaru Weak-Lensing Survey of Dark Matter Subhalos in the Coma Cluster}
\shortauthors{Okabe et al.}

\begin{document}


\title{Subaru Weak-Lensing Survey of Dark Matter Subhalos in the Coma Cluster : Subhalo Mass Function and Statistical Properties \altaffilmark{*}}

\altaffiltext{*}{Based on data collected from the Subaru Telescope and obtained from
 SMOKA,  operated by the Astronomy Data Center, National
Astronomical Observatory of Japan }


\author{Nobuhiro Okabe\altaffilmark{1,2}}
\author{Toshifumi Futamase\altaffilmark{3}}
\author{Masaru Kajisawa\altaffilmark{4}}
\author{Risa Kuroshima\altaffilmark{3}}

\email{nobuhiro.okabe@ipmu.jp}
with one command per each

\altaffiltext{1}{Kavli Institute for the Physics and Mathematics of the Universe (WPI), Todai Institutes for Advanced Study, University of Tokyo, 5-1-5 Kashiwanoha, Kashiwa, Chiba 277-8583, Japan}
\altaffiltext{2}{Institute of Astronomy and Astrophysics, Academica Sinica, PO Box 23-141, Taipei 106, Taiwan}
\altaffiltext{3}{Astronomical Institute, Tohoku University, Sendai 980-8578, Japan}
\altaffiltext{4}{Research Center for Space and Cosmic Evolution, Ehime University, Bunkyo-cho 2-5, Matsuyama 790-8577, Japan.}


\begin{abstract}
We present a 4 deg$^2$ weak gravitational lensing survey of subhalos in the very nearby Coma cluster using the Subaru/Suprime-Cam.
The large apparent size of cluster subhalos allows us to measure the mass of 32 subhalos detected in a model-independent manner, down to the order of $10^{-3}$ of the virial mass of the cluster.
Weak-lensing mass measurements of these shear-selected subhalos enable us to investigate
subhalo properties and the correlation between subhalo masses and galaxy luminosities
for the first time.
The mean distortion profiles stacked over subhalos show a sharply truncated feature which is well-fitted 
by a Navarro-Frenk-White (NFW) mass model with the truncation radius, 
as expected due to tidal destruction by the main cluster.
We also found that subhalo masses, truncation radii, and mass-to-light ratios decrease toward the cluster center.
The subhalo mass function, $dn/d\ln M_{\rm sub}$, in the range of 2 orders of magnitude in mass, 
is well described by a single power law or a Schechter function.
Best-fit power indices of $1.09^{+0.42}_{-0.32}$ for the former model and $0.99_{-0.23}^{+0.34}$ for the latter,
are in remarkable agreement with slopes of $\sim0.9-1.0$ predicted by the cold dark matter paradigm.
The tangential distortion signals in the radial range of $0.02-2\hMpc$ from the cluster center show a complex structure 
which is well described by a composition of three mass components of subhalos, the
NFW mass distribution as a smooth component of the main cluster, 
and a lensing model from a large scale structure behind the cluster.
Although the lensing signals are 1 order of magnitude lower than those for clusters at $z\sim0.2$, 
the total signal-to-noise ratio, S/N$=13.3$, is comparable to, or higher,
because the enormous number of background source galaxies compensates for the low
lensing efficiency of the nearby cluster.
\end{abstract}


\section{Introduction}

The cold dark matter (CDM) concordance cosmology has
 had considerable success in explaining various observations on a large scale, 
such as the cosmic microwave background \citep{WMAP07,WMAP09}.
It provides initial conditions for the hierarchical structure formation
involved in the mass assembly histories of halos, for high-resolution $N$-body simulations and analytical models.
In hierarchical clustering, less massive halos are accreted into more massive halos, which are then
subsequently eroded by effects combined with tidal stripping and dynamical friction of the host halo, 
 eventually becoming a smooth component.
Since galaxy clusters are the most massive virialized objects in the universe, 
the central regions of subhalos have survived 
under the over-density field until the recent epoch, and constitute their population. 
Numerical simulations of, and analytic approaches to CDM predict 
that subhalo mass functions at the intermediate and low mass scales follow a power law,
$dn/d\ln M_{\rm sub}\propto M_{\rm sub}^{-\alpha}$ with slopes of $\sim0.9-1.0$ 
\citep[e.g.,][]{Taylor04a,Taylor05b,Taylor05c,Oguri04,vandenBosch05,Diemand04,DeLucia04,Gao04,Shaw06,Angulo09,Giocoli10,Klypin11,Gao12,Wu13}.

Observations of cluster subhalo properties, such as mass function and spatial distribution,
provide us with a deeper understanding of the mass assembly history and
are the most stringent test of CDM predictions on scales of less than several Mpc.
A characteristic feature of the subhalo mass function 
is also critically important to constrain the nature of dark matter,
because it depends on the particle mass of dark matter.
Furthermore, a study of the correlation between galaxy properties and subhalo masses, 
incorporating different data-sets, sheds important insight on the physics of galaxy evolution associated with dark matter. 
Thus, it is of paramount importance to measure the mass function directly from observations 
without assuming a relationship between dark matter and luminous matter and the dynamical state of the system. 
It is difficult, though, to infer the masses of subhalos from visible matter, such as galaxies, 
because assumptions about the mass distribution extending beyond galaxies 
and dynamical state of the galaxies are required.
In this situation, weak gravitational lensing analysis plays an important role. 
Weak lensing analysis measures a coherent distortion pattern of background galaxy images 
caused by the gravitational field of the system and thus avoids any of the assumptions mentioned above \citep[e.g.,][]{Bartelmann01}.
However, the weak-lensing signal is obtained by averaging over a certain number of background galaxies, 
and thus, only the mass information over a scale of several arcminutes is obtained.  
Previous weak lensing studies or joint strong- and weak-lensing studies \citep{Broadhurst05,Okabe10b,Okabe11,Okabe13,Oguri10b,Oguri12,Umetsu08,Umetsu10,Umetsu11,Applegate12,Hoekstra12} 
mainly focused on clusters at redshift higher than approximately $0.15$, 
because of good lensing efficiency and high-quality imaging data obtained using wide-field cameras 
mounted on ground-based telescopes, 
such as the Subaru Prime Focus Camera \citep[Suprime-Cam;][]{Miyazaki02} on the 8.2m Subaru Telescope. 
However, it is very difficult to detect subhalos using weak lensing analysis of clusters at $z\sim0.2$ 
with masses on the order of $10^{12}h^{-1}M_\odot$, 
because the apparent truncation size ( $\sim0\farcm2-0\farcm5$ or less) of these subhalos
 is too low to be detected  in the lensing signal.
Stacking lensing studies for member galaxies help to overcome this disadvantage 
\citep{Natarajan04,Natarajan07,Natarajan09,Limousin05,Limousin07}.
It computes lensing distortion signals centered around member galaxies, 
and, thus, is independent of lensing selections, which increases 
an signal-to-noise ratio (S/N) in lensing signals due to the large sample size.
Hence, 
the stacking analysis enables us to measure a mean mass of subhalos associated with member galaxies.
However, a correlation between the luminosity and mass of the subhalos must be assumed 
to derive a subhalo mass function and/or conduct statistical studies of subhalos.

Weak-lensing studies of very nearby clusters ($z\simlt0.1$) overcome the problems described above.
In contrast to weak-lensing studies of clusters at $z\sim0.2$, 
there are three significant advantages in the analysis of dark matter subhalos.
First, a large apparent size enables us to easily resolve less massive subhalos inside the clusters.
Second, subhalos are sufficiently separated from the main cluster center and other subhalos to 
ignore their lensing contamination in subhalo mass measurements.
Third, a large angular scale provides a correspondingly large number of background galaxies, which 
leads to low statistical errors and compensates for low lensing efficiency to achieve a high S/N.
This last advantage also plays an important role in cluster mass measurements.
For example, the Coma cluster is at redshift $z_c=0.0236$, 
with a large apparent size  $\sim7$ times larger than that of clusters at $z\sim0.2$.
We thus use the area of $50-100$ square minutes or more in weak-lensing mass measurements of subhalos 
with masses  greater than $\sim10^{12}h^{-1}M_\odot$.
Indeed, \cite{Okabe10a} has demonstrated the power of weak-lensing analysis of the Coma cluster 
and discovered less massive subhalos.

Here we report the results of a 4.1 deg$^{2}$ weak gravitational lensing survey of subhalos in the Coma cluster 
by 18 pointing observations ($R_{\rm c}$ and $V$ bands) 
using the Subaru/Suprime-Cam to directly measure subhalo masses and their mass function. 
This paper is a continuation of our previous work \citep{Okabe10a}, which used archival Subaru/Suprime-Cam data with the $R_{\rm c}$ band. 
The archival data covers the central and the south-west regions (two pointings; see also Figure \ref{fig:pointing}),
with a total area of $\sim0.5$ deg$^{2}$.
Our new data significantly improves the quality of the weak-lensing analysis.
First, the data covers an area to the outskirts of the cluster, 
which enables us to study the radial dependence of subhalo properties.
Area fractions, for the previous and new data, respectively, account for $\sim10\%$ and $\sim80\%$ within $r_{200}$ inside of which the mean interior density is
$200$ times the critical mass density at the cluster redshift.
Second, the exposure time is deeper than 
the effective one used in the previous weak-lensing analysis ($\sim16$ minutes;Table \ref{tab:data}),
which increases the number of background galaxies and thus suppresses the noise for the intrinsic ellipticity.
Third, we used two filters to secure the background galaxies, 
avoiding contamination of unlensed member/foreground galaxies in our shear catalog.
Therefore, this new data 
enables us to conduct a systematic survey for cluster subhalos, for the first time.
We describe the details of data analysis in Section \ref{sec:data}, including shape measurements, 
background and member selections and modeling of background lensing signals.
In Section~\ref{sec:WLsubhalos}, we define the subhalos from lensing signals, measure model-independent projected mass, 
conduct stacked lensing analyses and evaluate 
systematic errors including the purity of the subhalo sample.
We also present a galaxy-galaxy lensing study for luminous member galaxies in Section~\ref{sec:stacked_lum}, 
which is complementary to and independent of the analyses in Section~\ref{sec:WLsubhalos}.
We describe measurement of the main cluster mass  in Section~ \ref{sec:main_mass}.
Finally, we discuss the subhalo mass function, subhalo properties, and future studies in Section \ref{sec:dis}.
The conclusions are stated in Section \ref{sec:con}.
Throughout this paper, we use the cosmology of $\Omega_{m0}=0.27$, $\Omega_\Lambda=0.73$ and $H_0=100h~{\rm km~s^{-1}~Mpc^{-1}}$. One arcmin corresponds to $20~h^{-1}{\rm kpc}$.

\section{Data Analysis} \label{sec:data}

\subsection{Survey Observation and Image Processing} \label{subsec:obs}

\begin{figure}
\epsscale{1}
\plotone{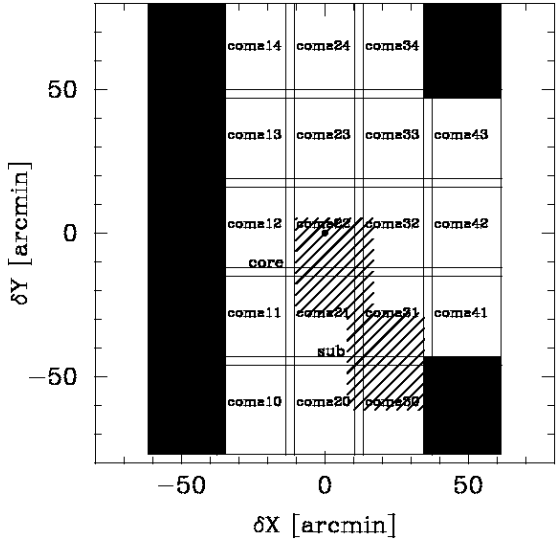}

\caption{Cadence design for the Coma cluster subhalo survey. 
The horizontal and vertical axes are R.A. and decl. with an offset distance from NGC 4874, in unit of arcmin.
The dataset name (Table \ref{tab:data}) is indicated in the middle of each pointing.
The pointings have an overlap of 2 arcmin, with a total survey area of $4.1$ deg$^2$.
Hatched regions represent areas of archival data (core and sub).
A summary of the imaging data is shown in Table \ref{tab:data}.
}
\label{fig:pointing}
\end{figure}

We observed the Coma cluster using the Suprime-cam \citep{Miyazaki02} at the Subaru 8.2-m telescope, 
in $R_{\rm c}$ and $V$ bands, in 2011 March and April.
The $R_{\rm c}$ band data is used for the wide-field weak lensing analysis, 
and combined with $V$ band data to minimize contamination of the member and foreground galaxies in the shear catalog.
The survey is covered by a mosaic of 18 pointings, specifically, coma10 ... coma43, 
as shown in Figure \ref{fig:pointing}.
Each pointing overlaps by 2 arcmin. 
The total survey area is  $4.1~{\rm deg}^{2}$.
A maximum projected radius from the brightest cluster galaxy, NGC 4874, reached $\sim100\farcm$ 
which is comparable to the cluster virial radius $\rvir$ (Section \ref{sec:main_mass}).
The $R_{\rm c}$ band data of coma30 was recollected due to the low number of background galaxies.
The typical exposure times for $R_{\rm c}$ and $V$ bands are $24.5$ and $\sim14.0$ minutes (Table \ref{tab:data}).
We also used two $R_{\rm c}$ imaging datasets retrieved from  Subaru archival data
(SMOKA\footnote{http://smoka.nao.ac.jp/index.jsp}).

We used the standard pipeline reduction software for the Suprime-Cam, 
SDFRED \citep{Yagi02,Ouchi04} modified for the new CCD, 
for flat-fielding, instrumental distortion correction, differential
refraction, point-spread-function (PSF) matching, sky subtraction and stacking. 
The seeing for each pointing is shown in Table \ref{tab:data}.
An astrometric calibration was performed using point sources from the Two Micron All Sky Survey catalog \citep{2MASS06}.
The typical residual values are no larger than the CCD pixel size.
Photometric calibration was carried out by fitting point sources detected in each dataset with 
stars from Sloan Digital Sky Survey (SDSS) DR8 photometry \citep{SDSSDR8}, 
taking into account the difference between their sensitivities.
The archival data obtained using the previous CCDs was reduced by the same procedure 
using the SDFRED for the previous CCDs.

\begin{table}
\caption{Imaging data} \label{tab:data}
\begin{center}
\begin{tabular}{ccccc}
\hline
\hline
Name\tablenotemark{a} & $R_{\rm c}$\tablenotemark{b}
     & $V$\tablenotemark{b}
     & Seeing\tablenotemark{c}
     & $\bar{r}_h^*$\tablenotemark{d} \\
     & [min]
     & [min]
     & [arcsec]
     & [arcsec]  \\
\hline
coma10    & 24.5
          & 13.75
          & 0.93
          & $0.48$ \\
coma11    & 24.5
          & 13.75
          & 0.65
          & $0.34$  \\
coma12    & 24.5
          & 13.75
          & 0.63
          & $0.32$ \\ 
coma13    & 24.5
          & 13.75
          & 0.63
          & $0.31$ \\
coma14    & 24.5
          & 13.75
          & 0.65
          & $0.33$ \\ 
coma20    & 24.5
          & 14.75
          & 0.75
          & $0.39$ \\
coma21    & 24.5
          & 13.75
          & 0.59
          & $0.29$ \\
coma22    & 24.5
          & 13.75
          & 0.75
          & $0.39$ \\
coma23    & 24.5
          & 13.75
          & 0.63
          & $0.33$ \\
coma24    & 24.5
          & 13.75
          & 0.57
          & $0.28$ \\
coma30\tablenotemark{e}    & 24.5
          & 13.92
          & 0.69
          & $0.36$ \\
coma30\tablenotemark{f}    & 24.5
          & 13.92
          & 0.67
          & $0.35$ \\
coma31    & 24.5
          & 13.75
          & 0.61
          & $0.30$  \\
coma32    & 24.5
          & 13.75
          & 0.71
          & $0.37$ \\
coma33    & 24.5
          & 13.75
          & 0.83
          & $0.54$ \\
coma34    & 24.5
          & 13.75
          & 0.70
          & $0.36$\\
coma41    & 24.5
          & 14.58
          & 0.72
          & $0.38$ \\
coma42    & 24.5
          & 14.58
          & 0.72
          & $0.38$ \\
coma43    & 24.5
          & 13.75
          & 0.76
          & $0.40$ \\
core\tablenotemark{g}  & 42.0
          & -
          &  0.81
          & $0.41$ \\
sub\tablenotemark{g}   & 16.0
          & -
          & 0.83
          & $0.41$ \\
\hline
\end{tabular}
\tablecomments{
\tablenotemark{a}: Dataset 
\tablenotemark{b}: Exposure times in $R_{\rm c}$ and $V$ bands, respectively.
\tablenotemark{c}: The seeing FWHM in unit of arcseconds, for $R_{\rm c}$ band.
\tablenotemark{d}: The median stellar half-light radius in unit of arcseconds, for $R_{\rm c}$ band.
\tablenotemark{e} $R_{\rm c}$ band data taken in 2011 March 1st.
\tablenotemark{f} $R_{\rm c}$ band data taken in 2011 March 30th.
\tablenotemark{g} Data retrieved from SMOKA.
}
\end{center}
\end{table}

\subsection{Weak Lensing Distortion Analysis}\label{subsec:wl}

The weak lensing measurements follow \cite{KSB}, referred to as the KSB+ method, 
which uses the IMCAT package 
with some modifications, similar to \cite{Umetsu10,Oguri12,Okabe13}.
Image ellipticity is measured from the weighted quadrupole moments of the surface brightness 
of objects detected in the $R_{\rm c}$ band imaging data (Table \ref{tab:data}).
The anisotropic PSF correction is conducted in the same manner as \cite{Okabe10a,Okabe10b,Okabe11,Okabe13}.
We select bright unsaturated stars in the half-light radius, $r_h$, and magnitude plane to 
estimate the stellar anisotropy kernel,
$q^*_{\alpha} = (P^*_{{\rm sm}})^{-1}_{\alpha \beta}e_*^{\beta}$, 
where $P_{\rm sm}^{\alpha\beta}$ is the {smear polarizability} matrix,  and
$e_{\alpha}$ is the image ellipticity. Quantities with an asterisk denote those for stellar objects. 
Following the KSB method, PSF anisotropy is corrected with the equation
\begin{equation} 
e'_{\alpha} = e_{\alpha} - P_{\rm sm}^{\alpha \beta} q^*_{\beta}. 
\label{eq:qstar}
\end{equation}
We estimate $q_*^{\alpha}(\btheta)$ at each galaxy position, $\btheta$, 
using a fitting function of second-order bi-polynomials of the vector $\btheta$ 
with iterative $\sigma$-clipping rejection. 
The data region is then divided into several rectangular blocks
based on the typical coherent scale of the measured PSF anisotropy pattern.
A number of tests were performed to assess the anisotropic PSF correction 
(see details in Appendix \ref{app:e_pattern}).
To estimate systematic residuals caused by imperfect PSF correction,
we computed an auto-correlation function for the stellar ellipticities
and a cross-correlation function for the ellipticities of galaxies and stars,
before and after the correction, respectively.
Although the auto correlation and the cross-correlation functions for raw ellipticities before the correction 
are highly corrected to the order of $10^{-5}-10^{-4}$, the residual/corrected ellipticities show no correlation,
which supports the accuracy of the anisotropic PSF correction.
x

Next,  the isotropic smearing effect of galaxy images is corrected to 
estimate the reduced distortion signal, $g_\alpha$, 
\begin{eqnarray}
g_\alpha&=& (P_g^{-1})_{\alpha\beta} e'_{\beta}, \label{eq:raw_g}
\end{eqnarray}
where $P^g_{\alpha\beta}$ is the pre-seeing shear polarizability tensor.
The measurement of $P^g_{\alpha\beta}$ is very noisy for individual faint galaxies 
because of its nonlinearity \citep{Bartelmann01}, which may result in 
a systematic bias in weak-lensing distortion measurements.
We therefore calibrate $P^g_{\alpha\beta}$ using the following procedures, 
in a similar way as \cite{Umetsu10} and \cite{Oguri12}.
The polarizability tensor is first computed as a scalar polarizability, 
$(P_{g})_{\alpha\beta}=\frac{1}{2}{\rm tr}[P_g]\delta_{\alpha\beta}$.
We then compute a median for $(P_{g})_{\alpha\beta}$ in $r_g$, 
with an adaptive grid to assemble as uniformly as possible.
Here, $r_g$ is the Gaussian smoothing radius used in the KSB method.
A sample of galaxies satisfies the following conditions to suppress the noise:  
a detection significance level of $\nu>30$,
a size condition of $r_h>\bar{r}_h^*+\sigma(r_h^*)$ and $r_g>\bar{r}_g^*+\sigma(r_g^*)$
and a positive raw $P_g$. 
Here, $\bar{r}_h^*$ ($\sigma(r_h^*)$) and $\bar{r}_g^*$ ($\sigma(r_g^*)$) are the median (rms dispersion)
of half-light radii and Gaussian smoothing radii for the stars selected above.
We interpolate the polarizability tensor for individual galaxies as a function of $r_g$.
A similar interpolation for the absolute value of the ellipticity, $|e|$, is also applied.
We use galaxies for the shear catalog with $\nu>10$ and the same size cut as in the calibration.
An rms error of the shear estimate, $\sigma_g$, is computed from 50 neighbors in the magnitude-$r_g$ plane.
We also assign the weight function for individual objects.
\begin{eqnarray}
w_g= \frac{1}{\sigma_g^2+\alpha^2} \label{eq:w_g}
\end{eqnarray}
where $\alpha$ is the softening constant variance representing the scatter due to the intrinsic ellipticity of the galaxies \citep[e.g.,][]{Hoekstra00,Hamana03,Okabe10b,Umetsu10,Oguri12}.
We choose $\alpha=0.4$, which is a typical value of the
mean rms $\sigma_g$ over the background sample, 
In the limit of $\sigma_{g} \ll \alpha$, individual galaxies are uniformly weighted.
On the other hand, noisy objects, such as fainter objects, are less weighted.

To check shear calibration, 
we use a number of realistic images for which the field-of-view is comparable to the Subaru/Suprime-cam
 (kindly provided by M. Oguri).
The mock images are generated with different seeing sizes (0.5–1.1 arcsec) and the Moffat profile 
with power slopes $3<\beta<12$, using  GLAFIC software \citep{Oguri10a}, 
as described in \cite{Oguri12}. We found that 
a multiplicative calibration bias, $m$ and an additive residual shear offset, $c$ in \cite{Heymans06} and \citep{Massey07} 
are $|m|\simlt 0.03$ and $|c|\simlt 2\times 10^{-4}$, respectively, for our typical seeing $\sim0\farcs7$.

We then combine the shear catalog constructed from individual images.
For the overlapping regions, since the same galaxies are detected in different images, 
we estimate weighted averages of their position and shear with $w_g$.
We compared reduced shear for $1.8\times10^5$ overlapping galaxies and confirmed that the deviation, 
$\Delta g_\alpha=(-2.16\times10^{-6}\pm6.6\times10^{-4},1.93\times10^{-6}\pm6.6\times10^{-4})$, is negligible.
Using this approach, the number of background galaxies is $\sim6.7\times10^5$.

\subsection{Photometry and Background Selection} \label{subsec:photo}

A secure selection of background galaxies in the color-magnitude plane was used 
because contamination by unlensed member or foreground galaxies in the shear catalog dilutes the weak-lensing signals, leading to an underestimation of the gravitational lensing mass, mainly for the central regions \citep{Broadhurst05,Okabe10b,Okabe13}.

Photometric catalogs were constructed from the mosaic images using SExtractor \citep{SExtractor}.
The SExtractor parameters are optimized for faint galaxies for shape measurements.
We compute the total magnitude for each object in the AB-magnitude
system using the MAG\_AUTO parameter and color using the MAG\_APER parameter.
For the color measurements, we degraded the seeing to the worst image.
The aperture diameter for the MAG\_APER parameter is $1.5$ times  the seeing FWHM.
The overlapping galaxies serve as a monitor of the offset in the magnitude.
We introduced an additional parameter in each data field to describe the offset using two magnitudes
and calibrated them simultaneously fitting bright objects with magnitudes  less than $22$ mag.
The measurement scatter for faint galaxies ($R_{\rm c}>24$~mag) is typically less than $0.1$ mag.
The magnitudes and colors for objects are estimated using weighting measurement errors.
We then match the shear and SExtractor catalogs.

\begin{figure}
\epsscale{1.}
\plotone{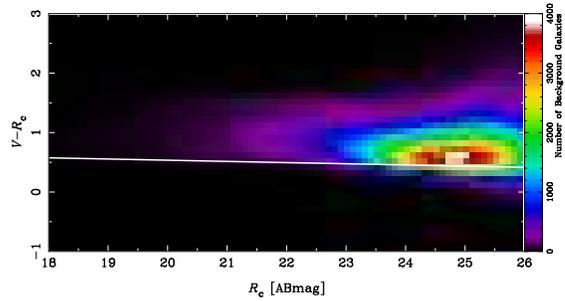}

\caption{Color-magnitude diagram. 
The color shows the number of background galaxies in each pixel ($0.1\times0.1$). 
The white solid line denotes the red-sequence of member galaxies, 
fitted to a linear function for the bright galaxies.
}
\label{fig:CMR}
\end{figure}

The red-sequence of member galaxies is fitted to a linear function, 
using luminous galaxies ($R_{\rm c}<18$~mag).
We then define the background galaxies with colors redder 
than the red-sequence in the magnitude range of $20$ mag $<R_{\rm c}<26$ mag (Figure \ref{fig:CMR}).
The number of background galaxies is reduced to $\sim6\times10^5$ after the color cut
but remains $30-60$ times higher than those of clusters at $z\sim0.2$, obtained by
previous studies using two path-band filters \citep{Okabe10b}. 
The number density, $n_g\simeq41.3~{\rm arcmin}^{-2}$, 
is also from two to eight times higher
than both those for clusters at $z\sim0.2$, and 
for our previous analysis of the Coma cluster \citep[][$n_g\simeq23{\rm~arcmin}^{-2}$]{Okabe10a}.
Thus, we can use a correspondingly large number of background source galaxies 
for nearby cluster weak-lensing analysis, for the following two reasons.
First, since the colors of red-sequence galaxies in clusters becomes more blue with decreasing redshifts,  
the number of galaxies, with colors are redder than those of member galaxies, increases.
Second, the large area encompassed by the nearby cluster increases the number of background galaxies.
Even if member galaxies are contained in the background catalog,
the dilution effect in lensing signals could be ignored because the ratio
of thousands of
member galaxies to the millions of  background galaxies is negligible.
Thus, weak-lensing analysis of a nearby cluster has a great advantage to compensate for low-lensing efficiency.

\subsection{Mean Lensing Depth}

Since the redshifts of individual galaxies in the shear catalog are not available,
we estimated the mean source redshift using a statistical approach.
The lensing signal depends on the source redshifts through the distance ratio. 
As a reference, we used the COSMOS photometric redshift catalog \citep{Ilbert09}
estimated by combining 30 broad, intermediate and narrow bands. 
Because the $R_{\rm c}$ band is not available in the COSMOS catalog, and
we converted magnitudes from $R_{\rm c}$ to $i'$ bands, based on the filter sensitivities of the Suprime-Cam.
The probability function of redshift, $dP_{\rm WL}/dz$, 
for our background galaxies selected by the color-magnitude plane (Section \ref{subsec:photo}) 
is computed by matching with the COSMOS photometric redshift, with a statistical weight of $w_g$.
The mean distance ratio is given by
\begin{eqnarray}
\langle D_{\rm ls}/D_{\rm s}\rangle=\int_{z_c}dz d P_{\rm WL}/dz D_{\rm ls}/D_{\rm s},
\end{eqnarray}
where $D_{\rm s}$ and $D_{\rm ls}$ 
are the angular diameter distance to the sources and between the cluster (lens) and the sources, respectively.
We obtain $\langle D_{\rm ls}/D_{\rm s}\rangle=0.9554$. 
The mean source redshift, $\langle z_s \rangle =0.61$, is slightly lower than that ($\langle z_s \rangle\sim0.7-0.8$) 
for clusters at $z\sim0.2$ \citep{Okabe13}, because we include many background galaxies at lower redshifts.
The mass estimate for nearby systems do not strongly 
depend on the redshift distribution of background sources.

\subsection{Luminous Member Galaxies} \label{subsec:member}

We defined luminous member galaxies identified in spectroscopic observations in order to compare the mass properties.
Luminous galaxies with a  magnitude  brighter than $i'<18~{\rm mag}$, were retrieved from SDSS DR8 \citep{SDSSDR8},
in a  $4\times4$ deg$^2$ region centered on NGC4874.
Furthermore, member galaxies were selected 
within the redshift range of $|z-z_c|<\sigma_v(1+z_c)/c$ and $\sigma=3000~{\rm kms^{-1}}$, 
where $c$ is the velocity of light.
To complete the catalog of member galaxies on the bright end, 
we also checked the redshifts of galaxies in NED\footnote{http://ned.ipac.caltech.edu/}.
If they  satisfied  the above conditions, they were added to the catalog. 
Luminosities of individual galaxies are estimated from apparent magnitudes 
using the k-correction for early-type galaxies, assuming a single redshift of $z_c$.

\subsection{Model of LSS Lensing} \label{subsec:LSSlens}

Weak-lensing mass measurements for clusters at low redshift suffer from lensing signals of
background galaxies between the cluster and the source redshifts, referred to as uncorrelated LSS lensing.
The three-dimensional, inhomogeneous mass distribution causes a locally strong shear pattern, 
which potentially gives biases in detection and mass measurements of localized objects, such as subhalos.
We therefore quantified the uncorrelated LSS lensing effect on each galaxy in the shear catalog,
using the luminosity and photometric redshift retrieved from SDSS DR8 \citep{SDSSDR8}, 
following \cite{Okabe10a}.
Galaxies are selected with magnitudes $i'<24$ mag and photometric redshifts between the cluster,
$z_{\rm ph}-z_c>\delta z=\sigma_{v}(1+z_c)/c \simeq 0.01$, and the source redshift,
where $z_{\rm ph}$ is the photometric redshift of each galaxy.
Galaxies spectroscopically identified as member galaxies (Section \ref{subsec:member}) were excluded.
Masses of individual galaxies are estimated using galaxy-galaxy lensing results from SDSS data \cite{Guzik02}.
Using the mass-to-light ratio in each band ($u'g'r'i'z'$) derived by stacked lensing analysis of galaxies,
the luminosity is converted into the mass.
The masses estimated with different bands are used to cross-check and calibrate systematic errors in
the mass-luminosity scaling relation utilized here.
Since  uncorrelated LSS lensing is obtained by integrating the effect of light deflections due to galaxies at different 
redshifts along the line-of-sight, the best-fit scaling relations in mass estimates for individual objects were used
in order to quantify an average LSS lensing effect.
The interior mass structure of each halo is assumed to be a universal mass profile found in numerical simulations,
 referred to as Navarro–Frenk–White \citep[NFW;][]{NFW96,NFW97}.
The NFW mass model is described by two parameters, the mass and concentration (see details in Appendix 
\ref{app:massmodel}).
It is well known that there is a correlation between mass and halo concentration \citep{Bullock01} 
predicted by the hierarchical structure formation scenario.
We use the mass-concentration relation obtained from recent numerical simulations based on {\it WMAP5} cosmology parameters \citep{Duffy08}
to describe the internal structure.
The tangential distortion signals of individual galaxies are computed on all source galaxies.
We found that the shear estimated in the $r'$ band is consistent with that in the $z'$ band,
but the estimates in the $u'g'i'$ bands are systematically different, 
as found in \cite{Okabe10a}.
The LSS lensing model based on the galaxy-galaxy lensing result 
is therefore defined with $g^{\rm (LSS)}_{\alpha}=(g^{r'}_{\alpha}+g^{z'}_{\alpha})/2$.
If groups or other clusters exist behind the cluster, this model would fail to incorporate those effects.
Since this would bias the mass measurement of the main cluster,
this is considered separately in Sections \ref{subsec:LSSsuperposed} and \ref{sec:main_mass}.
The LSS lensing model allows us to statistically estimate the lensing signals of 
{\it real} background structure from the observing data 
and the LSS bias in the mass measurement.
We thus use the LSS lensing model for the main analysis of this paper.

We also conduct the cluster and subhalo mass measurements 
taking into account the error covariance matrix of uncorrelated large-scale structure 
along the line-of-sight, \citep[e.g.,][]{Schneider98,Hoekstra03,Umetsu11,Oguri10b,Oguri11,Okabe13}, 
instead of the above LSS lensing model.
The LSS error covariance matrix is estimated from the weak-lensing power spectrum 
\citep[e.g.,][]{Schneider98,Hoekstra03} with {\it WMAP7} cosmology \citep{WMAP07}. 
Since we fully take into account both the LSS error matrix and 
the statistical noise caused by the intrinsic shapes of the galaxies 
and the noise in the shape measurement, 
this approach is complementary to the LSS lensing model.
However, the statistical error is dominated in subhalo mass measurements 
(see Section \ref{subsec:LSSCov}),
and it is difficult to identify {\it real} background structure using the error matrix.

\section{Weak-Lensing Analysis for Subhalos} \label{sec:WLsubhalos}

\subsection{Projected Distributions of Mass and Baryons} \label{subsec:maps}

We first make maps of the lensing convergence field ($\kappa(\btheta)$),
luminosity ($l(\btheta)$) and number density ($n(\btheta)$) of member galaxies 
and the model of the LSS lensing signal ($\kappa_{\rm LSS}(\btheta)$).
In order to identify subhalos in a model-independent way, 
the projected mass distribution is reconstructed following \cite{Kaiser93} 
with a Gaussian smoothing kernel.
The details of map-making are explained in Appendix \ref{app:mapmaking}.
We adopt various smoothing scales in the range of $r_{\rm sm}=1,\dots,5$ arcmin, 
stepped by $0.1$ for $1-2$ arcmin and $0.2$ for $2-5$ arcmin, 
to optimize the detection of subhalos with various mass properties. 
The definitions of subhalos are described in the next subsection. 
We present here the correlation between mass and luminous matter on the projected distribution.

\begin{figure*}
\epsscale{1.0}
\plotone{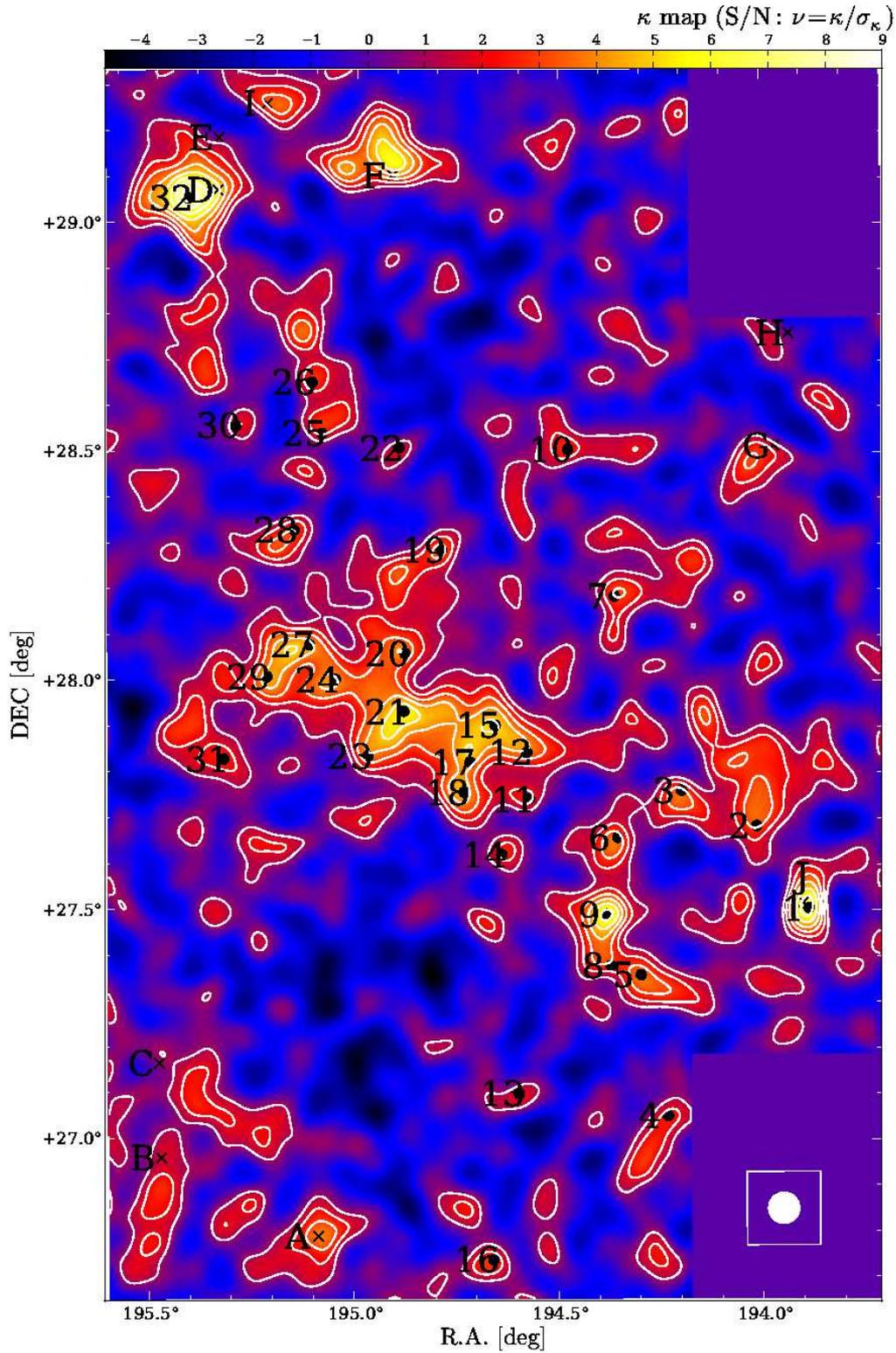}
\caption{Projected mass distribution with a smoothing scale of FWHM$=4\farcm$ 
and units of significance of $\nu=\kappa/\sigma_\kappa$. 
The shear is used without taking into account the LSS lensing effect.
The contours of significance start at $1\sigma$ with a step value of $1\sigma$.
The letters and numbers denote the names of known background systems (Table \ref{tab:bkg})
and the names of subhalos (Table \ref{tab:Msub}), respectively.
}
\label{fig:kappa}
\end{figure*}

Figure \ref{fig:kappa} shows the significance map, $\nu$, defined by $\kappa/\sigma_\kappa$, 
 with a smoothing scale of ${\rm FWHM}=4\farcm$ ($r_{\rm sm}=2\farcm4$), 
where the reconstruction error, $\sigma_\kappa$, is calculated over local background galaxies (see Appendix \ref{app:mapmaking}) 
where a typical value in this smoothing scale is $\sigma_\kappa\simeq7.7\times10^{-3}$.
The LSS lensing model was not taken into account.
The mass distribution in the central region, in which two cD galaxies NGC 4874 
($\alpha$,$\delta$)=($194^\circ.898, 27.^\circ959$)
and NGC 4889 ($195.^\circ034,  27.^\circ977$) exist, is elongated in the east and west directions.
Clumpy structures are found everywhere, but anisotropically distributed.
In particular, the projected distribution of clumpy structures is 
concentrated $30-60\farcm$ southwest of NGC 4874.
Some clumpy structures are associated with background groups 
in the literature (letters in Figure \ref{fig:kappa} and Table \ref{tab:bkg}).

\begin{figure*}
\epsscale{1.1}
\plotone{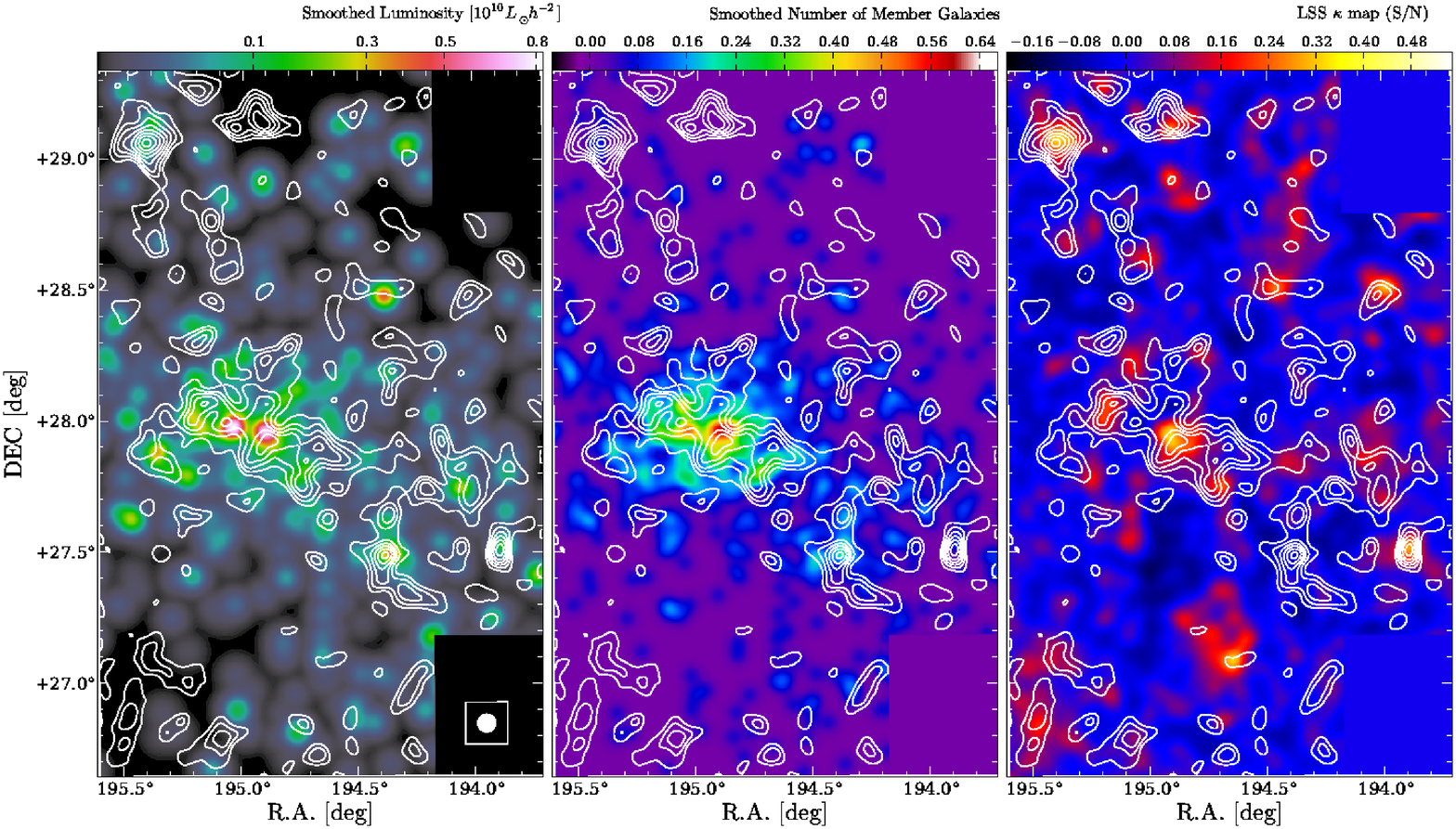}
\caption{Luminosity map for member galaxies spectroscopically identified in the SDSS DR8 and NED catalog ($i'<18$; Sec \ref{subsec:member}). The contours of the lensing $\kappa$-field are overlaid in units of $1\sigma$ reconstruction error (Figure \ref{fig:kappa}), without taking the LSS lensing model into account.
Middle : density map for member galaxies.
Right  : mass map of the LSS lensing model estimated from galaxy-galaxy lensing using photometric redshifts and luminosities for individual galaxies (Sec \ref{subsec:LSSlens}).
}
\label{fig:lummaps}
\end{figure*}

The left and middle panels of Figure \ref{fig:lummaps} 
show maps of luminosity and the number density of member galaxies, 
overlaid with the contours of the projected mass distribution.
The mass and galaxy distribution are clearly correlated with each other.
The right panel of Figure \ref{fig:lummaps} is the convergence map of the LSS lensing model.
The S/Ns of the LSS lensing map are at most $0.4\sigma$, 
and thus, the LSS lensing effect accounts for a small fraction of the observed signal.
However, the LSS model fails to describe the lensing signals around some groups behind the cluster (Table \ref{tab:bkg}).
The reason is likely that the estimation of the LSS lensing effect is based on galaxy-galaxy lensing
which fails to take group-scale or cluster-scale structures into account.
This is further elaborated in the Section \ref{subsec:LSSsuperposed}.

\begin{table*}
\caption{Known background systems appearing in the mass maps}\label{tab:bkg}
\begin{center}
\begin{tabular}{lccc}
\hline
ID\tablenotemark{a}  & Name 
                    & $z_{\rm phot}$\tablenotemark{b}
                    & Reference \\
\hline
A & MaxBCG J195.08820+26.78870 & $0.162$ & \cite{Koester07} \\
B & GMBCG  J195.47315+26.95810 & $0.219$ & \cite{Hao10}\\
C & MaxBCG J195.47907+27.16429 & $0.208$ & \cite{Koester07}\\
D & GMBCG J195.34791+29.07201  & $0.189$ & \cite{Hao10}\\
E & MaxBCG J195.34617+29.18616 & $0.170$ & \cite{Koester07}\\
F & NSC J125939+290715         & $0.189$ & \cite{Gal03}\\
G & GMBCG J193.96542+28.51557  & $0.257$ & \cite{Hao10}\\
H & MaxBCG J193.92901+28.76123 & $0.259$ & \cite{Koester07} \\
I & SDSSCGB 06685 & $0.183$ & \cite{McConnachie09} \\
J & WHL J125535.3+273104 & $0.418$ & \cite{Wen09} \\
\hline
\end{tabular}
\tablecomments{
\tablenotemark{a}:The identification of background systems in Figure \ref{fig:kappa}. 
\tablenotemark{b}:Photometric redshifts.
}
\end{center}
\end{table*}

To quantify the correlations shown in the maps, 
we compute the pixel-to-pixel coefficients 
between the mass maps ($\kappa(\btheta)$) and 
the luminosity ($l(\btheta)$) and density ($n(\btheta)$) maps for member galaxies.
The resultant coefficients for both $\langle \kappa l \rangle$ and $\langle \kappa n \rangle$ 
change from $0.57\pm0.08$ ($7\sigma$) to $0.16\pm0.02$ ($8\sigma$) with a decrease in spatial resolution.
Here, the errors are estimated by bootstrap re-sampling with 200 realizations of $\kappa$ maps,
describing that noise peaks are accidentally correlated with smoothed luminous distributions.
In short, the correlation between mass and member galaxy distributions is at the level of $7\sigma$-$8\sigma$.
We also computed the coefficients between the mass map and the LSS lensing map ($\kappa_{\rm LSS}(\btheta)$),
and found high correlation, 
$0.51\pm0.08$ ($6\sigma$) for $r_{\rm sm}=5\farcm$ 
and $0.28\pm0.02$ ($14\sigma$) for $r_{\rm sm}=1\farcm$, respectively.
The significance level is higher with an increase in resolution, 
indicating that the LSS lensing signal caused by small background objects creates a local shear pattern.

\begin{figure}
\plotone{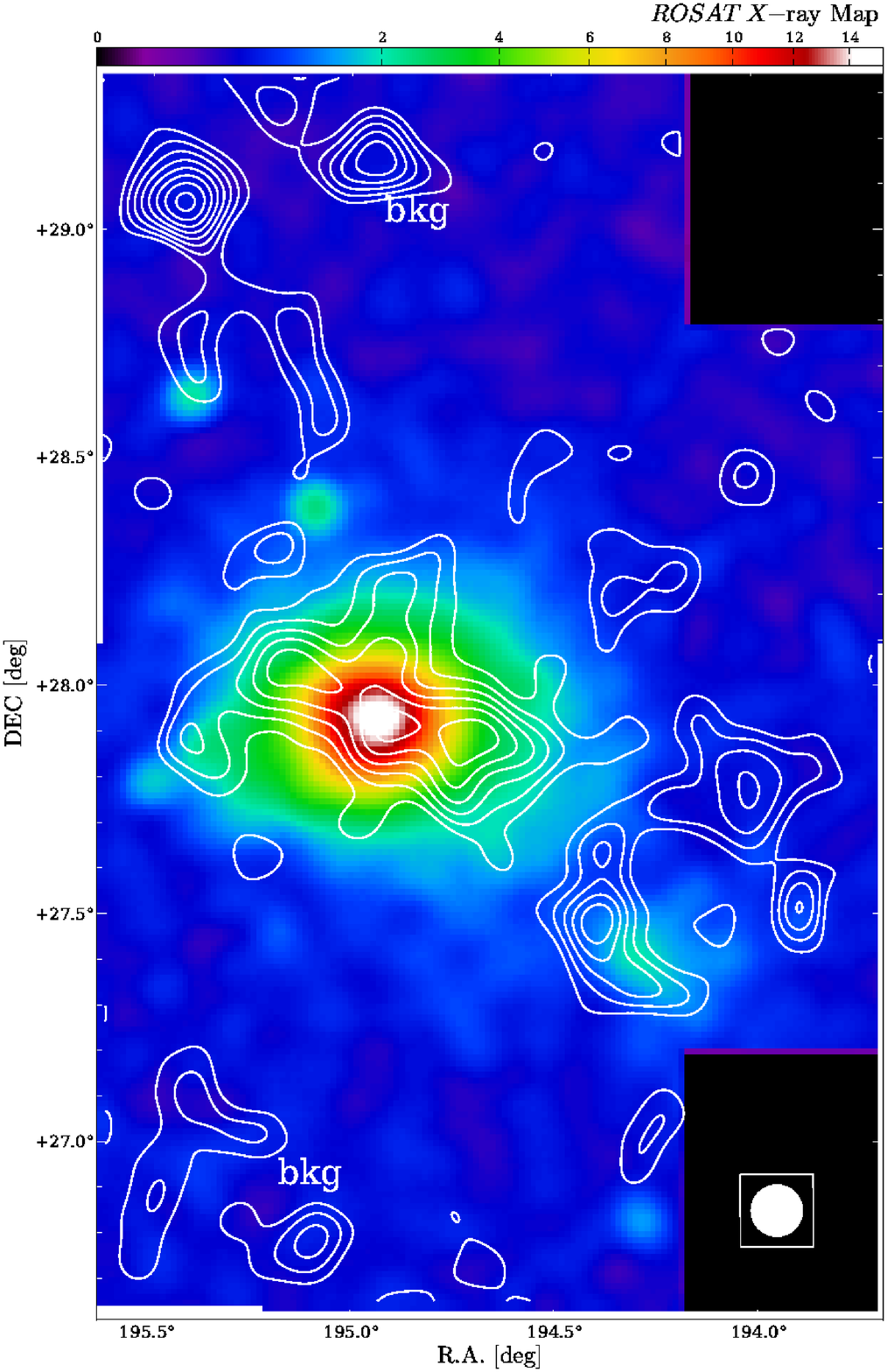}
\caption{ X-ray surface brightness distribution in the 0.1-2.4 keV band from {\it ROSAT} X-ray satellite.
The contours of the mass map are overlaid with ${\rm FWHM}=8\farcm3$, taking into account the LSS lensing model.
The contour level starts at $1\sigma$ and increases in steps of $1\sigma$. 
}
\label{fig:rosat}
\end{figure}

As shown in Figure \ref{fig:rosat}, the X-ray surface brightness distribution of {\it ROSAT} X-ray satellite 
shows an elongated X-ray distribution in the central region and 
an excess X-ray flux associated with the NGC4839 group in the southwest direction \citep{Briel92,White93,Neumann01}.
Mass contours are overlaid with a smoothing scale of FWHM$=8\farcm33$ to compare with the diffuse emission detected with a 
large PSF of the {\it ROSAT}.
Although the diffuse X-ray emission  from the NGC4839 group is associated with a clumpy mass structure,
all mass structures are not necessarily correlated with X-ray features.
This point is further discussed in Section \ref{sec:dis_fut}.

\subsection{Selection and Mass Measurements of Subhalos} \label{subsec:subhalos}

We explore, in a model-independent way, subhalo candidates 
by finding peaks in the mass maps reconstructed using several smoothing scales.
As described in Sec \ref{subsec:maps}, 
maps of the observed lensing signals are correlated with those of the LSS lensing model (Figure \ref{fig:lummaps}).
In order to securely identify cluster subhalos and accurately measure their masses,
it is crucial to minimize the contamination by the LSS lensing effect.
We therefore calibrate the reduced shear with an approximate form of $g_{\alpha}^{\rm (corr)}=g_\alpha-g_\alpha^{\rm LSS}$
to eliminate the LSS lensing effect along the line-of-sight as much as possible. 
Mass reconstructions are then repeated using the calibrated shear catalog.

Subhalo candidates with  peaks above a threshold in the mass maps are selected.
The mass maps are represented as the convolution of the lensing distortion pattern of a 
cluster mass distribution with smoothing kernels.
Therefore, the Gaussian smoothing scales used for the mass reconstruction vary from $1\farcm$ to $5\farcm$ 
in order to optimize for the detection of subhalos with various mass properties.
Here, the pixelized $\kappa$ field  changes slightly using the reconstruction kernels, 
similar to top-hat or wavelet filters.

We use a significance level, $\nu\equiv\kappa/\sigma_\kappa$, for the selection of subhalo candidates,
where $\kappa$ and $\sigma_\kappa$ are the dimensionless surface mass density and 
the reconstruction error, respectively.
Since the variance and skewness of the $\nu$ histogram in the pixels depend on the smoothing scale, 
we identify subhalo candidates above a threshold set at three times the standard deviation.
The threshold of significance  in the highest resolution corresponds to $\nu>3.4$.
We first identify subhalo candidates at various smoothing scales.
Then, two peaks appearing between two different smoothing scales are matched with the condition $d<{\rm FWHM}$, 
where $d$ is the distance between the two peaks which appeared in different scales, 
and ${\rm FWHM}$ is the full width and half the maximum of the larger smoothing scale. 
This process results in 49 subhalo candidates.
We note that \cite{Okabe10a} used a mass map with single smoothing scale (FWHM=$2\farcm$) and applied the lower threshold.
Therefore, two of the seven subhalo candidates in the previous paper \cite[][numbers 6 and 8]{Okabe10a} are below a more conservative threshold of this analysis.

Since we minimized the LSS lensing contribution by applying the galaxy-galaxy lensing model,
eight known background objects (Table \ref{tab:bkg}) are below the thresholds selected. 
However, the model does not perfectly describe the full LSS lensing effect. 
Three other peaks associated with the known background objects (Table \ref{tab:bkg}) 
are detected with the above conditions. 
One is the background object ``I'' and two peaks are around the object ``F'' (see Figure \ref{fig:kappa}). 
These objects are likely to be groups because the lensing signals are stronger 
than what is expected from the luminosity of a single galaxy.
Furthermore, there is a possibility that background groups are accidentally superimposed with cluster subhalos,
giving a systematic bias on mass estimates of subhalos.
This point is discussed in Section \ref{subsec:LSSsuperposed}.

Next, we measure the model-independent projected masses \citep[][see also Appendix \ref{app:M2D}]{Clowe00}
 for shear-selected subhalo candidates.
This measurement has several important advantages.
First, a large number of background galaxies are available,
because a projected mass within a circular aperture radius is computed by integrating source galaxies outside the radius.
The measured projected mass is a cumulative function of radius.
Thus, this approach suppresses the random noise relevant to the intrinsic ellipticity, 
compared to a tangential distortion profile, 
which averages the tangential component of all background galaxies residing in radial bins.
Second, since the measurement subtracts the background mass density surrounding subhalos,
the contribution of the main cluster mass distribution to subhalo masses is excluded.
Third, the mass density of subhalos is expected to be close to zero outside of the tidal radius, and the
measured aperture mass corresponds to the subhalo mass itself.
If the mass density profile follows the universal NFW profile \citep{NFW96,NFW97}
without any truncation radii, the aperture mass is higher than the spherical one \citep{Okabe10b}.
As expected from tidal destruction, 
the radial profile of the projected mass is saturated outside the truncation radii, $r_t$.
We measure projected masses for all the candidates. 
Since the smoothing kernel for the mass reconstructions gives rise to centroid uncertainties of the candidates, we determine the central position by choosing maximal lensing signals 
within a $8\farcm\times8\farcm$ box where the center is aligned with the map peak position.
For accurate mass measurements of subhalos with a variety of sizes, 
it is important to explore truncation radii where the projected mass profile is saturated.  
We systematically compute projected mass profiles by changing the background annulus 
and then statistically determining the truncation radii.
Here, the inner radius changes from $0\farcm7$ to $14\farcm5$ in steps of $0\farcm2$ and the width is fixed at $3\farcm$.
The projected mass $M_{\rm 2D}$ is computed from saturated values, taking into account the error covariance matrix.
The measurement method is detailed in Appendix \ref{app:M2D}.
The same analysis was repeated for different background widths 
which showed that the result does not significantly change.
Mass measurements used a considerably large number of source galaxies ($4\times10^3-2\times10^4$).
The number is comparable or less than that for main clusters at $z\sim0.2$ \citep[e.g.,][]{Okabe10b}
for which the background number densities are $n_g\sim5-20~{\rm (arcmin^{-2})}$.
Less massive subhalos which are detected inside more massive ones 
should be excluded in order to avoid double-counting these subhalos. 
We count the $i$th subhalo using two conditions of the radius $r_{{\rm t},i}>r_{{\rm t},j}$ 
and the subhalo mass $M_{{\rm 2D},i}>M_{{\rm 2D},j}$ ($i \neq j$).
The number of candidates is then reduced from 49 to 39 using this procedure.
As mentioned above, the LSS model fails to fully explain the lensing signals of background systems, 
especially on group scales. 
Furthermore, since there is a possibility to detect mass structures behind the cluster,
we conservatively select the candidates hosting spectroscopically identified member galaxies 
within their truncation radii as the cluster subhalos. 
Having applied these limitations,  32 peaks are identified as dark matter subhalos. 
Three candidates are associated with the background systems (Table \ref{tab:bkg}). 
Four candidates have no optical counter: they  
are located around $\sim70\farcm$ in the south-east direction and the north-west direction, respectively.

\begin{figure*}
\plotone{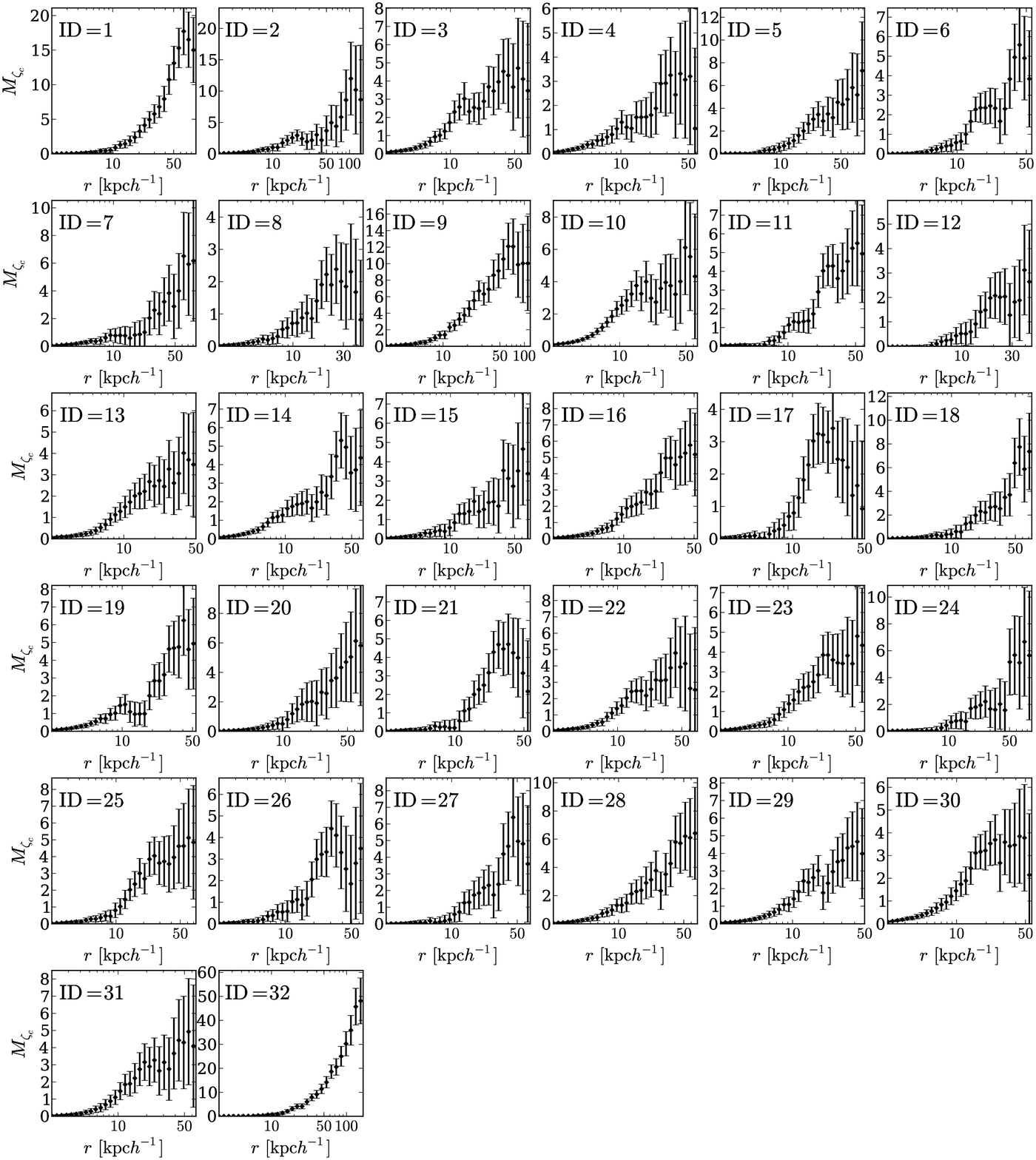}
\caption{Radial profiles of projected mass for subhalos, showing that the masses are saturated at their outer radii.
Numbers in the top-left corner denote subhalo names (see also Figure \ref{fig:kappa}).}
\label{fig:Mzeta}
\end{figure*}

These 32 subhalos are labeled by integers, in the order of right ascension. 
The resulting subhalo masses, $M_{{\rm 2D}}$, range 
from $\sim2\times10^{12}\hMsol$ to $\sim5\times10^{13}\hMsol$ (Table \ref{tab:Msub}).
As shown in Figure \ref{fig:Mzeta}, 
the radial profiles of the projected mass clearly show saturation at some outer radii.
The subhalos are widely distributed from the northeast to the southwest in the sky (Figure \ref{fig:kappa}).
Interestingly, the direction connecting between the Coma cluster and
A1367 which are parts of the Coma supercluster \citep{Gregory78} agrees roughly with the subhalo distributions.
Several massive subhalos are associated with well-known, 
spectroscopically identified groups in the cluster \cite[e.g.,][]{Mellier88,Adami05b}.
Galaxies or groups associated with subhalos are summarized with references in Table \ref{tab:Msub}. 
The cD galaxies, NGC4874 and NGC4889, are associated with subhalos ``21'' and ``24'', respectively.
The mean mass ratio reported in this paper compared to the previous paper for overlapping subhalos is 
$\langle M_{\rm new}/M_{\rm old}\rangle=1.02\pm0.54$.
We also measured the projected masses for two subhalos with peaks below the threshold in this analysis. The mean mass ratio is $\langle M_{\rm new}/M_{\rm old}\rangle=0.74\pm0.66$.
Since the number density of background galaxies in the previous analysis is about half of that reported in this analysis,
we cannot rule out the possibility that these peaks are actually above the threshold.

\begin{table*}
\caption{The properties of Subhalos} \label{tab:Msub}
\begin{center}
\begin{tabular}{lcccccl}
\hline
ID\tablenotemark{a}             & $(\rm R.A., Decl)$\tablenotemark{b}
               & $M_{\rm 2D}$\tablenotemark{c}
               & $\nu$\tablenotemark{d}
               & Representative Galaxies\tablenotemark{e}\\
               & (deg)
               & ($10^{12}h^{-1}M_\odot$)
               &
               &
               &  \\                 
\hline
1\tablenotemark{$\dagger$} & (193.885, 27.505)
   & $15.42\pm2.79$ 
   & $5.98$ 
   & NGC4807  \\
2 & (194.011, 27.685)
   & $8.79\pm4.69$ 
   & $3.55$ 
   & NGC4816 Group\tablenotemark{$\sharp$}\\
3 & (194.197, 27.763)
   & $3.71\pm1.08$ 
   & $4.61$ 
   & SDSS J125645.42+274638.0 \\
4 & (194.232, 27.053)
   & $2.89\pm1.08$ 
   & $3.51$ 
   & SDSS J125647.00+270324.9 \\
5 & (194.298, 27.360)
   & $5.00\pm2.34$ 
   & $3.86$ 
   & 2MASX J12571076+2724177 \\
6 & (194.355, 27.660)
   & $2.52\pm1.27$ 
   & $4.45$ 
   & G12 Group\tablenotemark{$\ddagger$} \\
7 & (194.361, 28.187)
   & $5.99\pm2.84$ 
   & $3.80$ 
   & UGC08071, 2MASX J12572841+2810348  \\
8  & (194.372, 27.380)
   & $1.87\pm0.73$ 
   & $3.54$ 
   & 2MASX J12573148+2723048 \\
9 & (194.381, 27.493)
   & $12.11\pm2.52$ 
   & $6.45$ 
   & NGC4839 Group\tablenotemark{$\sharp$}, G4 Group\tablenotemark{$\natural$}, NGC4842, X-ray subhalo\tablenotemark{$\flat\flat$} \\
10 & (194.477, 28.507)
   & $3.24\pm0.75$ 
   & $3.42$ 
   & 2MASX J12575392+2829594 \\
11 & (194.572, 27.745)
   & $4.13\pm0.85$ 
   & $4.03$ 
   & 2MASX J12581922+274543 \\
12 & (194.579, 27.846)
   & $2.02\pm0.78$ 
   & $3.87$ 
   & SDSS J125818.20+275054.5  \\
13 & (194.597, 27.101)
   & $2.70\pm0.77$ 
   & $3.61$ 
   & 2MASX J12581552+2705137 \\
14 & (194.640, 27.623)
   & $4.51\pm1.27$ 
   & $3.53$ 
   & NGC4853  \\
15 & (194.656, 27.905)
   & $2.96\pm1.44$ 
   & $6.90$ 
   & NGC4839 Group\tablenotemark{$\sharp$} \\
16 & (194.659, 26.738)
   & $5.03\pm1.06$ 
   & $4.19$ 
   & SDSS J125839.93+264534.2\\
17 & (194.718, 27.825)
   & $3.13\pm0.74$ 
   & $4.94$ 
   & G9 Group\tablenotemark{$\natural$}, SA 1656-030\tablenotemark{$\flat$}\\
18 & (194.732, 27.759)
   & $6.48\pm2.03$ 
   & $4.47$ 
   & G8 Group\tablenotemark{$\natural$}\\
19 & (194.790, 28.288)
   & $4.66\pm1.26$ 
   & $4.74$ 
   & SDSS J125914.99+281503.6 \\
20 & (194.879, 28.062)
   & $2.90\pm1.58$ 
   & $4.16$ 
   & 2MASX J12593141+2802478 \\
21 & (194.882, 27.936)
   & $4.29\pm1.06$ 
   & $7.23$ 
   & NGC4874(cD),part of G1 Group\tablenotemark{$\natural$}, X-ray subhalo 2\tablenotemark{$\sharp\sharp$}\\
22 & (194.895, 28.511)
   & $4.50\pm1.90$ 
   & $3.54$ 
   & 2MASX J12594129+2830257 \\
23 & (194.971, 27.837)
   & $3.75\pm1.04$ 
   & $4.26$ 
   & J194.9353+27.83393\tablenotemark{$\ddagger$}, SA 1656-054\tablenotemark{$\flat$},X-ray subhalo 3\tablenotemark{$\sharp\sharp$} \\
24 & (195.052, 28.005)
   & $5.20\pm2.40$ 
   & $4.71$ 
   & NGC4889(cD),part of G1 Group\tablenotemark{$\natural$} X-ray subhalo 1\tablenotemark{$\sharp\sharp$}\\
25 & (195.086, 28.542)
   & $3.86\pm0.95$ 
   & $3.93$ 
   & 2MASX J13002268+2834285 \\
26 & (195.111, 28.654)
   & $2.75\pm0.79$ 
   & $4.43$ 
   & SDSS J130037.14+283950.9 \\
27 & (195.115, 28.080)
   & $4.28\pm1.74$ 
   & $6.24$ 
   & SDSS J130030.95+280630.2,part of G7 Group\tablenotemark{$\natural$}\\
28 & (195.155, 28.331)
   & $5.70\pm1.68$ 
   & $3.68$ 
   & NGC4896 \\
29 & (195.220, 28.010)
   & $3.64\pm1.30$ 
   & $4.31$ 
   & NGC 4908, NGC 4908 Group \\
30 & (195.300, 28.558)
   & $3.12\pm0.66$ 
   & $4.03$ 
   & SDSS J130114.96+283118.3  \\
31 & (195.325, 27.830)
   & $2.97\pm1.42$ 
   & $3.41$ 
   &  G4 Group\tablenotemark{$\natural$}, NGC4919  \\
32\tablenotemark{$\dagger\dagger$} & (195.421, 29.054)
   & $45.95\pm7.57$ 
   & $8.35$ 
   & G15 Group\tablenotemark{$\natural$},IC 4088,2MASX J13014399+2859587\\
\hline
\end{tabular}
\tablecomments{
\tablenotemark{a} : Subhalo name
\tablenotemark{b} : Weak-lensing center in units of deg 
\tablenotemark{c} : Subhalo mass in units of $10^{12}h^{-1}M_\odot$.
\tablenotemark{d} : Maximum signal-to-noise ratio appearing in the mass maps ($\kappa$).
\tablenotemark{e} : Name of representative galaxies or optical groups.
\tablenotemark{$\dagger$} : Possibly an overlapped background structure, WHLJ125535.3+273104 (Table \ref{tab:bkg}).
\tablenotemark{$\dagger\dagger$} : Possibly an overlapped background structure, GMBCG J195.34791+29.07201 (Table \ref{tab:bkg}).
\tablenotemark{$\sharp$} : \cite{Mellier88}
\tablenotemark{$\natural$} : \cite{Adami05b}
\tablenotemark{$\flat$} : \cite{Conselice99}
\tablenotemark{$\ddagger$} : \cite{Adami09}
\tablenotemark{$\sharp\sharp$} :\cite{Andrade-Santos13}
\tablenotemark{$\flat\flat$} :\cite{Briel92}
}
\end{center} 
\end{table*}

\subsection{Stacked Lensing Analysis for Subhalos} \label{subsec:stacked_subhalos}

Next, we conducted a stacked lensing analysis for the subhalo candidates, 
which is complementary to the projected mass measurement.
The power of the stacked lensing technique is to reduce the random noise due to intrinsic ellipticities
by increasing the number of source galaxies.
Tangential profiles, even for small and less massive subhalos, can be computed and 
their average parameters can be determined with lower measurement errors. 

\begin{figure*}
\epsscale{1.2}
\plotone{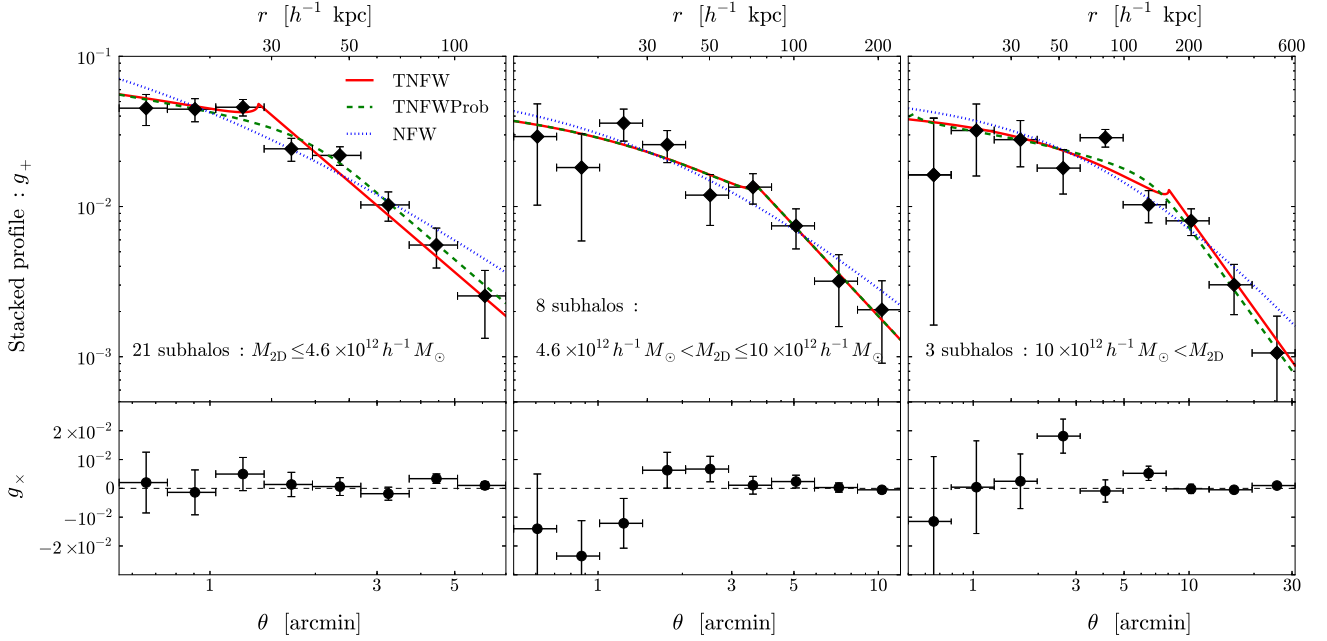}
\caption{ Mean distortion profiles obtained by azimuthally averaging the 
measured galaxy ellipticities for 32 subhalos.
The subsamples are selected with  model-independent projected masses.
Left : 21 subhalos with $M_{\rm 2D} \le 4.6\times10^{12}\hMsol$.
Middle : 8 subhalos with $4.6\times10^{12}\hMsol<M_{\rm 2D} \le 4.6\times10^{12}\hMsol$.
Right : 3 subhalos with $10\times10^{12}\hMsol\le M_{\rm 2D}$.
The profile slopes drastically change at the truncation radii.
The red solid, green dashed, and blue dotted curves are the best-fit TNFW, TNFWProb, and NFW models, respectively.
The NFW and TNFWProb models  adequately describe a sharp truncation,
while the NFW model for less massive (left) and massive (right) subsamples is strongly disfavored.
The best-fit truncation radius increases with an increasing mass of the subsample.  
}
\label{fig:stacked_g+_massbin}
\end{figure*}

We first divide the subhalos into three subsamples based on the model-independent projected masses 
of $M_{\rm 2D} \leq4.6\times10^{12}\times 10^{12}\hMsol$, 
$4.6\times10^{12}\hMsol \leq M_{\rm 2D} \leq10^{13}\hMsol$ and $10^{13}\hMsol\leq M_{\rm 2D}$.
The mass thresholds are chosen by a subhalo mass function which is described in Sec \ref{sec:mass_func}.
The number of subhalos are 21, 8, and 3, progressing from less massive to more massive subhalos.
Figure \ref{fig:stacked_g+_massbin} shows that the tangential component
is positive (top panel) and the $45^\circ$ rotated component is positive and negative 
in random order (bottom panel). 
The mean of the $45^\circ$  rotated component over the radial range 
is consistent with a null signal, within the error of the mean.
A sharply truncated feature is found in the stacked signal of the tangential profile.
Outside the breaks, the profiles are proportional to $\theta^{-2}$, which indicates 
that the mass density becomes zero.
We emphasize that such a feature was not found in massive clusters \citep[e.g.,][]{Okabe10b,Okabe13} 
but was identified in the stacked lensing profile for subhalos in our previous paper \citep{Okabe10a}.
We did not apply any rescaling procedures to the radial bins corresponding to the lensing signals,
because this mass weight scheme biases the mass estimates, as described by \cite{Okabe13}.
Here, an off-centering effect \citep{Yang06} in the lensing signals from the main cluster mass is negligible 
because the mean projected distance from the cluster center is much larger than the maximum radius for the plots.
The stacked lensing profiles are then fit with NFW, TNFW and TNFWProb models (Appendix \ref{app:massmodel}). 
Here, the TNFW model is a truncated NFW \citep{Takada03,Okabe10a}.
The TNFW model is an extreme case of truncation models, where
the mass density outside the truncation radius is zero as described in Appendix \ref{app:massmodel}.
The TNFWProb model is the TNFW model taking into account a probability function for the truncation radius
which is assumed to be Gaussian with the mean, $\langle r_t \rangle$, and the standard error $\sigma_{r_t}$.
Given this function, we measure a mean subhalo mass $\langle M_{\rm sub} \rangle$.
In the process of fitting the model, we propagate systematic errors by possible 
background structures around subhalos ``1'' and ``32'', which is described in \ref{subsec:LSSsuperposed}.
As expected from the clear truncation feature, 
the mean tangential profiles are well fitted using the TNFW and TNFWProb models (Figure \ref{fig:stacked_g+_massbin}).
The best-fit subhalo masses and truncation radii are listed in Table \ref{tab:mass_stacked}.
The best-fit mass and truncation increase with increasing model-independent projected masses.
If the subhalo sample was entirely from false peaks, these characteristic features could not be found.
We compute the significance probability, $Q(\nu/2,\chi^2_{\rm min}/2)$, 
that the data shows as a poor fit, as the observed value of $\chi^2_{\rm min}$ by chance.
The NFW models for the lowest- and highest- 
mass samples are rejected with a significance level of $Q=4\%$.
Thus, the NFW model is inadequate to describe the tangential shear profile with breaks.
The mean ratio between the best-fit subhalo masses and the mean projected mass of the subsamples, $\langle M_{\rm 2D} \rangle$,
are $\langle \langle M_{\rm 2D} \rangle / M_{\rm sub} \rangle=1.02\pm0.12$ for the TNFW model 
and $\langle \langle M_{\rm 2D} \rangle / \langle M_{\rm sub}\rangle \rangle=1.06\pm0.15$ for the TNFWProb model, respectively,
These two models are in good agreement.

\begin{table*}
\caption{Best-fit mass parameters for TNFW and TNFWProb models for lensing-selected subhalos} \label{tab:mass_stacked}
\begin{center}
\begin{tabular}{l@{\hspace{-0.3mm}}c@{\hspace{-0.3mm}}ccccc@{\hspace{-0.2mm}}c@{\hspace{-0.5mm}}cc}
\hline
Sub-sample\tablenotemark{a}         & $N_{\rm sub}$\tablenotemark{b}
               & $M_{\rm sub}$\tablenotemark{c} 
               & $r_{t}$\tablenotemark{c} 
               & $\langle M_{\rm sub}\rangle$\tablenotemark{d}  
               & $\langle r_{t} \rangle$\tablenotemark{d} 
               & $\sigma_{r_{t}}$\tablenotemark{d} 
               & $\langle L_{i'} \rangle$\tablenotemark{e}  
               & S/N\tablenotemark{f} 
               & ${\mathcal P}_{\rm fake}$\tablenotemark{g} \\
               &
               & {\scriptsize ($10^{12}h^{-1}M_\odot$)}
               & {\scriptsize ($\hkpc$)}
               & {\scriptsize ($10^{12}h^{-1}M_\odot$)}
               & {\scriptsize ($\hkpc$)}
               & {\scriptsize ($\hkpc$)}
               & {\scriptsize($10^{10}h^{-2}L_{i',\odot}$)}
               & 
               &  \\
\hline
{\scriptsize $-4.6\times10^{12}\hMsol$}\tablenotemark{$\dagger$}    & $21$
                  & $2.91_{-0.29}^{+0.28}$
                  & $27.48_{-1.91}^{+2.43}$
                  & $3.53_{-0.44}^{+0.49}$
                  & $35.57_{-4.53}^{+4.78}$
                  & $9.97_{-4.38}^{+5.41}$
                  & $2.11$
                  & $13.69$
                  & $0.019$ \\
{\scriptsize $(4.6-10)\times10^{12}\hMsol$}   & $8$
                  & $5.93_{-1.11}^{+1.43}$
                  & $72.79_{-15.07}^{+25.42}$
                  & $5.95_{-1.12}^{+1.66}$
                  & $73.18_{-12.20}^{+33.06}$
                  & $<1.75$
                  & $5.24$
                  & $8.67$                 
                  & $0.011$ \\
{\scriptsize $10^{13}\hMsol-$}\tablenotemark{$\dagger$}  & $3$
                  & $26.72_{-4.10-5.88}^{+4.28}$
                  & $161.15_{-22.33}^{+57.25}$
                  & $23.13_{-6.33-5.06}^{+7.37}$
                  & $127.11_{-35.22}^{+71.86}$
                  & $33.31_{-18.95}^{+18.39}$
                  & $7.49$
                  & $5.35$
                  & $2\times10^{-5}$ \\
\hline
$0-20\farcm$   & $11$
               & $3.05_{-0.58}^{+0.56}$
               & $35.10_{-4.26}^{+5.28}$ 
               & $3.05_{-0.62}^{+1.49}$
               & $35.66_{-4.65}^{+23.02}$
               & $<19.71$
               & $5.35$
               & $8.25$
               & $0.082$ \\
$20\farcm-40\farcm$             & $10$
                  & $5.00_{-0.65}^{+0.74}$
                  & $49.29_{-7.66}^{+8.76}$
                  & $5.00_{-0.65}^{+0.73}$ 
                  & $49.29_{-11.84}^{+9.77}$
                  & $<19.43$
                  & $3.47$
                  & $10.70$
                  & $0.062$ \\
$40\farcm-60\farcm$   & $8$
                  & $5.43_{-1.33-0.73}^{+1.04}$ 
                  & $65.08_{-19.51}^{+10.55}$
                  & $4.84_{-1.09-0.17}^{+1.24}$
                  & $49.56_{-11.42}^{+13.78}$
                  & $14.62_{-7.81}^{+9.62}$
                  & $1.52$
                  & $8.39$
                  & $0.023$ \\
$60\farcm-80\farcm$   & $3$
                  & $30.29_{-3.23-1.75}^{+3.21}$ 
                  & $209.69_{-13.29}^{+2.99}$
                  & $30.27_{-4.12-1.75}^{+5.22}$ 
                  & $209.31_{-39.86}^{+4.87}$
                  & $<19.00$
                  & $6.16$
                  & $7.62$ 
                  & $5\times10^{-5}$ \\
\hline
\end{tabular}
\tablecomments{
\tablenotemark{a} : Name of subsamples for subhalos in the stacked lensing analysis
\tablenotemark{b} : Number of subhalos
\tablenotemark{c} : Best-fit subhalo mass and truncation radius for the TNFW model
\tablenotemark{d} : Best-fit subhalo mass, and the average and standard error of the truncation radius for the TNFWProb model
\tablenotemark{e} : Average luminosity for associated galaxies, estimated by weighting tangential distortions, $g_+$
\tablenotemark{f} : Signal-to-noise ratio for the tangential distortion profile
\tablenotemark{g} : Probabilities that the TNFW mass and truncation radius represent false subhalos 
are within $1\sigma$ contours for the best-fit values of observed subhalos
\tablenotemark{$\dagger$} : The NFW model is strongly disfavored
}
\end{center} 
\end{table*}

\begin{figure*}
\epsscale{1}
\plotone{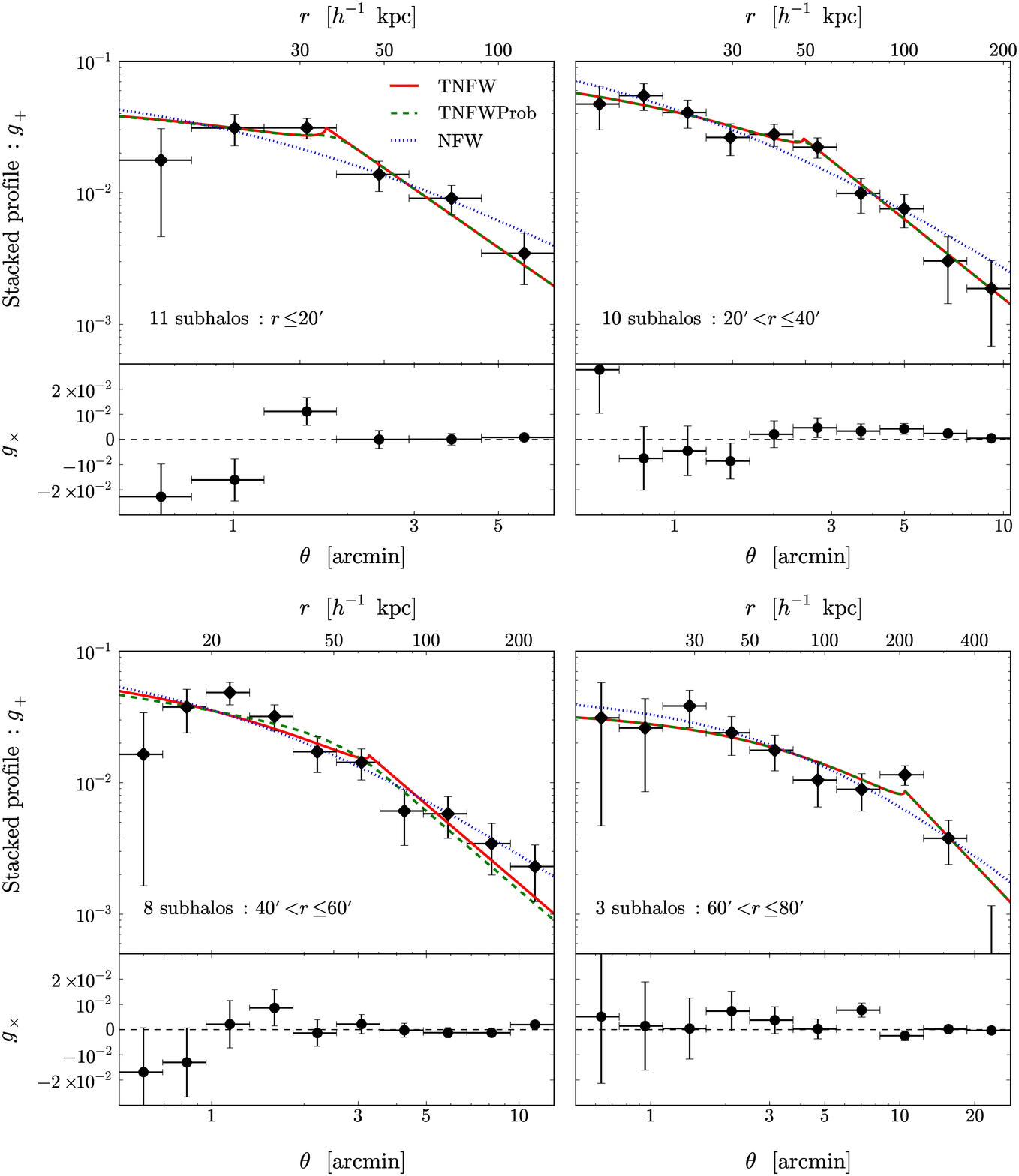}
\caption{Mean distortion profile for four subsamples of 32 subhalos, 
selected by the projected cluster-centric radii of $0-20\farcm$ (top-left), $20-40\farcm$ (top-right), 
$20-40\farcm$ (bottom-left) and $60-80\farcm$ (bottom-right).
The red solid, green dashed and blue dotted curves are the best-fit TNFW, TNFWProb, and NFW models, respectively.
}
\label{fig:stacked_g+_rbin}
\end{figure*}

Next, we repeat the stacked lensing analysis for 
four subsamples divided by the projected cluster-centric radii 
($0-20\farcm$,$20-40\farcm$,$40-60\farcm$, and $60-80\farcm$).
Since tidal destruction predicts that the truncation radii are statistically 
correlated with the three-dimensional radius, 
a stacked procedure averages out line-of-sight positions for subhalos.
This provides information regarding the dependence of mean subhalos size on the cluster-centric radius. 
Figure \ref{fig:stacked_g+_rbin} displays the mean tangential profiles with clear breaks.
The TNFW and TNFWProb models give a better to fit the stacked profiles (Table \ref{tab:mass_stacked}).
Although the NFW model fit is acceptable ($Q>10\%$), the TNFW and TNFWProb are preferred 
based on comparing the goodness-of-fit of each model.
The mean truncation radius increases as the projected radius from the cluster center increases.
The mean mass ratios are $\langle \langle M_{\rm 2D} \rangle / M_{\rm sub} \rangle=0.97\pm0.11$, 
and $\langle \langle M_{\rm 2D} \rangle / \langle M_{\rm sub}\rangle \rangle=1.00\pm0.16$ 
for the TNFW and TNFWProb models, respectively.

Previous papers \citep[e.g.,][]{Natarajan07,Natarajan09,Limousin05,Limousin07} estimated 
subhalo masses by galaxy-galaxy lensing method using a model of a pseudo-isothermal elliptical mass distribution 
(PIEMD) of which three-dimensional mass density is given by 
$\rho \propto (1+r^2/r_{\rm core}^2)^{-1}(1+r^2/r_{\rm cut}^2)^{-1}$.
Here, the core radius $r_{\rm core}$ is at the order of 100 pc \citep{Limousin05} 
and $r_{\rm cut}$ is the truncation radius. 
The tangential shear for the PIEMD model \citep{Natarajan07} 
falls as $\gamma_t \propto r^{-1}$ in the transition region ($r_{\rm core}<r<r_{\rm cut}$) and $\gamma_t \propto r^{-2}$ 
in the outer region ($r_{\rm cut}<r$). 
We also tried to fit the stacked lensing signals with the PIEMD model. 
Here we assume that the core radius is one-hundredth of the truncation radius, 
because there is no data on scales of sub kpc and thus we cannot constrain it. 
We also assume a spherical model for the simplicity.
The PIEMD model gave a poor fit because the slope of the model in the transition region ($r_{\rm core}<r<r_{\rm cut}$) 
is different from the observed profile for the mass scales of our subhalos.

\subsection{Systematic Errors}

In this section, we assess various systematic errors on the subhalo analysis, such as 
a projection effect on subhalo mass measurements (Section \ref{subsec:LSSsuperposed}),
LSS error covariance matrix (Section \ref{subsec:LSSCov}),
a probability of spurious peaks (Section \ref{subsec:stacked_fake}), selection criteria of subhalos (Section \ref{subsubsec:selection}) 
and stacking procedure (Section \ref{subsubsec:stacking}).
They are critically important for further discussion of subhalo properties such as a mass function 
(Section \ref{sec:mass_func}; construction of Figure \ref{fig:mass_func}). 
Each systematic error would have an independent effect on the mass function.
For instance, 
the projection effect would lead to a bias in subhalo mass measurements (the $x$-axis of the mass function).
A contamination of spurious peaks and selection criteria would change a shape of the mass function (the $y$-axis),
especially on small mass scales.

\subsubsection{Projection Effect} \label{subsec:LSSsuperposed}

The projection effect on lensing mass measurement of cluster subhalos,
caused by background groups accidentally superimposed along the line-of-sight, is examined here.
Although LSS modeling is quantified based on scaling relations between mass and luminosities,
it fails to fully describe massive background structures, such as groups or clusters.
This effect would lead to a bias in mass estimates, 
possibly changing the mass of the x-axis in the mass function
(Section \ref{sec:mass_func}; Figure \ref{fig:mass_func}).
As shown in Figure \ref{fig:kappa}, possible background structures, J and D (Table \ref{tab:bkg}), 
are located within two subhalo regions labeled ``1'' and ``32'' (Table \ref{tab:Msub}), respectively.
Although it is in principle very difficult to discriminate between them from the observed lensing signal,
a difference between the expected mass density profiles enables us to assess a contribution from
background structures in the observed lensing signal.
Since interior subhalos are tidally destroyed by their parent halos, 
it is expected that the mass density profile outside the tidal radius sharply declines. 
On the other hand, virialized background groups or clusters do not show such a feature 
as long as there is no neighboring massive halo.
Indeed, tangential distortion profiles for individual groups or clusters and stacked profiles
show a clear curvature as a characteristic signature of the NFW prediction and no evidence of a truncation feature
\cite[e.g.,][]{Johnston07,Okabe10b,Okabe13,Umetsu11,Oguri12,Taylor12}.
Thus, fitting models to tangential distortion profiles helps us to 
discriminate between subhalos and background objects.
As mentioned above, the profile for less massive subhalos is very noisy 
and slightly changed by the choice of radial bins, 
because the number of source galaxies is small in proportion to the area surrounding less massive subhalos.
We therefore concentrate on computing 
distortion profiles for three massive subhalos (``1'', ``9'' and ``32''), 
with masses greater than $10^{13}\hMsol$.

\begin{figure*}[t]
\epsscale{1.2}
\plotone{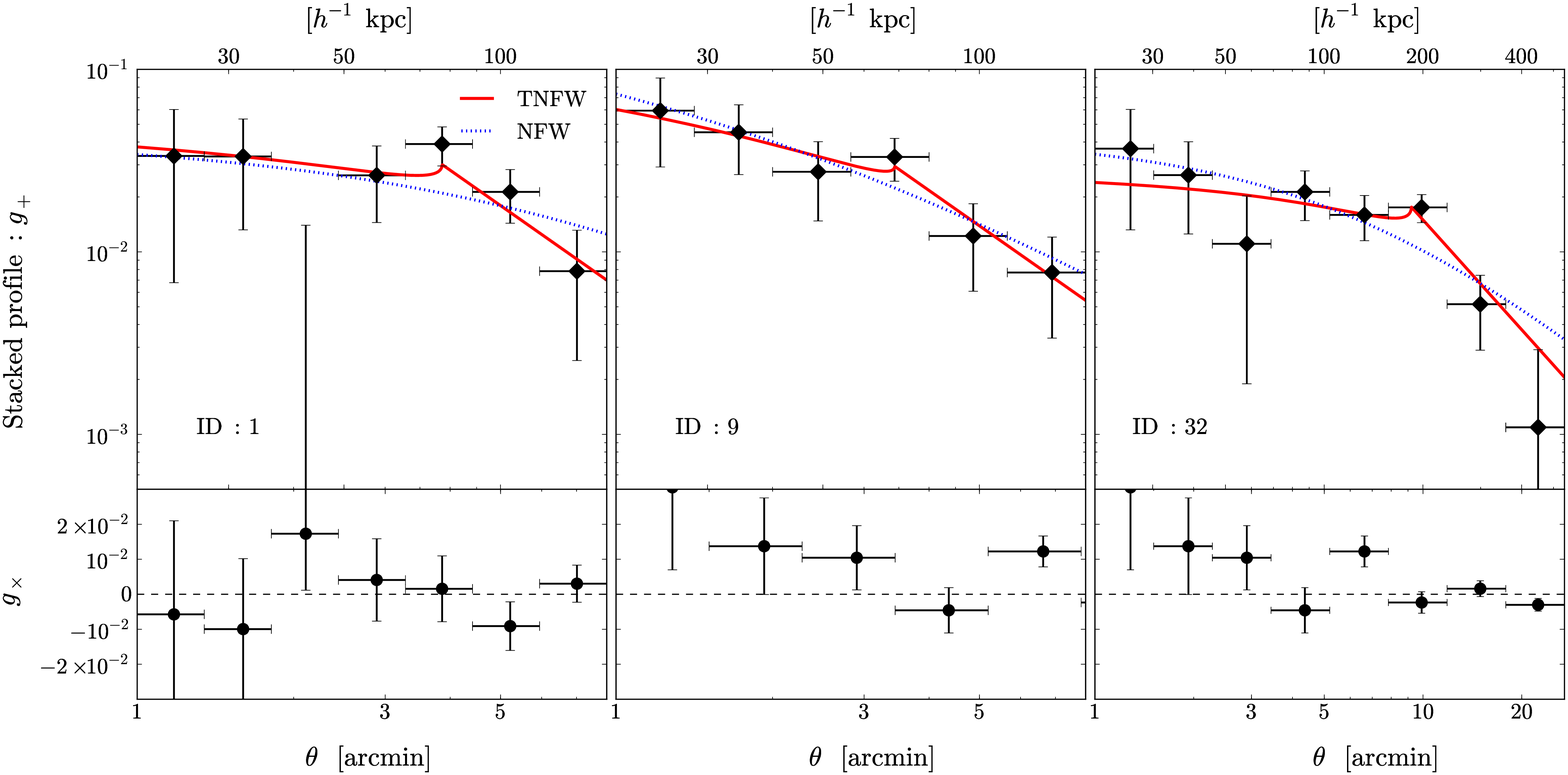}
\caption{Radial profiles of the tangential shear component (top panel), $g_+$, and the $45^\circ$ rotated
component (bottom panel), $g_\times$, for three massive subhalos (``1'', ``9'', and ``32''; from left to right).
The tangential signals sharply decline outside the truncation radii. 
The red solid and blue dotted lines are the best-fit for TNFW and NFW models, respectively.
}
\label{fig:g+_massive}
\end{figure*}

Figure~\ref{fig:g+_massive} displays breaks in the tangential shear profiles.
The slope follows $\propto \theta^{-2}$ outside the break, as shown by stacked lensing analysis 
(Section \ref{subsec:stacked_subhalos}).
The off-centering effect \citep{Yang06} of the main cluster mass on the lensing signal is negligible 
because of large separation from the cluster center.
We first fit the TNFW model as a model of subhalos to the tangential distortion profiles.
We also tried to fit a truncated singular isothermal sphere \citep[TSIS;][]{Okabe10a} model 
to the profile but found a poor fit for massive subhalos.
The best-fit masses and truncation radii are shown in Table \ref{tab:mass_subhalos}.
We found that the best-fit values do not change significantly by a choice of radial bins. 
The solid lines for the best-fit values describe the profiles with the breaks (Figure \ref{fig:g+_massive}) well.
Next we fit the NFW model to the data and then obtain larger minimum reduced $\chi^2_{\rm min}$ values than those for the TNFW model.
The significance probabilities, $Q$, for the NFW model are $0.12$, $0.74$, and $0.08$ 
for  subhalos ``1'' , ``9'', and ``32'', respectively.
When we adopt the threshold of $10\%$, 
the NFW model for  subhalo ``32'' is unacceptable and for subhalos ``1'' and ``9'' are acceptable although 
$Q$ for subhalo ``1'' is close to the threshold.
If observed lensing signals around subhalos ``1'' and ``32'' were explained only by background objects,
the profiles should be well described by the NFW model. 
We repeat the tangential fits using photometric redshifts $z_{\rm phot}=0.418$ and $0.189$ 
as redshifts of background objects around subhalos ``1'' and ``32'',
and obtain the virial masses for the NFW model, 
$M_{\rm vir}^{(1)}=12.11_{-4.93}^{+11.06}\times10^{14}h^{-1}M_\odot$, and
$M_{\rm vir}^{(32)}=3.98_{-0.91}^{+1.06}\times10^{14}h^{-1}M_\odot$, respectively.
In this mass scale, 
no clear truncation radius was found in the tangential profiles \citep[e.g.,][]{Okabe10b,Oguri12}.
Thus, it implies that all lensing signals cannot be explained solely by background objects.
We then fit using a combined model of the TNFW for subhalos and NFW for backgrounds,
where we assume the mass and concentration relation for backgrounds \citep{Duffy08}
and fix the truncation radius derived by fitting  the TNFW model.
Although measurement errors of subhalo masses become larger, 
we find that the best-fit subhalo masses are decreased by 30\%-40\%.
We therefore add these differences to the second error in Table \ref{tab:mass_subhalos}, 
as the systematic error, 
and propagate them into the stacked lensing analysis (Section \ref{subsec:stacked_subhalos};
 Table \ref{tab:mass_stacked}).

\begin{table*}
\caption{Best-fit masses and truncation radii for three massive subhalos using the tangential distortion profiles} \label{tab:mass_subhalos}
\begin{center}
\begin{tabular}{lccccc}
\hline
ID\tablenotemark{a}         & $M_{\rm t}$\tablenotemark{b}
           & $r_{t}$\tablenotemark{c}
           & $\chi_{{\rm min}}^2/{\rm d.o.f}$\tablenotemark{d} 
           & S/N\tablenotemark{e}
           & $\chi_{\rm min}^2/{\rm d.o.f}$ (NFW)\tablenotemark{f} \\
           & ($10^{12}h^{-1}M_\odot$)
           & ($\hkpc$)
           &
           &
           & \\
\hline
1          & $14.26_{-2.53-5.55}^{+2.37}$
           & $77.22_{-3.81}^{+2.74}$
           & $5.38/3$
           & $6.17$
           & $8.79/4$ \\
9          & $11.05_{-1.84}^{1.83}$
           & $68.70_{-9.49}^{+5.54}$
           & $0.59/3$
           & $5.99$
           & $1.97/4$ \\
32         & $47.65_{-5.81-13.42}^{+5.81}$
           & $184.38_{-16.65}^{+14.75}$
           & $3.74/5$
           & $8.33$
           & $11.31/6$ \\
\hline
\end{tabular}
\tablecomments{
\tablenotemark{a} : Name of subhalos (Table \ref{tab:Msub}).
\tablenotemark{b} : Best-fit mass, in units of $10^{12}\hMsol$.
\tablenotemark{c} : Best-fit truncation radii, in units of $\hkpc$.
\tablenotemark{d} : Reduced chi-square for the best-fit truncated NFW (TNFW) model (d.o.f is the degrees of freedom).
\tablenotemark{e} : Signal-to-noise ratio for the tangential distortion profile.
\tablenotemark{f} : Reduced chi-square for the best-fit NFW model. All are higher than those of the TNFW model. 
In particular, the significance probability, $Q$, for the NFW model of subhalo '32' 
is less than 10\%, indicating that the profile is not well fitted by the NFW model. 
}
\end{center} 
\end{table*}

\subsubsection{LSS Error Covariance Matrix} \label{subsec:LSSCov}

An alternative approach to take into account LSS lensing effect in weak-lensing mass measurements
is to use the error covariance matrix of uncorrelated large-scale structure along the line-of-sight 
\citep[e.g.,][]{Schneider98,Hoekstra03}, instead of the LSS lensing model. 
Here, We estimate the error covariance matrix $C_{ij}=C_{g,ij}+C_{{\rm LSS},ij}$ in the $i$- and $j$-th radial bin,
where $C_{g,ij}=\sigma_g^2\delta_{ij}$ is a diagonal matrix of the uncertainty caused by the intrinsic
shapes of the galaxies and the noise in the shape measurement, and $C_{{\rm LSS},ij}$ is calculated by 
the weak-lensing power spectrum \citep[e.g.,][]{Schneider98,Hoekstra03} with {\it WMAP7} cosmology \citep{WMAP07}.
The diagonal component of LSS error covariance matrix, $\sigma_{\rm LSS}=C_{{\rm LSS}}^{1/2}$, 
is lower than the statistical error $\sigma_g$ for $r\simlt50'$ and comparable to those for $r\simgt50'$, 
respectively. 
Thus, the statistical error is denominated in the radial range of the subhalo mass measurements.
We computed the stacked tangential shear profiles from the shear catalog without LSS lensing model.
As the truncation position in the tangential shear profiles does not change significantly, 
the best-fit truncation radii agree within $2\%$ with those in Table \ref{tab:mass_stacked}.
As the lensing signals at the truncation radii become 
higher than those estimated from the shear catalog with the LSS lensing, 
the subhalos masses for mass and radial bins become $\sim10\%$ and $\sim13\%$ higher.
In other words, our LSS model corrects the LSS lensing bias by $\sim10\%$.
The measurement uncertainties of the subhalo mass and the truncation radius, 
estimated with the error covariance matrix,
are consistent with those estimated from the statistical error, because the statistical error is dominated.

\subsubsection{Probability of Spurious Peaks} \label{subsec:stacked_fake}

To measure a reliable subhalo mass function, 
it is of critical importance to statistically rule out 
the possibility that the subhalo candidates are actually spurious peaks. 
The peak finding method always suffers from the presence of spurious peaks.
It is therefore of vital importance to quantify the number and properties of spurious peaks
in order to confirm the purity of {\it real} subhalos.
Especially, if our subhalo catalog included spurious peaks, 
a shape of a subhalo mass function (Sec \ref{sec:mass_func};Figure \ref{fig:mass_func}) would be changed.
For this purpose, we create 200 bootstrap data-sets 
generated by randomly swapping reduced shear at fixed positions,
repeat the map making process and then identify artificial false peaks which satisfy the same conditions
except for the spectroscopic information of the galaxies.
The number of spurious peaks is $5.32\pm2.23$ for each realization. 
Thus, we cannot completely rule out a contamination in the subhalo catalog.
It is, however, difficult to quantify the purity of subhalos 
comparing the number of detected subhalo candidates and false peaks,
because we excluded several subhalo candidates taking into account 
the cloud-in-cloud problem and the background groups.
Stacked lensing analysis enables us to measure the mean parameters even for spurious peaks,
although individual measurements of spurious peaks are very difficult due to high statistical noise.
Comparing the statistical properties of spurious peaks allows us to discuss the purity of the subhalo catalog.
We generated 500 bootstrap replications of stacked tangential profiles using the catalog 
of artificial false peaks. 
Here, the number of spurious peaks and radial bins are the same as 
those for subsamples in stacked lensing analysis of subhalos (Sec \ref{subsec:stacked_subhalos}). 
The mean mass and truncation radius for spurious peaks 
are estimated by fitting with the TNFW model.
We found that $38\%-45\%$ of profiles for spurious peaks gives a poor fit 
($Q<10\%$) or are ill-constrained.
This indicates that the profile shapes are different from those of observed subhalos.
Indeed, the stacked lensing analysis for shear-selected subhalos 
shows that the best-fit truncation radius depends on the projected mass and the cluster-centric radius
(Figures \ref{fig:stacked_g+_massbin} and \ref{fig:stacked_g+_rbin}).
If our sample consisted entirely of spurious peaks, such a clear dependence could not be found.
To make a more robust conclusion, 
we estimate the probability that the parameters for spurious peaks
accidentally coincide with those for observed subhalos within 1 $\sigma$ uncertainty, 
based on Monte Carlo re-distributions of the best-fit values with the covariance matrix of the measurement errors.
The false probabilities for individual subsamples in stacked lensing analysis , ${\mathcal P}_{\rm fake}$, 
are from $10^{-3}\%$ to $\sim8\%$ (Table \ref{tab:mass_stacked}).
Multiplying the number of subhalos by the false probability in each sub-sample, 
the expected numbers of spurious subhalos are less than unity.
Therefore, we conclude that our sample of subhalos has a high degree of purity.

\subsubsection{Selection Criteria} \label{subsubsec:selection}

A choice of the threshold in the S/N for mass maps results in one of the main systematic errors,
because we cannot completely rule out the possibility that 
peak heights in mass maps are accidentally above or below
the threshold due to reconstruction errors of finite sampling of background galaxies. 
Assuming Poisson fluctuations of the reconstruction noise, the S/N changes by
$\delta {\rm (S/N)} = \sqrt{\delta N_{\rm bkg}/N_{\rm bkg}} = N_{\rm bkg}^{-1/4}\simeq0.25$, 
where $N_{\rm bkg}$ is an effective number density in the smoothed mass map with the highest resolution. 
We repeated the same analysis with different thresholds.
The sample numbers become 24 and 49 with higher and lower thresholds, respectively.
This systematic error 
is taken into consideration to compute a subhalo mass function in Sec \ref{sec:mass_func}.

\subsubsection{Stacking Procedure} \label{subsubsec:stacking}

We investigate whether the stacking method gives systematic errors,
because the mean tangential profile stacked over subhalos with various truncation radii
would blunt the break feature. 
We make synthetic weak shear catalogs of subhalos using the analytic TNFW model and the intrinsic ellipticity.
Here, the number of background sources is the same as that observed.
The parameters of the TNFW model for individual subhalos are generated from a Gaussian distribution.
The mean and standard error of subhalo masses and the truncation radii for simulated samples are shown in Table \ref{tab:sim_stacked}. 
We assume that the coefficient between the subhalo mass and the truncation radius is 0.7
and that the halo concentration is 1, for the sake of simplicity.
We compute 500 samples in each stacked profile and fit them with the TNFW and TNFWProb models.
As shown in Table \ref{tab:sim_stacked}, the mean tangential profiles are able to recover the input values.
In TNFWProb model fitting, since the mean and standard error of the truncation radii are degenerate,
$\langle r_t \rangle$ are not well constrained in some cases, resulting in a large 
 mean measurement error.

\begin{table*}
\caption{Results of stacked lensing analysis using a mock shear catalog} \label{tab:sim_stacked}
\begin{center}
\begin{tabular}{ccccccccc}
\hline
$N_{\rm sub}$\tablenotemark{a}  & $\langle M_{\rm sub}^{\rm input} \rangle$\tablenotemark{b} 
               & $\sigma(M_{\rm sub}^{\rm input})$\tablenotemark{b} 
               & $\langle r_{t}^{\rm input} \rangle$\tablenotemark{b} 
               & $\sigma(r_{t}^{\rm input})$\tablenotemark{b} 
               & {\scriptsize  $\left< \frac{M_{\rm sub}^{\rm TNFW}}{M_{\rm sub}^{\rm input}} \right>$}\tablenotemark{c} 
               & {\scriptsize  $\left< \frac{r_t^{\rm TNFW}}{r_t^{\rm input}}  \right>$}\tablenotemark{c}
               & {\scriptsize  $\left< \frac{M_{\rm sub}^{\rm TNFWProb}}{M_{\rm sub}^{\rm input}} \right>$}\tablenotemark{d} 
               & {\scriptsize  $\left< \frac{r_t^{\rm TNFWProb}}{r_t^{\rm input}}  \right>$}\tablenotemark{d} \\
\hline
$21$              & $3.5$
                  & $1.0$
                  & $35$
                  & $10$
                  & $1.01\pm0.09$ ($0.11$)
                  & $1.10\pm0.12$ ($0.21$)
                  & $1.03\pm0.12$ ($0.15$)
                  & $1.12\pm0.18$ ($0.29$)\\
$8$               & $7.0$
                  & $2.0$
                  & $70$
                  & $20$
                  & $0.92\pm0.21$ ($0.21$)
                  & $0.96\pm0.36$ ($0.29$)
                  & $1.04\pm0.23$ ($0.23$)
                  & $1.11\pm0.44$ ($8$)\\
$3$               & $20.0$
                  & $5.0$
                  & $130$
                  & $40$
                  & $1.02\pm0.26$ ($0.21$)
                  & $1.08\pm0.49$ ($0.35$)
                  & $1.03\pm0.16$ ($0.20$)
                  & $1.02\pm0.21$ ($0.6$) \\
$11$          & $3.0$
              & $1.0$
              & $35$
              & $20$
              & $0.98\pm0.28$ ($0.17$)
              & $1.18\pm0.38$ ($0.79$) 
              & $1.02\pm0.18$ ($0.22$)
              & $1.11\pm0.40$ ($5$) \\
$10$          & $5.0$
              & $1.0$
              & $50$
              & $20$
              & $0.92\pm0.17$ ($0.18$)
              & $0.99\pm0.33$ ($0.37$)
              & $1.01\pm0.19$ ($0.21$)
              & $1.09\pm0.35$ ($13$) \\
$8$           & $5.0$
              & $1.0$
              & $50$
              & $15$
              & $0.93\pm0.17$ ($0.19$)
              & $1.00\pm0.35$ ($0.36$)
              & $1.02\pm0.18$ ($0.22$)
              & $1.11\pm0.41$ ($5$) \\
$3$           & $30.0$
              & $5.0$
              & $200$
              & $20$
              & $1.04\pm0.26$ $(0.24)$
              & $1.10\pm0.28$ $(0.52)$
              & $1.04\pm0.14$ $(0.15)$
              & $1.03\pm0.11$ $(11)$ \\
$64$   & $1.0$
       & $0.5$
       & $60$
       & $20$
       & $0.91\pm0.27$ ($0.47$)
       & $1.00\pm0.20$ ($0.77$) 
       & $1.18\pm1.18$ ($1.00$) 
       & $1.04\pm0.75$ ($6$) \\   
\hline
\end{tabular}
\tablecomments{
\tablenotemark{a} : Number of simulated subhalos.
\tablenotemark{b} : Mean and standard error of the mass ($10^{12}\hMsol$) and truncation radius ($\hkpc$) 
for simulated subhalos.
\tablenotemark{c} : Mean ratio of  outputs to  inputs for the TNFW model. Errors shown are the standard deviation based on 500 realizations. The values in  brackets are the mean measurement uncertainties.
\tablenotemark{d} : Mean ratio of  outputs to  inputs for the TNFWProb model.}
\end{center} 
\end{table*}

\section{Cluster Galaxy - Galaxy Lensing Analysis} \label{sec:stacked_lum}

\begin{figure}
\epsscale{1.2}
\plotone{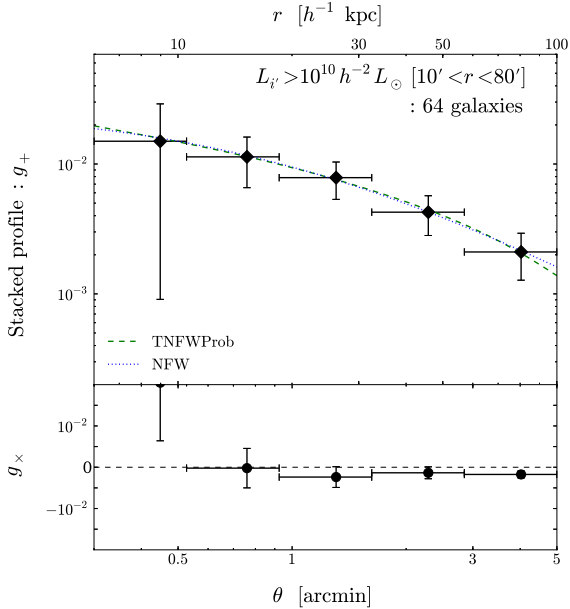}
\caption{Mean distortion profile for 64 luminous member galaxies with luminosities  more than
$10^{10}h^{-2}L_\odot$. The red solid, green dashed and blue dotted curves are the best-fit TNFW, TNFWProb, and NFW models, respectively.
}
\label{fig:stacked_g+_lumgal}
\end{figure}

Galaxy-galaxy lensing analysis for member galaxies selected solely by their luminosities 
provides us with complementary  and important information regarding cluster subhalos, 
because the sample is unbiased with respect to the lensing definition of subhalos (Sec \ref{sec:WLsubhalos}).
We use member galaxies with luminosities in the $i'$ band  larger than $10^{10}h^{-2}L_{i',\odot}$.
We compute a stacked tangential distortion profile as a function of the radius from luminous member galaxies. 
Since the cluster field is crowded, 
neighboring luminous galaxies may lead to serious contamination in the stacked lensing profile, 
if they are not sufficiently separated.
We thus need to determine the outermost radius of the profile in order to minimize  lensing contamination from  
neighboring luminous members.
We estimate the histogram of projected distances between the luminous member galaxies.
The outermost radius is chosen to be $5\farcm$ by applying 
a threshold that the mean number of neighboring luminous galaxies is less than unity.
We compile 64 luminous member galaxies located in the projected cluster-centric radius of $10\farcm<r<80\farcm$.
Figure \ref{fig:stacked_g+_lumgal} shows the mean tangential profile.
A sharply truncated profile is not found, in contrast to lensing-selected subhalos (Figures 
\ref{fig:stacked_g+_massbin} and \ref{fig:stacked_g+_rbin}).
The NFW and TNFWProb models are then applied to describe the profile, and these two models (Table \ref{tab:mass_lumstacked}) give an acceptable fit. 
The best-fit tangential profile for the TNFWProb model is similar to that for the NFW model,
because the intrinsic distribution of $r_t$ makes a sharply truncated profile blunt.
The mean virial radius for the NFW model, $273.87_{-45.98}^{+58.80}\hkpc$, is larger than
the mean truncation radius $\langle r_{t} \rangle=60.84_{-15.34}^{+7.62}\hkpc$ for the TNFWProb model.
On average, three luminous galaxies are inside the mean virial radius of luminous galaxies.
In other words, the mass distribution of subhalos associated with luminous galaxies overlap each other.
The NFW model is therefore unlikely to represent halos associated with luminous member galaxies.
On the other hand, the TNFWProb model gives a large scatter of the truncation radius compared to the mean value,
$\sigma_{r_t}/\langle r_t \rangle\sim37\%$. 
A broad distribution of the truncation radius smooths the truncation feature in the mean tangential profile for 
a large sample of less massive subhalos,
which makes it difficult to resolve the subhalo size.
To confirm this explanation, we conducted a stacked lensing analysis using a mock shear catalog 
(Table \ref{tab:sim_stacked}) in the same way as Section \ref{subsubsec:stacking}. 
We found that the truncation feature in the stacked profile is obscured and the mock simulation recovers the input values.  
Although we also conducted fitting stacked profiles for subsamples 
divided by luminosities or projected distances from the cluster center, 
only the upper limits can be derived.
We also checked less luminous galaxies with luminosities  less than $10^{10}h^{-2}L_\odot$, 
but found significant contamination from neighboring luminous or less luminous galaxies 
in the mean tangential profile.

Previous studies \citep{Natarajan04,Natarajan07,Natarajan09,Limousin05,Limousin07} 
conducted galaxy-galaxy lensing studies for clusters at $z\simgt0.2$ using single-band images.
As described by \citep{Broadhurst05} and \citep{Okabe10b,Okabe13}, 
lensing signals would be significantly diluted by a contamination 
of unlensed member galaxies in the shear catalog.
Their catalog for background source galaxies using the single filter would 
suffer from a contamination of member galaxies.
It is thus difficult to make a fair comparison between our result and the previous studies.

\begin{table*}
\caption{Best-fit mass profile parameters for NFW and TNFWProb models, obtained by galaxy-Galaxy lensing for 
luminous member galaxies} \label{tab:mass_lumstacked}
\begin{center}
\begin{tabular}{lcccccccc}
\hline
Number\tablenotemark{a}                       & $\langle M_{\rm sub}\rangle$\tablenotemark{b}
               & $\langle r_{t} \rangle$\tablenotemark{b}
               & $\sigma_{r_{t}}$\tablenotemark{b}
               & $M_{\rm vir}^{\rm NFW}$\tablenotemark{c} 
               & $c_{\rm vir}$\tablenotemark{c}
               & $\langle L_{i'} \rangle$\tablenotemark{d}
               & S/N \\\tablenotemark{e}
               & ($10^{12}h^{-1}M_\odot$)
               & ($\hkpc$)
               & ($\hkpc$)
               & ($10^{12}h^{-1}M_\odot$)
               & 
               & ($10^{10}h^{-2}L_\odot$)          
               &  \\
\hline
$64$              & $1.10_{-0.40}^{+0.40}$ 
                  & $60.84_{-15.34}^{+7.62}$
                  & $22.56_{-10.64}^{+19.47}$
                  & $2.39_{-1.01}^{+1.89}$
                  & $22.99_{-9.81}^{+20.25}$
                  & $2.25$
                  & $5.63$ \\
\hline
\end{tabular}
\tablecomments{  
\tablenotemark{a}: Number of luminous member galaxies selected by luminosities ($L_{i'}>10^{10}h^{-2}L_\odot$)
and  cluster-centric radii ($10\farcm\le r \le 80\farcm$).
\tablenotemark{b} : Best-fit mass, the average and  standard error of truncation radius distribution for the TNFWProb model.
\tablenotemark{c} : Best-fit virial mass and halo concentration for the NFW model.
\tablenotemark{d} : Average luminosity $\langle L_{i'} \rangle$.
\tablenotemark{e} : Signal-to-noise ratio for the tangential distortion profile.
}
\end{center} 
\end{table*}

\section{Main Cluster Mass Measurement} \label{sec:main_mass}

\begin{figure}
\epsscale{1.2}
\plotone{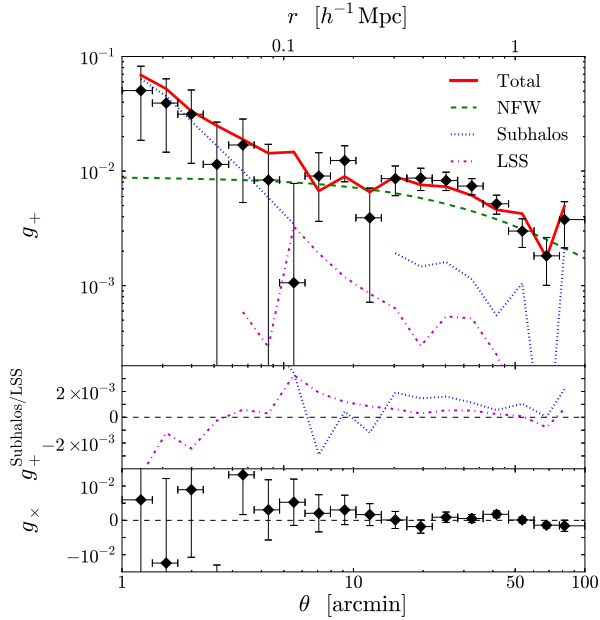}
\caption{Top panel: The tangential distortion component, $g_+$, 
 with respect the projected cluster-centric radius, in the range of  $1-100$ arcmin, is
estimated by azimuthally averaging the measured galaxy ellipticities.
The green dashed, blue dotted and magenta dashed-dotted lines are 
the best-fit NFW profile as the smooth mass component of the main cluster, 
lensing signals expected from subhalos and the LSS lensing model, respectively.
The total lensing signal (red solid line) of the three components is consistent with the observed distortion profile.
Middle panel: The tangential profiles for subhalos (blue dotted) and the LSS lensing model (magenta dashed-dotted),
are the same as that in the top panel, except for the use of a linear scale.  
Bottom panel: The $45^\circ$ rotated component, $g_\times$, is consistent with a null signal.}
\label{fig:g+_main}
\end{figure}

A tangential distortion profile, $g_+$, with respect to the cluster center, 
is a powerful tool to estimate the cluster mass \citep{Okabe10b}.
The tangential distortion profile as a function of the projected cluster-centric radius 
is computed by azimuthally averaging the measured galaxy ellipticities.
Here, we assume that the cluster center is at the central position of the subhalo ``21'' 
associated with the cD galaxy (NGC4874).
The top panel of Figure \ref{fig:g+_main} shows a complex feature 
of the lensing profile in the radial range of $1\farcm-100\farcm$, extending over 2 orders of magnitude.
Here, the data points are calculated using the shear catalog without the LSS modeling.
The lensing signal changes from $\mathcal{O}(10^{-1})$ to $\mathcal{O}(10^{-3})$ as the radius increases.
As expected from the low lensing efficiency of the nearby cluster, 
the lensing signal is 1 order of magnitude lower than that for massive clusters at $z\sim0.2$ \citep[e.g.,][]{Okabe10b}.
However, the S/N reaches S/N$\simeq 13.3$ thanks to a remarkably large number 
of background galaxies, which is comparable to or higher than those of clusters at $z\sim0.2$ \citep{Okabe10b}.
This high S/N validates weak-lensing analysis for low redshift clusters ($z\simlt 0.1$) which have been overlooked for a long time.
We also find that the $45^\circ$ rotated component, $g_\times$ (bottom panel of Figure \ref{fig:g+_main}), 
which is a non-lensing mode serving as a monitor of systematics errors, 
is 1 order of magnitude smaller than the lensing mode, $g_+$, which is consistent with a null signal.

The tangential distortion contains complete information for the lensing signals, including
the smoothed mass component of the main cluster, 
the interior substructure (Section \ref{subsec:subhalos}) 
and LSS lensing signals behind the cluster (Section \ref{subsec:LSSlens}). 
To understand the profile, 
we computed the tangential shear profiles for 32 shear-selected subhalos and the LSS lensing model, illustrated by the blue dotted and magenta dashed-dotted lines in the top and middle panels of Figure \ref{fig:g+_main}.
The S/N for the subhalos is S/N $\simeq 4.4$, 
accounting for $33\%$ of the total distortion signal. 
This indicates that the profile highly resolves the lensing signal from the interior substructure 
by the large apparent size of the cluster.
The observed signal in the central region ($r\simlt 5\farcm$) is dominated by the subhalo ``21''.
The lensing signals in $5\farcm \simlt r\simlt12\farcm$ and in $r\sim70\farcm$ are depressed by prominent subhalos.
As for LSS lensing, the signal-to-noise ratio is S/N$\simeq 1.3$.
Here, since the LSS lens modeling for possible background groups, ``F'' and ``I''
(Table \ref{tab:bkg} and Figure \ref{fig:kappa}), has failed significantly, 
we estimated the lensing distortion pattern from the NFW profile determined by the tangential profile for this object
and found that this background group depresses the observed lensing signal at $r\sim70\farcm$.

We fit a single NFW model to the distribution profile corrected using the LSS lensing model.
The lensing signals from the subhalos gradually change from positive to negative in the central region ($\simlt 12\farcm$).
These absolute values account for a large fraction of the total lensing signals,
which makes it difficult to discriminate between signals from the subhalos and the main cluster based on the central signals.
Indeed, when a single NFW model fits the data,
the best-fit halo concentration is systematically changed 
($\Delta c_{\rm vir}\simlt1$) by a choice of the innermost radius.
To avoid the subhalo bias in cluster mass measurements, we estimate the radial range for the fitting,  
by requiring that the fraction of the absolute value of subhalo signals to observed signals is less than $30\%$.
We fit the tangential profile between $13\farcm$ ($\sim260\hkpc$) and $64\farcm$ ($\sim1.3\hMpc$) with a single NFW model. 
Here, the physical scale of the innermost radius is comparable to those for massive clusters at $z\sim0.2$ 
\citep{Okabe10b}.
The resultant masses at different overdensities are listed in Table \ref{tab:mass_lit}.
The virial mass and concentration are $\Mvir=8.42_{-2.42}^{+4.17}\times10^{14}h^{-1}\hMsol$
and $\cvir=3.57_{-1.12}^{+1.54}$, respectively.

We also fit the profile in the full range with 
the NFW model for the smooth matter component and a central point mass contribution of the brightest cluster galaxy (BCG).
We obtain the point mass $M_{\rm pt}=5.67_{-0.32}^{+0.32}\times10^{12}\hMsol$, 
which is consistent with the projected  mass measurement of  subhalo ``21'' (Table \ref{tab:Msub}).
The summation of the virial mass and the point mass is $M_{\rm tot}=8.21_{-1.98}^{+2.99}\times10^{14}\hMsol$. 
The total mass is in agreement with the estimated virial mass  derived using the tangential fit for
the radial range to minimize subhalo contributions.

We next repeat fitting with the NFW model as the smooth component, 
by fully taking into account lensing signals from all shear-selected subhalos and the LSS lensing model.
The best-fit profile for the smooth component is shown in the green dashed line in Figure \ref{fig:g+_main}.
The total signal (red solid line) from three different components of the smooth NFW profile, subhalos and LSS lens model 
describes  the observed signals remarkably well.
The summation of the virial mass and subhalo masses, $M_{\rm tot}=8.18_{-2.02}^{+3.78}\times10^{14}\hMsol$,
is in good agreement with the virial mass $\Mvir$ (Table \ref{tab:mass_lit}).  
A singular isothermal sphere (SIS) model is strongly disfavored as the smoothed mass component,
returning a goodness-of-fit statistic of $\chi^2_{\rm min}/{\rm d.o.f.}=59.3/11$.

We also conduct the cluster mass measurement using the LSS error covariance matrix \citep[e.g.,][]{Schneider98,Hoekstra03},
as described in Section \ref{subsec:LSSCov}.
Although the error covariance matrix does not significantly change the result of subhalo mass measurement, 
the situation for the cluster mass measurement is slightly different.
The diagonal component of the LSS error covariance matrix becomes comparable to the statistical error for $r\simgt50'$.
The S/N estimated from the covariance matrix in the tangential shear profile becomes smaller
$S/N\simeq 7.5$ from the case of the statistical errors, 
because the LSS errors in different radial bins are correlated.
The S/N is consistent with \cite{Hoekstra03}.
We use the shear catalog without the correction of the LSS model in the full radial range of the cluster 
and apply the NFW model for the smooth matter component and a central point source of the BCG.
The best-fit mass is $M_{\rm tot}=8.80_{-3.74}^{+7.59}\times10^{14}\hMsol$. The upper and lower errors become 
larger by $\sim 80\%$ and $\sim40\%$, respectively.

\begin{figure}
\epsscale{1.2}
\plotone{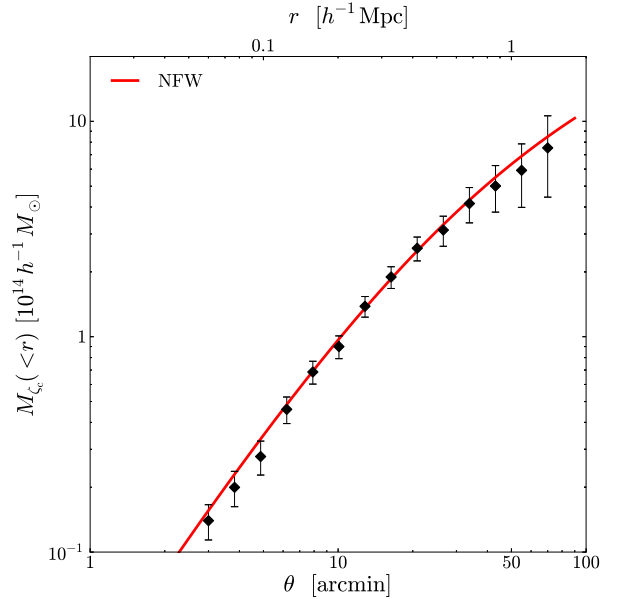}
\caption{$M_{\zeta_{\rm c}}$ profile as a function of the projected cluster-centric radius,
estimated by azimuthally averaging  galaxy ellipticities with a correction factor from the LSS lens model.
The red solid line is the best-fit NFW profile.
}
\label{fig:Mzeta_main}
\end{figure}

The projected mass ($M_{\zeta_c}$) measurement ($\zeta_c$ statistics) for the main cluster 
is less sensitive to lensing signals from subhalos,
because it estimates a cumulative profile. 
It is thus complimentary to the tangential fit. 
Figure \ref{fig:Mzeta_main} shows the $M_{\zeta_c}$ profile calculated with the fixed background annulus of $70\farcm-90\farcm$.
The background region is inside the best-fit virial radius ($r_{\rm vir}=96.67_{-10.32}^{+13.89}$ arcmin) derived from the tangential shear fit.
Following \cite{Okabe08}, we fit the $\zeta_c$ profile with a single NFW model, 
taking into account the error covariance matrix.
The best-fit profile is shown by the red solid line.
The best-fit values (Table \ref{tab:mass_lit}) are consistent with those for the tangential shear fit.

We compare the best-fit mass and concentration with results in the literature (Table \ref{tab:mass_ref}). 
The mass measurements in this study are consistent with our previous analysis \citep{Okabe10a}.
The statistical precision of the new mass estimates is improved by four times thanks to the huge number of background galaxies.
\cite{Gavazzi09} conducted fitting 
the tangential distortion profile in the range of $0\farcm35\simlt r\simlt 35\farcm$ 
using the NFW model including and excluding priors on mass and concentration relations, using CFHT/Megacam data.
In that study, the LSS lensing effect was not accounted for.
The best-fits, regardless of priors, are compatible within their large errors, with the present study.
However, their lensing signals \citep[Figure 3 in][]{Gavazzi09} differ from the results in the present study (Figure \ref{fig:g+_main}).
Their profile ($0\farcm35\simlt r\simlt 35\farcm$) is well described by a single NFW model.
The lensing signals in the same radial range in the present study are dominated by prominent subhalos in $r\simlt12\farcm$ and by
the smooth mass component in $12\simlt r\simlt35\farcm$, respectively.
We conduct the tangential fit using the profile computed with the same radial bins as \cite{Gavazzi09} and a single NFW model 
and obtain only an upper limit on $\Mvir<4\times10^{16}\hMsol$ because of an inadequate model.
\cite{Kubo07} carried out weak-lensing analysis using SDSS data.
The best-fit $M_{200}$ \citep{Kubo07} derived by fitting 
the tangential profile up to $10\hMpc$ with a single NFW model is three times higher than that observed here. 
The background LSS lensing effect was not accounted for in that study.
Since their outermost radius ($10\hMpc$) is five times higher than our best-fit virial radius, 
their mass would be overestimated by mass distribution outside the cluster. 
The dynamical mass estimates of $\Mvir$ and $M_{200}$ \citep{Rines03,Lokas03} agree with our best-fits, 
although their concentration is three times higher than estimates in the present study.

\begin{table*}
\caption{Main cluster mass estimates} \label{tab:mass_lit}
\begin{center}
\begin{tabular}{lccccccccc}
\hline
Fitting Method\tablenotemark{a}   & $M_{\rm vir}$ 
                 & $c_{\rm vir}$ 
                 & $M_{\rm 200}$ 
                 & $c_{\rm 200}$ 
                 & $M_{\rm 500}$  
                 & $M_{\rm 1000}$
                 & $M_{\rm 2500}$ \\
                 & ($10^{14}\hMsol$)
                 &
                 & ($10^{14}\hMsol$)
                 &
                 & ($10^{14}\hMsol$)
                 & ($10^{14}\hMsol$)
                 & ($10^{14}\hMsol$) \\
\hline
$g_+$ profile   & $8.42_{-2.42}^{+4.17}$
                & $3.57_{-1.12}^{+1.54}$
                & $6.23_{-1.58}^{+2.53}$
                & $2.55_{-0.84}^{+1.17}$
                & $3.89_{-0.76}^{+1.04}$
                & $2.47_{-0.37}^{+0.44}$
                & $1.15_{-0.22}^{+0.22}$ \\
${\zeta_c}$ profile   & $8.31_{-1.82}^{+2.42}$
                        & $3.24_{-0.67}^{+0.80}$
                        & $6.08_{-1.20}^{+1.51}$
                        & $2.30_{-0.50}^{+0.61}$
                        & $3.67_{-0.60}^{+0.69}$
                        & $2.27_{-0.33}^{+0.36}$
                        & $1.00_{-0.18}^{+0.18}$ \\
\hline
\end{tabular}
\tablecomments{  
\tablenotemark{a} : profiles for fitting.
}
\end{center} 
\end{table*}

\begin{table*}
\caption{Mass and concentration previously reported in the literature} \label{tab:mass_ref}
\begin{center}
\begin{tabular}{lcccc}
\hline
Reference      & $M_{\rm vir}$
               & $c_{\rm vir}$
               & $M_{200}$
               & $c_{200}$ \\
               & ($10^{14}\hMsol$)
               &
               & ($10^{14}\hMsol$)
               & \\
\hline
WL : \citep{Kubo07}    & -
                 & -
                 & $18.8^{+6.5}_{-5.6}$
                 & $3.84^{+13.16}_{-1.84}$ \\
WL : w/o priors\tablenotemark{a} \citep{Gavazzi09} & $4.27^{+8.47}_{-2.45}$
                 & $6.7^{+4.1}_{-3.3}$
                 & $3.57_{-1.47}^{+3.01}$ 
                 & $5.0_{-2.5}^{+3.2}$ \\
WL : w/ priors\tablenotemark{a} \citep{Gavazzi09} &  $7.77^{+11.69}_{-4.27}$
                 & $4.9^{+1.7}_{-1.4}$
                 & $6.79^{+4.27}_{-2.45}$
                 & $3.5^{+1.1}_{-0.9}$ \\
WL : \citep{Okabe10a}  & $8.92^{+20.05}_{-5.17}$ 
                 & $3.50^{+2.56}_{-1.79}$
                 & $6.61^{+12.06}_{-3.63}$ 
                 & $2.50^{+1.94}_{-1.34}$ \\
Dynamics : \citep{Rines03}                 & 
                 & -
                 & $7.85$
                 & - \\
Dynamics : \citep{Lokas03} & $8.45\pm3.15$
                     & $9.4$
                 & -
                 & - \\
\hline
\end{tabular}
\tablecomments{  
\tablenotemark{a} priors with and without the mass-concentration relation.
}
\end{center} 
\end{table*}

\section{Discussion} \label{sec:dis} 

\subsection{Subhalo Mass Function} \label{sec:mass_func}

\begin{figure}
\epsscale{1.2}
\plotone{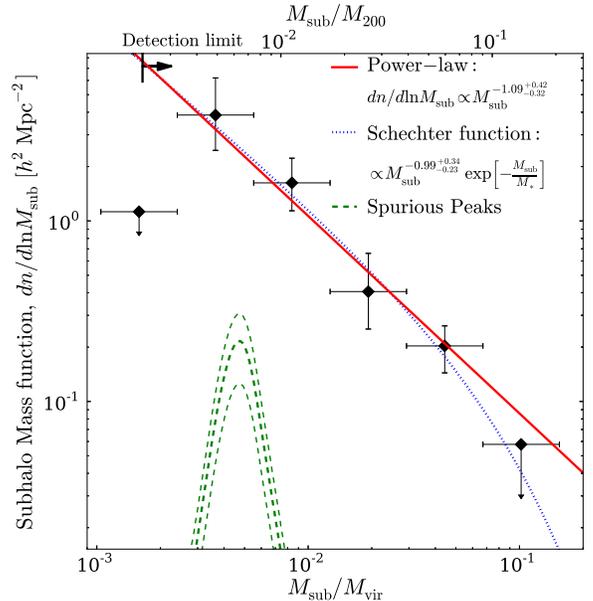}
\caption{Subhalo mass function spanning 2 orders of magnitude of subhalo masses.
The red solid and blue dotted lines are the best-fit power-law model
and Schechter function, respectively. 
The best-fit powers are in remarkable agreement with CDM predictions.
Green dashed lines are the mass function for spurious peaks.
The thick and thin dashed lines are the best-fit and the $68\%$ C.L. uncertainty, respectively.
}
\label{fig:mass_func}
\end{figure}

The subhalo mass function is computed from Monte Carlo redistributions of subhalo masses 
taking into account both measurement uncertainties and the systematic error on the applied threshold 
(Section \ref{subsubsec:selection}).
Figure \ref{fig:mass_func} shows the subhalo mass function.
The error bars are based on both measurement and systematic errors. 
The resulting subhalo mass function covers over 2 orders of magnitude in mass. 
The number of subhalos decreases as their mass increases, 
while it is significantly decreased in the low-mass end ($M_{\rm sub}/M_{\rm vir}\sim 10^{-3}$) 
because the detection limit appeared in the mass map. 
To access the purity, we also compute the mass function of spurious peaks, $dN_{\rm fake}/d\ln M_{\rm fake}$.
Here, we calculate the mass function for spurious peaks, using the probability distribution of best-fit masses,
$M_{\rm fake}$, derived by stacked lensing analysis (Section \ref{subsec:stacked_fake}). 
The number of false peaks as a function of the mass is given by 
$N(M_{\rm fake})=N_{\rm fake}\sum_i N_{\rm sample,i}P_i(M_{\rm fake})/\sum_i N_{\rm sample,i}$,
where $N_{\rm fake}$ is the total number of spurious peaks and $P_i(M_{\rm fake})$ and $N_{\rm sample,i}$
are the probability distribution and the subhalo number for the subsamples of the mass bin, respectively.
The probability distribution, $P_i(M_{\rm fake})$, 
is calculated by the best-fit masses taking into account the measurement uncertainty.
The green dashed lines show a single peak of the mass function for spurious peaks.
The functional form is different from the observed mass function.
The peak height of spurious peaks is 1 order of magnitude lower than the observed mass function in the same mass range.
Stacked lensing analysis of false peaks (Section \ref{subsec:stacked_fake}) disfavors the contamination of spurious peaks in the sample of subhalos. Even if they exist, the contamination level is negligible for a study of the mass function.

We fit the subhalo mass function with single power law model \citep[e.g.,][]{Gao12},
\begin{eqnarray}
dn/d\ln M_{\rm sub}\propto M_{\rm sub}^{-\alpha} \label{eq:powerlaw}
\end{eqnarray} 
and a Schechter function \citep[e.g.,][]{Schechter76,Shaw06}, 
\begin{eqnarray}
dn/d\ln M_{\rm sub}\propto M_{\rm sub}^{-\beta} \exp(-M_{\rm sub}/M_{\rm *}). \label{eq:sch}
\end{eqnarray} 
The mass function is modified from these analytical functions 
because of finite measurement errors for the subhalo masses. 
This corrects the modeling for the so-called Eddington bias.
The model of the mass function is described by the convolution between the analytical forms and the errors, 
$d n_{\rm model}/d \ln M_{\rm sub} = \int d n/d \ln x p(x,M_{\rm sub}) d x/ \int p(x,M_{\rm sub}) dx$.
Here, we assume a Gaussian probability function,
$p(x,M_{\rm sub})=\Sigma_i \exp\left(-(x-M_{{\rm sub},i})^2/2/\sigma_{M,i}^2\right)/(2\pi \sigma_{M,i}^2)^{1/2}$, 
where $M_{{\rm sub},i}$ and $\sigma_{M,i}$ are the mass estimate and the error for $i$-th subhalo, respectively.
The cutoff mass, $M_{\rm *}$, in the Schechter function is sensitive to abundance at the high-mass end. 
However, since the abundance of massive subhalos is small, it is not well constrained,
$M_{\rm *}/M_{\rm vir}=0.089_{-0.064}^{+0.135}$.
We are therefore unable to discriminate between the single power law and the Schechter function. 
The best-fit power indices, which characterize the shape of the function at the intermediate and low ranges, 
are in good agreement 
($\alpha=1.09^{+0.42}_{-0.32}$ and $\beta=0.99_{-0.23}^{+0.34}$).
We also computed a subhalo mass function including four subhalo candidates with no optical counter
and obtain the best-fit 
$\alpha=1.15_{-0.32}^{+0.38}$ and $\beta=0.99_{-0.24}^{+0.38}$.
For further verification, 
we excluded the most and least massive of the massive subhalos to construct a mass function 
and found that the best-fit slope values 
do not significantly change.
The best-fit slopes are in remarkable agreement with CDM predictions $\sim0.9-1.0$ from numerical simulations 
\citep[e.g.,][]{Diemand04,DeLucia04,Gao04,Shaw06,Angulo09,Giocoli10,Klypin11,Gao12} 
and analytical models \citep[e.g.,][]{Taylor04a,Oguri04,vandenBosch05,Giocoli08a}. 
A recent high-resolution numerical simulation study \citep{Gao12} found that 
the slope of mass function in the range of $10^{-6}<M_{\rm sub}/M_{200}<10^{-3}$ gives $\alpha=0.98$.

The mass fraction for observed subhalos is estimated as $f_{\rm sub}=\sum_i M_{\rm sub,i}/M_{\rm vir}=0.226^{+0.111}_{-0.085}$ 
with the tangential fit for the main cluster and $f_{\rm sub}=0.229_{-0.064}^{+0.078}$ with $\zeta_c$ fit, respectively.
\cite{Shaw06} estimated the mean mass fraction as a function of the virial 
mass, $\langle f_{\rm sub}\rangle=0.14\pm0.02(\Mvir/8\times10^{14}\hMsol)^{0.44\pm0.06}$.
The mass fraction for the Coma cluster is larger than the mean fraction, $\langle f_{\rm sub} \rangle=0.16$, 
estimated using the best-fit virial mass. 
We also calculated the mass fraction within $r_{200}$, $f_{\rm sub,200}=0.222\pm0.077$.

\subsection{Correlation between Subhalo Masses and Truncation Radii}

It is interesting to investigate the correlation between the subhalos' masses and truncation radii,
because they are both free parameters in stacked lensing analysis.
We compile the stacked lensing results divided by mass and cluster-centric bins and find a tight correlation of 
$M_{\rm sub}\propto r_t^{1.18_{-0.09}^{+0.10}}$ and $M_{\rm sub}\propto r_t^{1.19_{-0.16}^{+0.17}}$ 
for the TNFW and TNFWProb models, respectively. 
Considering a functional form of the NFW model, $M_{\rm NFW}(<x)\propto \log(1+x)-x/(1+x)$ (Equation (\ref{eq:MNFW}))
where $x=r/r_s$ is the radius normalized by a scale radius, 
the best-fit slope values imply that the mass loss occurs in the subhalo outskirts beyond the scale radius 
as long as the internal structure does not change during movement in the host halo.

\subsection{Radial Dependence of Subhalo Properties} 

\begin{figure*}
\epsscale{1.}
\plottwo{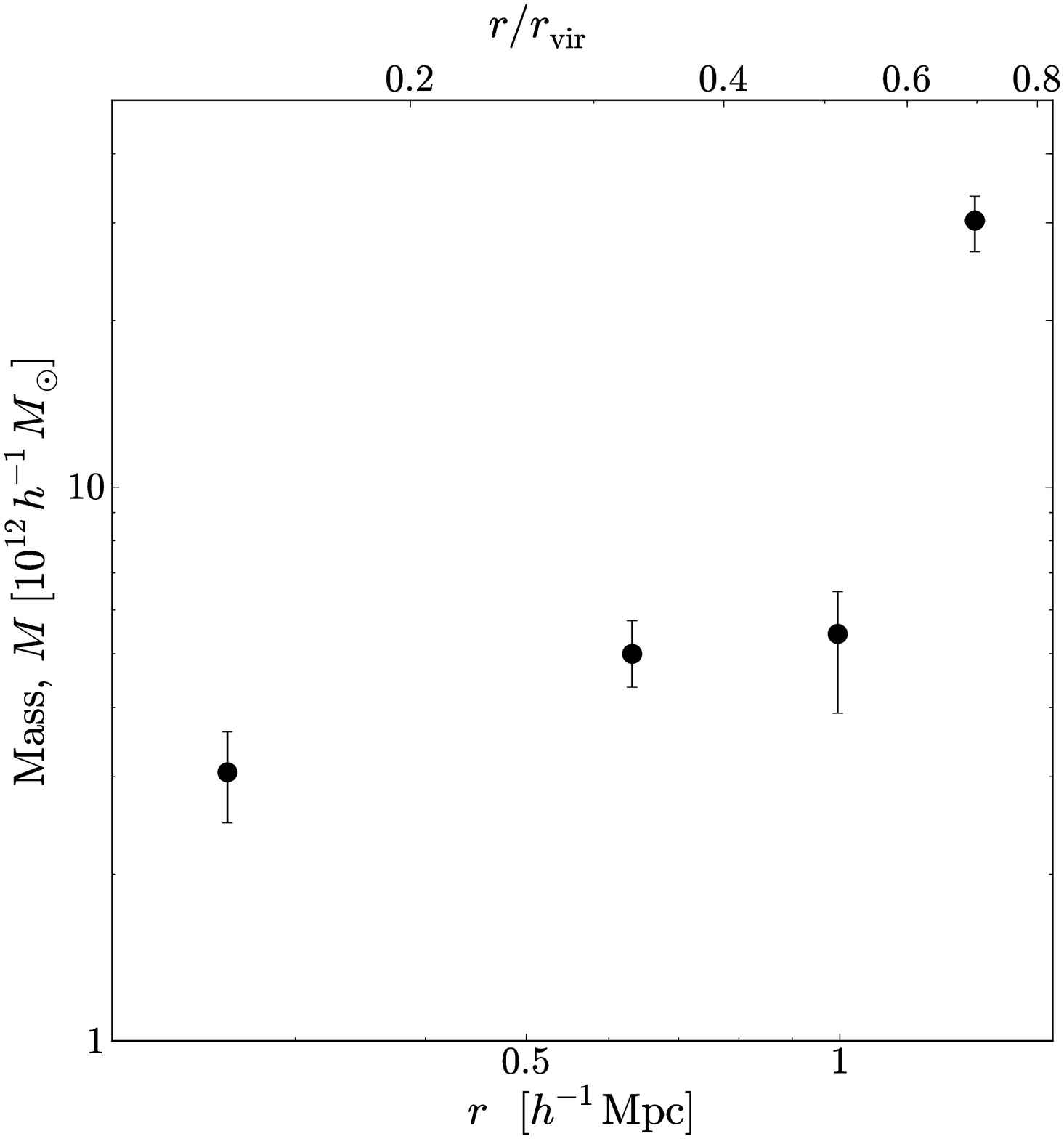}{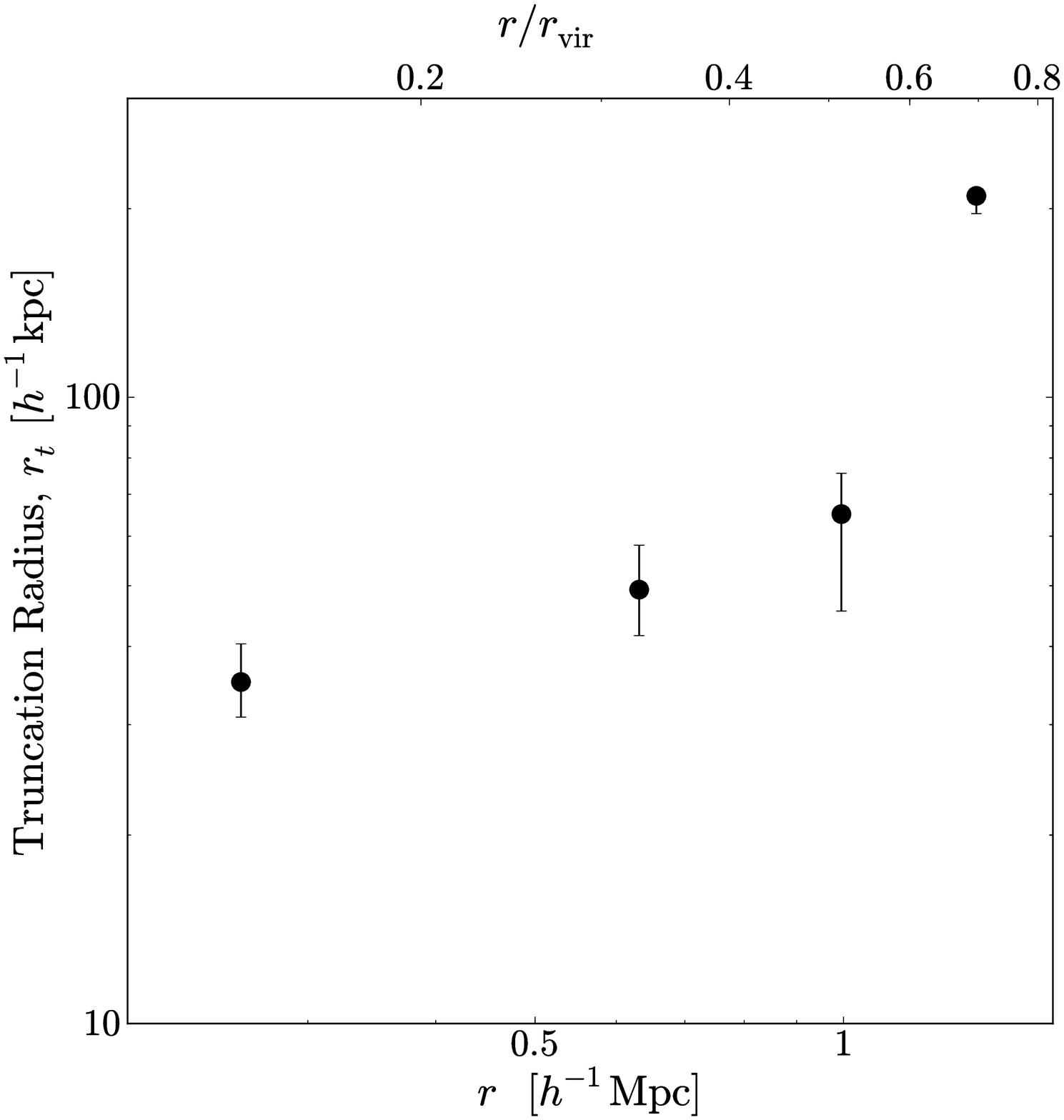}
\caption{Mass (left) and  truncation radius (right) profiles 
of subhalos as a function of the projected cluster-centric radius. 
The black circles represent the masses and truncation radii of subhalos using stacked weak-lensing analysis for lensing-selected subhalos, showing that they monotonically increase with an increasing radius.
}
\label{fig:Msub_vs_r}
\end{figure*}

Subhalos captured by more massive halos are subject to dynamical friction, losing their angular momentum 
and subsequently falling inward the center.
Simultaneously, their masses are reduced by the tidal force which increases 
with an increasing radius from the cluster center.
The subhalos in the central region have been affected by the tidal field for a longer time than 
those on the outskirts. 
It is thus expected that the subhalo mass and truncation radius are an increasing function of cluster radius.
The survey of subhalos using the wide-field imaging data allows us to study the radial dependence of their properties.
For this purpose, we use the projected position of shear-selected subhalos.
It is  difficult to constrain the pericenter radius and a line-of-sight position of the subhalo.
In order to reduce these uncertainties, 
we compute the mean subhalo masses and truncation radii derived from stacked lensing analyses 
for the subsample divided by their positions from the cluster center (Section \ref{subsec:stacked_subhalos}).
The mean projected distance of the subsample is estimated as a weighted average of projected distances from the cluster center for the individual subhalos. 
The weight function is given by the tangential distortion signals with respect to the subhalo center.
The left and right panels of Figure~\ref{fig:Msub_vs_r} 
display a clear radial dependence of subhalo masses and truncation radii, 
as expected from tidal destruction.
The subhalo masses and radii gradually increase out to $\sim1\hMpc$ and 
drastically rise due to massive subhalo ``32'' on the outskirts in the range of $1.2-1.6\hMpc$.
We fit the subhalo mass profile with a functional form of 
$\log(M_{\rm sub}/M_{\rm pivot})=A+B\log(r/r_{\rm pivot})$ 
where $M_{\rm pivot}=10^{12}\hMsol$ and $r_{\rm pivot}= 1 \hMpc$ and obtain, 
$A=2.66\pm0.08$ and $B=1.45\pm0.12$.
The best-fit parameters for the inner three bins are $A=1.86\pm0.23$ and $B=0.55\pm 0.26$.
The subhalo size profile is fitted with $\log(r_{t}/r_{t,{\rm pivot}})=A+B\log(r/r_{\rm pivot})$ 
and $r_{t,{\rm pivot}}=1\hkpc$.
The best-fit parameters for all, and for the three inner bins,
are $A=4.95\pm0.04$ and $B=1.18\pm0.08$, and $A=4.07\pm0.26$ and $B=0.38\pm0.24$, respectively.
The best-fit slope values for the mass and truncation radius are 
significantly changed by the presence of the massive subhalo on the outskirts ($60\farcm<r<80\farcm$).
The similar trend on the half mass radius for subhalos in simulated clusters are found by \cite{Limousin09}.

The tidal radius of subhalos is generally defined by a competition between 
the differential tidal forces of the host halo potential and the acceleration toward the subhalos.
Equivalently, this condition can be rewritten as a balance between the average density of subhalos and the host halo, 
${\bar \rho}_{\rm sub}=\eta{\bar \rho}_{\rm main}$, where $\eta$ is an efficiency factor. 
For instance, $\eta=3$ for a point mass case on a circular orbit, $\eta=2$ for the Roche limit, 
and $\eta=2-d\ln M_{\rm main}/d\ln R$ in the case of extended mass profiles of subhalos and the host halo, 
\citep[e.g.,][]{Tormen98,Taylor04a,Gan10} in a linear regime of $M_{\rm sub}/M_{\rm main}\ll 1$ and $r_t/R\ll 1$, 
where $R$ is the pericenter radius from the cluster center.
The minimum subhalo size is determined by the pericenter radius.
Although we cannot constrain the pericenter radius of subhalos from the current position of the subhalos, 
it is interesting to compare the density ratio, $\eta$, with these trial approximations.
We calculate the mean densities for subhalos, ${\bar \rho}_{\rm sub}=M_{\rm sub}/r_t^3$, and for the host halo, 
${\bar \rho}_{\rm main}=M_{\rm NFW}(<R)/R^3$, where $M_{\rm NFW}$ is 
a spherical NFW mass enclosed within the three-dimensional radius.
Here, we use the best-fit NFW model and the projected cluster-centric radius for subhalos.
The mean density for subhalos is higher than that for the cluster mass.
The density ratios, $\eta\sim10-40$ for shear-selected subhalos (Section \ref{subsec:stacked_subhalos}) and $\sim10$ for luminous galaxies 
(Section \ref{sec:stacked_lum}), are comparable to each other, 
but they are higher than expected by the linear regime assuming that the current position is the pericenter.
If the discrepancy could be explained by a difference of the positions, 
shear-selected subhalos and luminous galaxies would be located inward. 
On the other hand, the subhalo mass implies that the detected subhalos are remnants of group-scale structure.
Large and massive subhalos would not be described  by the linear regime.

\subsection{Surface Number Density for  Subhalos}

\begin{figure}
\epsscale{1.2}
\plotone{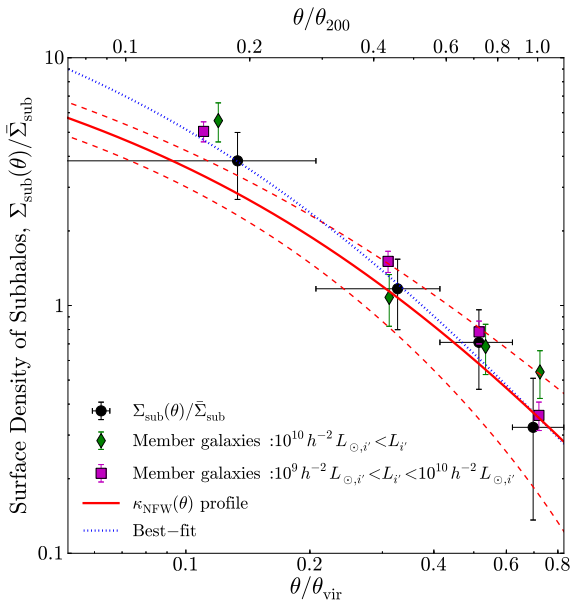}
\caption{Surface density (black circles) profile of subhalos. 
The best-fit profile is shown by a blue dotted line.
For comparison, a red solid line shows the surface mass density for the total mass of the main cluster, normalized with the number density at $r_{200}$. Red dashed lines are the $68\%$ confidence uncertainties for the total mass. Green diamonds and magenta squares denote the surface mass densities for member galaxies of which luminosities are $L_{i'}>10^{10}h^{-2}L_{\odot,i'}$ and $10^{9}h^{-2}L_{\odot,i'}<L_{i'}<10^{10}h^{-2}L_{\odot,i'}$, respectively.
}
\label{fig:Nsub_vs_r}
\end{figure}

It is important to estimate a surface number density profile for subhalos 
to understand the evolution of subhalos in the cluster.
Figure~\ref{fig:Nsub_vs_r} shows the surface number density profile of subhalos 
normalized by the mean surface density. The errors for the surface number density are assumed to be Poisson noise. 
It is clear that the surface number density increases while decreasing the cluster-centric radius.
For comparison, we compute the surface number density profiles of member galaxies for which the luminosities are 
$L_{i'}>10^{10}h^{-2}L_{\odot,i'}$ and $10^{9}h^{-2}L_{\odot,i'}<L_{i'}<10^{10}h^{-2}L_{\odot,i'}$.
Both profiles are similar to that of subhalos.
To quantify these distributions, we assume the spherical symmetric NFW distribution for the subhalos and member galaxies and fit the profiles. 
The surface number density profile is specified by three parameters including normalization, the concentration and the virial radius.
We here use the virial radius determined by the tangential shear fit (Section \ref{sec:main_mass}).
The best-fit concentrations are $\cvir=5.73\pm4.46$ for subhalos,
$\cvir=5.97\pm3.28$ for luminous galaxies and $\cvir=5.35\pm1.00$ for less luminous galaxies.
All best-fit concentrations agree with each other.
We also compute the surface mass density profile for the total mass from the best-fit NFW parameters (Section \ref{sec:main_mass}). 
Here, the normalization is set to be the surface density of the subhalos at $r_{200}$.
The best-fit concentrations for the subhalos and luminous member galaxies do not differ from that for the main mass,
while the less luminous galaxies are more centrally concentrated.
Recent numerical simulations \citep[e.g.,][]{Ghigna00,Diemand04,DeLucia04,Gao04,Gao04b,Nagai05,Gao12} and analytical models 
\citep[e.g.,][]{Taylor05c,Zentner05} have shown that 
the radial distribution of subhalos is less concentrated than that of the total mass, 
because the subhalos lose their mass more efficiently in the inner regions of the main halo. 
The local surface density of subhalos is noisy and there is an uncertainty in their position along the line-of-sight.
The Coma cluster contains the famous NGC4839 group in the southwest central region (Figure \ref{fig:kappa}), and thus
the presence of subhalos in the southwest direction from the cluster center might significantly affect the surface profile.
We thus need further studies to measure the subhalo distribution and 
compare these with the dark matter distribution of the main cluster, 
using a large sample of clusters, especially nearby clusters.

\subsection{Mass-to-light Ratio} 
\begin{figure*}
\plottwo{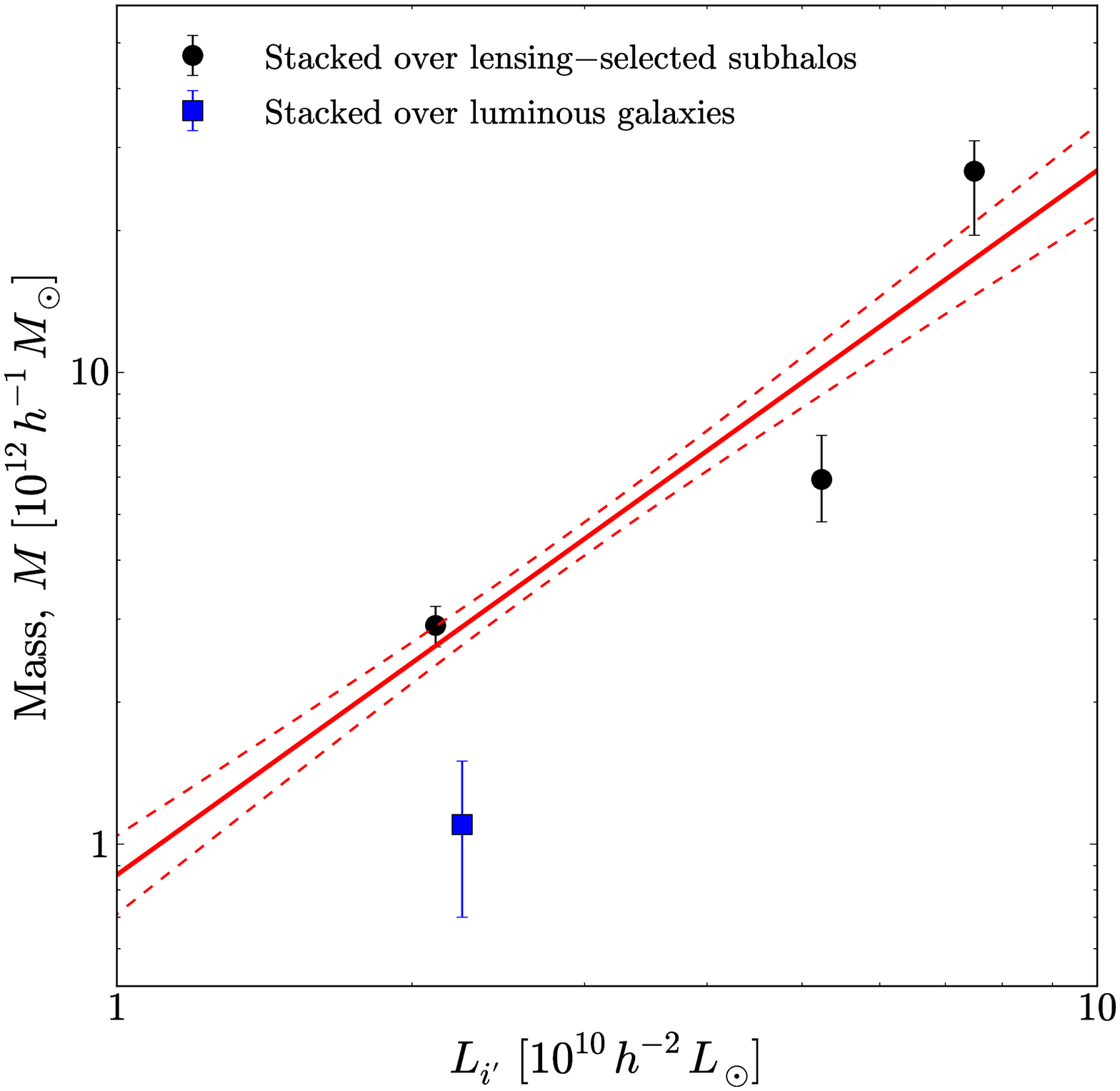}{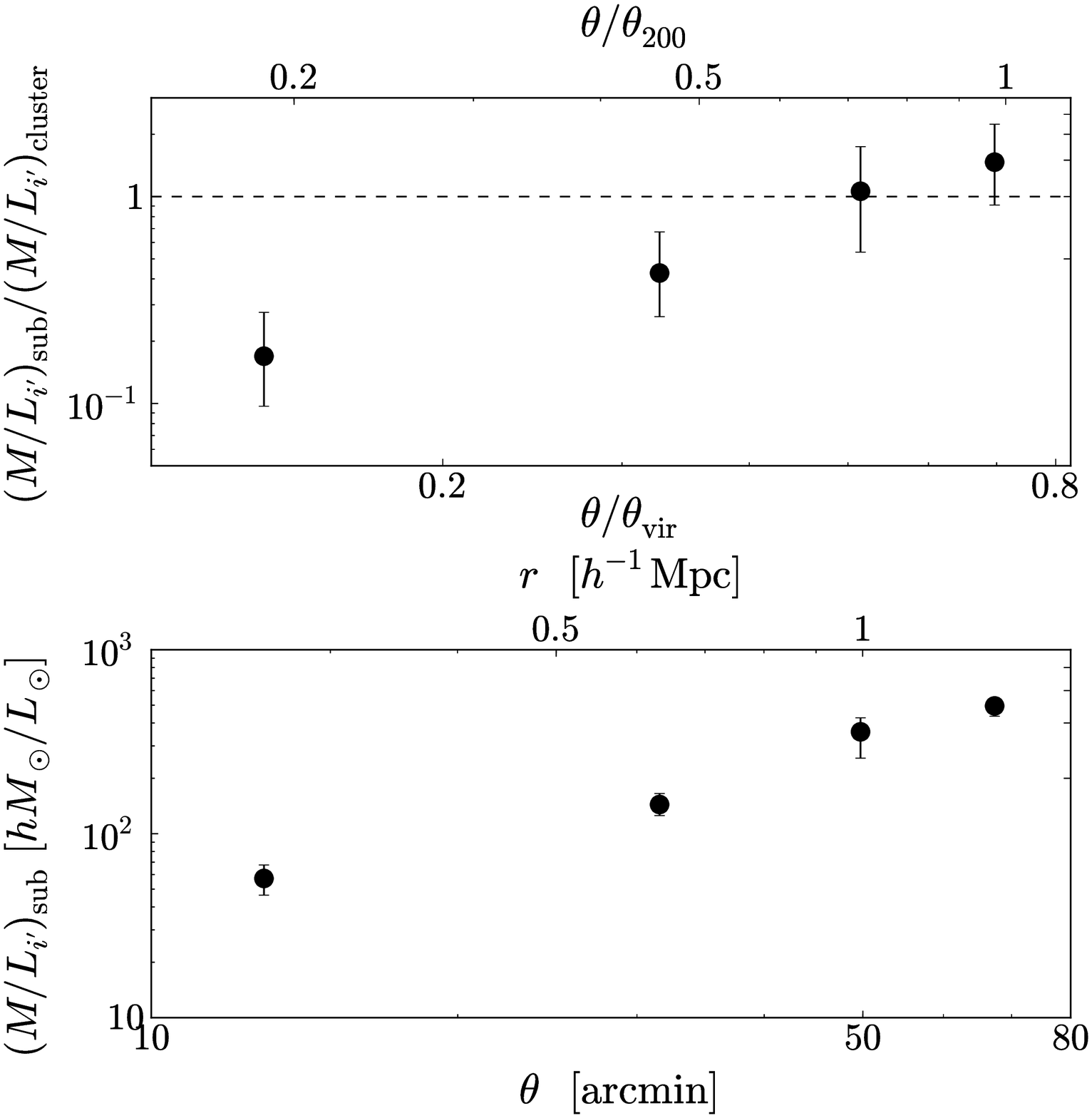}
\caption{Left panel: The mass and luminosity relation for shear-selected subhalos (black circles) and luminous member galaxies (blue squares). The mass of shear-selected subhalos is from stacked lensing analysis of subsamples divided by model-independent masses.
The red solid and dashed lines are the best-fit relationships and 68\% confidence level uncertainties, respectively.
Right panel: The radial dependence of the mass-to-light ratio obtained from the stacked lensing results for subhalo subsamples 
divided by their cluster-centric radius. The top panel shows the mass-to-light ratio normalized by the cluster mass-to-light ratio,
as a function of the radius normalized by the cluster virial radius ($r_{\rm vir}$ or $r_{200}$).
}
\label{fig:Msub_vs_lum}
\end{figure*}

The evolution of galaxies is profoundly affected by their surrounding environments, such as the presence of dark matter halos.
The environmental processes \citep[e.g.,][]{Boselli06} in an overdensity region such as groups or clusters
consume cold gasses in galaxies through star-bursts triggered by mergers, 
tidal interaction with other galaxies, and ram-pressure stripping by the gas, 
which leads to a halt in star formation.
Based on hierarchical structure formation scenarios, 
some cluster galaxies have spent a long time in group-scale environments, 
before being captured in their current host halo.
These group-scale environments may play an important role in the evolution of the galaxies, 
rather than that of the cluster environment \citep{Zabludoff98}.
On the other hand, subhalo masses and sizes depend on their initial properties, 
infall epochs and subsequent evolution in the cluster halo.
Therefore, study of a galaxy-dark matter connection may provide insights into how galaxies 
form and evolve with different mass properties.
We compare two independent quantities of weak-lensing masses of subhalos and luminosities for associated galaxies.
It is well established observationally that there is a correlation between luminosity, velocity dispersion, and 
scale length for early type galaxies, the so-called fundamental plane \citep{Djorgovski87,Dressler87}.
These studies estimate dynamical mass tracing the gravitational potential of a galaxy at small scales.
Weak-lensing analysis measures the mass of dark matter where the distribution extends beyond the galaxies.
Thus, our approach is complementary to previous studies \citep[e.g.,][]{Cappellari06,Cody09}.
We compile two stacked lensing results of shear-selected subhalos divided by individual masses (TNFW;Section \ref{subsec:stacked_subhalos}) 
and luminous member galaxies (Section \ref{sec:stacked_lum}). 
Here, we use the truncated mass rather than the NFW model.
The luminosity of each sample is estimated by an average over all galaxies associated with the subhalos, 
with a weight of the tangential distortion signals.
We plot the correlation between the subhalo masses and the galaxy luminosities, 
in the left panel of Figure \ref{fig:Msub_vs_lum}.
The luminosity ranges between $10^{10}-10^{11}h^{-2}L_\odot$, 
indicating that member galaxies in the subsamples are mainly composed of elliptical galaxies.
The mass increases with increasing luminosity.
To quantify this trend, we fit with $\log(M_{\rm sub}/M_{\rm pivot})=A+B\log(L_{i'}/L_{\rm pivot})$, 
where $L_{\rm pivot}=10^{10}h^{-2}L_\odot$. 
The best-fit slope, $B=1.49\pm0.16$, gives a positive slope at an $8\sigma$ level.
The normalization is $A=-0.15\pm0.19$. 
The data points show a large amount of scatter.
To understand the scatter, further careful study using other parameters of galaxies is needed to constrain 
the fundamental plane between the subhalo masses and the galaxy properties.
We convert the scaling relation into the mass-to-light ratio, 
$M/L=86.1^{+18.1}_{-15.0}(L_{i'}/10^{10}h^{-2}L_\odot)^{0.49\pm0.16} [hM_\odot/L_{i',\odot}]$. 
\cite{Limousin09} have investigated the scaling relation between the luminosity and the total mass in simulated clusters and 
found $M_{\rm tot}\propto L^{1.431\pm0.119}$ in the massive cluster ($\Mvir=1.3\times10^{15}h_{70}^{-1}M_\odot$) at $z=0$, which is in a good agreement with our result.
We compare this analysis to the mass-to-light ratio determined by dynamical masses.
\cite{Cappellari06} estimated the scaling relation of $M/L=(2.35\pm0.19)(L_{I}/10^{10}L_{I,\odot})^{0.32\pm0.06}$ for the SAURON sample.
Our normalization is higher by 1 order of magnitude than dynamical estimates. 
It is likely due to a difference in mass measurements,
because weak-lensing analysis measures the mass of dark matter halos extending beyond galaxy scales.
Similar results have been reported by \cite{vanUitert11}, who showed that weak-lensing masses within $r_{200}$ are about 10 times larger than dynamical masses.

The right panel of Figure \ref{fig:Msub_vs_lum} clearly shows decreasing mass-to-light ratio toward the cluster center, 
similar to the radial dependence of the mass and truncation radius (Figure \ref{fig:Msub_vs_r}).
Here, we use the TNFW mass for subsamples divided by the projected distance from the cluster center (Section \ref{subsec:stacked_subhalos}).
The luminosity is estimated in the same way as for subsamples divided by mass bins.
We fit the form of $\log(M/L/(M/L)_{\rm pivot})=A+B\log(r/r_{\rm pivot})$ to quantify the radial dependence, where
$(M/L)_{\rm pivot}=1hM_\odot/L_\odot$ and $r_{\rm pivot}=1\hMpc$. We obtain $A=5.76_{-0.13}^{+0.13}$ and $B=1.35_{-0.08}^{+0.08}$.
The upper panel shows the mass-to-light ratio of subhalos normalized by the cluster mass-to-light ratio, 
$M/L=337.4^{+140.2}_{-92.5}h~[M_\odot/L_{i',\odot}]$, within the virial radius derived from the tangential fit.
The mass-to-light ratios for subhalos on the outskirts ($0.7\simlt\theta/\theta_{\rm vir}\simlt1$) are close to unity, 
while the ratios in the central region account for $17\%$.
This feature is explained by a scenario where the dark matter subhalos are more subjected to mass loss due to tidal truncation 
than luminous galaxies which tend to be in the central region of subhalos.
Furthermore, since the mean luminosity increases toward the cluster center, 
it is also associated with  galaxy evolution.
Similar trends are suggested by numerical simulations \citep{Springel01}.
They found the that the median mass-to-light ratio gradually increases out to $1.5-2\hMpc$ and is saturated beyond that point.
\cite{Gao04b} also found a similar result, where the median mass-to-light ratio for subhalos is saturated beyond $r_{200}$.
The saturated values \citep{Springel01} are $\sim15\%-20\%$ of the cluster mass-to-light ratio in the $B$-band.
The discrepancy in these ratios might be 
caused by a difference in the subhalo mass range between their simulations and our catalog,
because we used shear-selected subhalos and they selected cluster galaxies and associated subhalos.

This study suggests that the mass-to-light ratio for cluster subhalos 
depends on both the luminosity (mass) and the cluster-centric radius.
To derive more robust conclusions, will require a study using a larger sample of nearby cluster lensing analyses.
However, the present results suggest that an assumption of the constant scaling relation between the mass and the luminosity gives a systematic bias on mass measurements and their statistical properties.

\subsection{Future Studies} \label{sec:dis_fut}

We present a direct observation of the dark matter subhalo mass function using weak gravitational lensing analysis,
which is the first evidence for consistency with CDM predictions on cluster sub-scales.
It is thus an important step toward studying subhalo mass functions and properties with a large sample of clusters 
to make stringent tests of the nature of dark matter and the details of structure formation.

Although the subhalo mass function is well described by the single power law or the Schechter function,
it is difficult to discriminate between the two functions because the abundance of high-mass subhalos is low.
The subhalo mass function stacked over a large sample of clusters will enable us to make a more robust determination of the functional form.
Furthermore, the shape of the mass function has a characteristic feature 
depending on the masses of the other components of dark matter, if any.  
Thus, the subhalo mass function allows us to constrain the nature of dark matter and  structure formation.
Hierarchical structure formation predicts that the subhalo mass function depends on host halo mass 
\cite[e.g.,][]{Gao04,vandenBosch05,Shaw06}.
Less massive halos form first  at higher redshifts where the mean background mass density is higher. 
Subhalos captured by less massive halos  efficiently lose their mass in the high density environment. 
These subhalos are furthermore exposed to destruction over a longer time. 
Less massive halos are therefore expected to contain fewer subhalos. 
To investigate the parent mass dependence, an increase of sample of clusters is essential.
The systematic survey for nearby clusters with these properties 
will increase the total number of dark matter subhalos on the order of a few hundred or more and improve the statistical accuracy. 
Furthermore, finer weak-lensing resolution of nearby clusters will enable us to 
conduct  principal component analyses of the properties of dark matter halos/subhalos \citep[e.g.,][]{Jeeson-Daniel11,Skibba11,Wong12}.
Analytical models such as the halo occupation distribution \citep[e.g.,][]{Seljak00,Cooray02}, the abundance matching \citep[e.g][]{Vale04},
and the conditional luminosity function \citep[e.g.,][]{vandenBosch03} would be helpful to understand the galaxy-dark matter connection.

The near future for multi-wavelength study of subhalos 
will give us direct and important information on the long-standing problem of the interplay between dark halos 
and baryons of member galaxies and gasses.
We investigated the correlation between galaxy luminosities, subhalo masses and their projected radius from the cluster center.
It would also be interesting to investigate the correlation with ages, 
star formation rates and specific star formation rates of galaxies.
\cite{Smith12} have shown an anisotropic spatial distribution of the galaxies age and found 
that the older population of galaxies distribute around subhalo ``17''. 
This might suggest that stellar population properties would vary from subhalo-to-subhalo because 
some cluster galaxies spent a long time in group scale environments before being captured by the cluster.
Further systematic studies using other data-sets, such as stellar masses, star formation rate, specific star formation rate and galaxy types, will provide us with information regarding cluster galaxy evolution.

The {\it ROSAT} X-ray surface brightness distribution (Fig \ref{fig:rosat}) shows
that all  shear-selected subhalos do not contain X-ray extended structures.
This can be understood in terms of observational and/or physical effects.  
First, relatively small X-ray sources are unresolved by the poor PSF of the {\it ROSAT}.
Recent observation \citep{Andrade-Santos13} using {\it Chandra} and {\it XMM-Newton} satellites with high resolution
resolves three X-ray subhalos in the central region that are associated with subhalos ``21'',''23'', and ``24'' , respectively.
Subhalo masses were estimated under the assumption of hydrostatic equilibrium 
and are in good agreement with our mass estimates.
However, X-ray subhalos in other shear-selected subhalos are not found even with high resolution data,
which is due to various physical processes on gas initially bound in the subhalos.
\cite{Tormen04} has pointed out using numerical simulations that the gaseous components,
 which is collisional matter, 
are easily destroyed by ram pressure stripping and hydrodynamic instabilities.
Accordingly, the lifetime of the X-ray subhalos is much shorter than those of dark matter subhalos.
Furthermore, the temperature of the intracluster medium in the central region ($\sim8$keV) is too high to be trapped by subhalos.
Since X-ray observation at the outskirts requires stable and low X-ray backgrounds,
the {\it Suzaku} $X$-ray satellite has a great advantage to search for gas components 
associated at the outskirts of the subhalos \cite[e.g.,][]{Kawaharada10,Walker12,Sato12,Ichikawa13} 
rather than the {\it ROSAT}.
\cite{Simionescu13} has measured the temperature profile out to $70\farcm\sim1.4\hMpc$ using the {\it Suzaku}.
The temperature at the outskirts drops down to $2$keV. 
The sound velocity at this temperature, $c_s\sim720(k_BT/2~{\rm keV})^{1/2}~{\rm km~s^{-1}}$, is lower than
the escape velocity, $v_{\rm esc}\sim1500~{\rm km~s^{-1}}$, expected from the most massive subhalo, number  ``32''.
The enhancement of gas distribution is thus expected to be detected in this region.
However, the {\it Suzaku} pointings do not fully cover the whole area of the cluster. 
The X-ray follow-up observation of the subhalo regions provides us with  important information regarding the gas evolution
and the interplay with subhalos \citep{Tozzi01},  
and resolves possible systematics on the temperature measurement by gas clumpy structures at the outskirts \citep{Nagai11}.
The thermal Sunyaev Zeldovich effect (SZE) observation with different sensitivity from the X-ray 
is also powerful for  gas studies. Indeed, \cite{PlanckCOMA12} has shown 
that the SZE map with FHWM$=10\farcm$ \cite{PlanckCOMA12} 
detected the excess flux around the NGC 4839 group (subhalo ``9''), similar to the {\it ROSAT} X-ray image.

\section{Conclusions} \label{sec:con}

We conducted a weak-lensing survey of subhalos in the very nearby Coma cluster, 
with 18 pointing Subaru/Suprime-Cam observations, covering 4 deg$^2$
and measure the mass of 32 subhalos down to the order of $10^{-3}$ of the virial mass.
We quantified systematic issues relevant to lensing signals from the large-scale structure behind the cluster 
and the probability of spurious peaks.
Our findings are summarized as follows:

\begin{itemize}
\item Weak-lensing analysis for the very nearby Coma cluster offers three important advantages to 
      study cluster subhalos.
      First, the large apparent size of the subhalos enables us to easily resolve the truncation radii for
      less massive subhalos.
      Second, the large apparent area covering the subhalos provides us with a correspondingly large number of background galaxies, 
      which leads to low statistical errors, compensates for low lensing efficiency and achieves a high S/N.
      Third, subhalos mass measurements do not suffer from contamination in lensing signals from the main cluster and other subhalos,
      because the subhalos are well separated. 
      It is acknowledged that this analysis is at a disadvantage in distinguishing cluster subhalos from background group structures, 
      although the LSS lensing effect was accounted for in this analysis. 
      Spectroscopic or photometric redshifts of galaxies are essential to make a secure selection of subhalos. 
      A difference in the models of subhalos and background groups/clusters is also helpful to assess the observed lensing signals.
     
\item Reconstructed mass maps are associated with the projected distributions of member galaxies and the LSS lensing 
      model at $\sim7\sigma-14\sigma$, 
      suggesting that the observed shear catalog contains complete information regarding the mass structure of, and behind, the cluster.

\item We discovered 32 cluster subhalos by applying thresholds of peaks which appeared in the mass maps.
      We estimate the model-independent projected masses of subhalos, $\sim2-50\times10^{12}\hMsol$, where 
      the smooth mass component for the main cluster has been subtracted. 

\item Stacked lensing analysis for samples divided by subhalo masses and cluster-centric radii
      shows a sharply truncated profile. The profile is proportional to $g_+\propto \theta^{-2}$ outside the truncation radii.
      This feature is well described by truncated NFW (TNFW) profile rather than the universal NFW profile without any truncations,
      as expected based on a tidal destruction model.
      For the two subsamples with the most and least massive of the subhalos, the NFW model is strongly disfavored.
      The stacked lensing masses are consistent with model-independent masses for the individual subhalos. 

\item The cluster galaxy-galaxy lensing analysis for luminous member galaxies shows a curved tangential shear profile
      which is well fitted by the NFW model or the summation of the truncated NFW model 
      considering a Gaussian distribution of the truncation radius. However, the NFW model is unlikely for subhalo models,
      because the best-fit virial radius is too large for the subhalo model.

\item   The subhalo mass function $dn/d\ln M_{\rm sub}$ is computed taking into account the systematics of subhalo selection.
        The mass function in the range of 2 orders of magnitude in mass 
        is well described by a single power law $\propto M_{\rm sub}^{-\alpha}$ 
        or the Schechter function $\propto M_{\rm sub}^{-\beta} \exp(-M_{\rm sub}/M_{\rm *})$.
        These best-fit slope $\alpha=1.09^{+0.42}_{-0.32}$ and $\beta=0.99_{-0.23}^{+0.34}$
        are in a remarkably good agreement with CDM predictions $\sim0.9-1.0$ \citep[e.g.,][]{Giocoli10,Gao12}.
        This is the first evidence of consistency with CDM predictions on cluster sub-scales.

\item The subhalo masses, truncation radii, and mass-to-light ratios decrease toward the cluster center, 
      as expected from tidal destruction. The galaxy luminosities associated with subhalos depend on both their mass 
      and the cluster-centric radius.

\item The tangential distortion signals, $g_+$, in the range of $\sim0.02-2\hMpc$ show a complex structure 
      which is well described by three mass components of the smooth mass distribution of the NFW model, subhalos, and LSS lensing model.
      Although the lensing signals are 1 order of magnitude lower than those for clusters at intermediate redshifts, $z\sim0.2$ \citep{Okabe10b}, 
      the total S/N, S/N$\simeq 13.3$, is comparable to them or higher because of 
       a correspondingly large number of background galaxies ($\sim6\times10^5$). 
      The signal-to-noise ratios for subhalos and LSS lensing models are S/N$\simeq 4.4$ and $\simeq 1.3$, respectively.
      The $45^\circ$ rotated component, $g_\times$, is consistent with a null signal.

\end{itemize}

\section*{Acknowledgments}

We are grateful to N. Kaiser for developing the IMCAT package and making it publicly available. 
We thank Keiichi Umetsu, Masahiro Takada, Takashi Hamana, Eiichiro Komatsu, Yuki Okura, 
Neal Dalal, Marceau Limousin, Richard Massey, Hao-Yi Wu, Priya Natarajan, 
and Jean-Paul Kneib for helpful comments and/or discussions.
We also thank Alan Lefor for English correction.
This work was supported 
by World Premier International Research Center Initiative (WPI Initiative), MEXT, Japan.

\appendix

\section{Anisotropic PSF correction} \label{app:e_pattern}

The anisotropic PSF correction is critically important for weak-lensing analysis of clusters 
with low lensing efficiency, such as the Coma cluster. Systematic residual ellipticities remaining after the correction will introduce a systematic bias in estimates of cluster and subhalo masses. We conducted five tests to assess the anisotropic PSF correction.
As described in Section \ref{subsec:wl}, we estimated $q_*^{\alpha}(\btheta)$ patterns by fitting
the stellar ellipticity components with the function of second-order bi-polynomials of the vector $\btheta$,
and applied it to galaxy ellipticities.

\begin{figure}
\epsscale{1}
\plotone{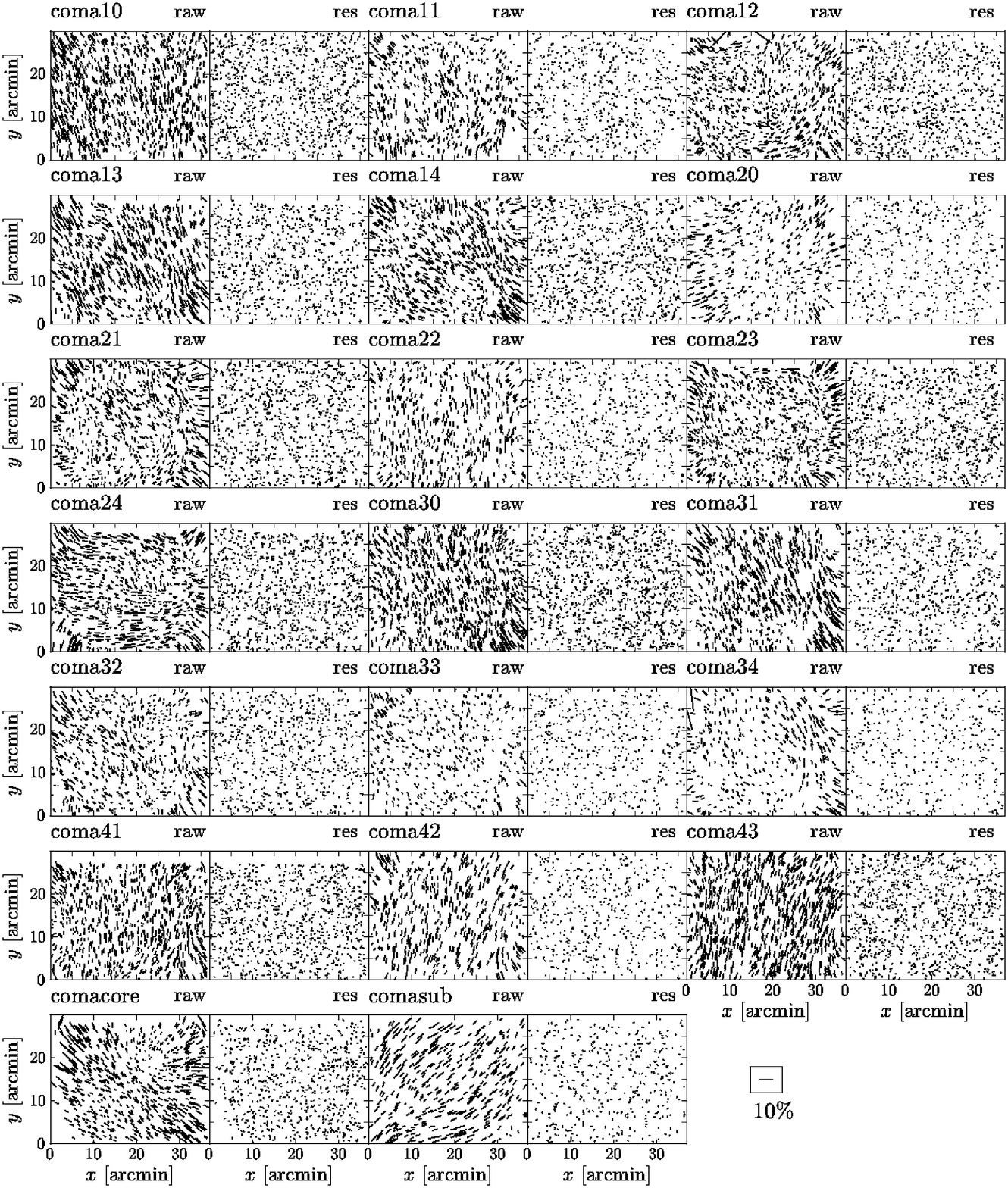}
\caption{Pattern of stellar ellipticity before and after the anisotropic PSF 
correction for individual pointings.
The data name is shown in the top left corner.
The left and right panels show the raw $(e^{*,{\rm raw}}_1,e^{*,{\rm raw}}_2)$ and residual $(e^{*,{\rm res}}_1, e^{*,{\rm res}}_2)$ stellar ellipticities, respectively.
}
\label{fig:e_pattern}
\end{figure}

To test the model, we first  compared  the observed $q_*^{\alpha}(\btheta)$ distribution patterns with those of the model by rapidly alternating the two distributions on a monitor to allow visual identification of significant differences, since the human eye is sensitive to differences in rapidly changing images. We also investigated spatial distributions of the stellar ellipticity components, 
$e_*^{\alpha}(\btheta)$, before and after the anisotropic PSF correction, as shown in Figure \ref{fig:e_pattern}.
Although there is a large-scale coherent pattern of raw stellar ellipticities (left panel of Figure \ref{fig:e_pattern}),
the residual stellar ellipticities after the correction visually confirmed the random distribution patterns 
(right panel of Figure \ref{fig:e_pattern}).

\begin{figure}
\epsscale{1}
\plottwo{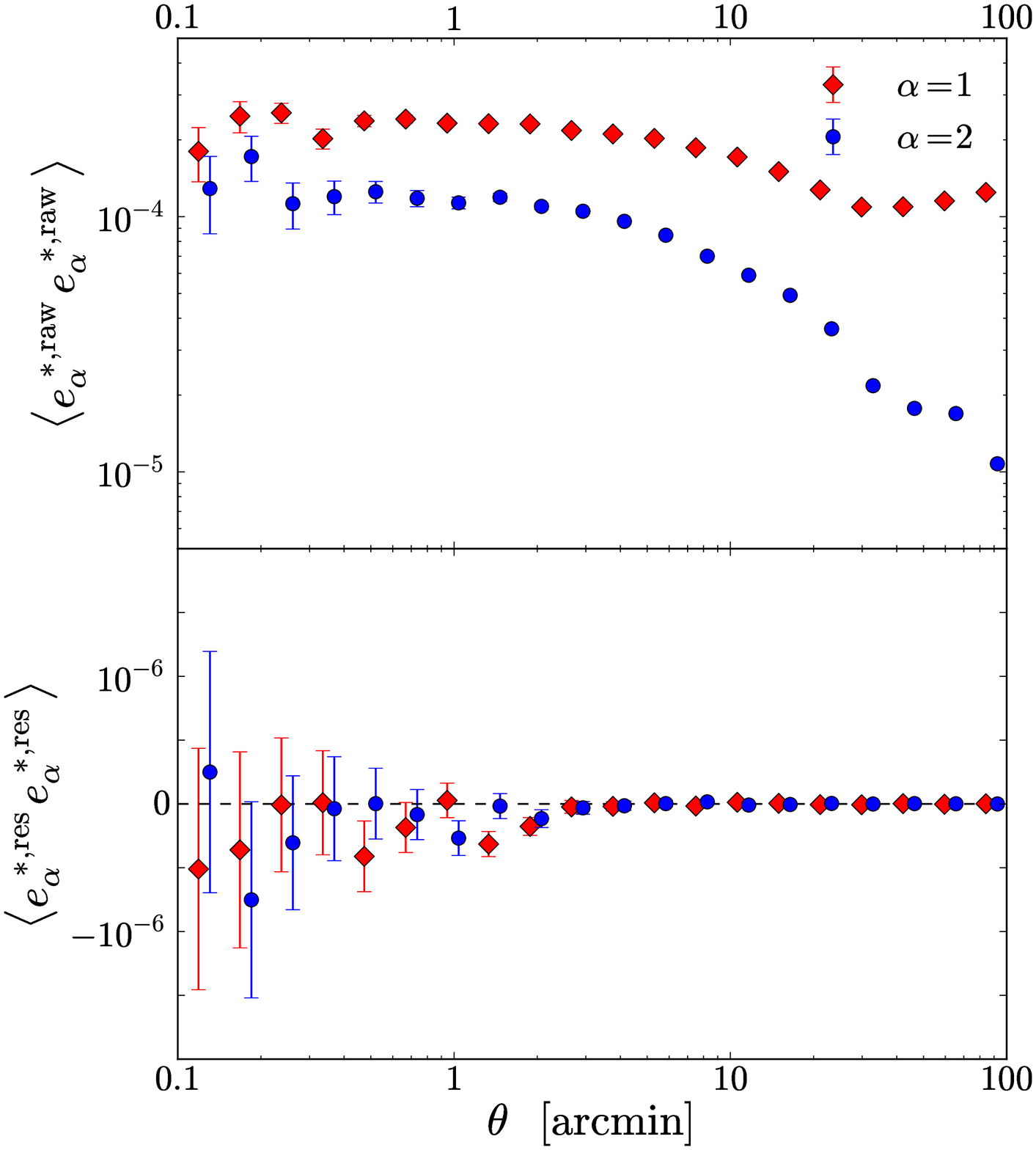}{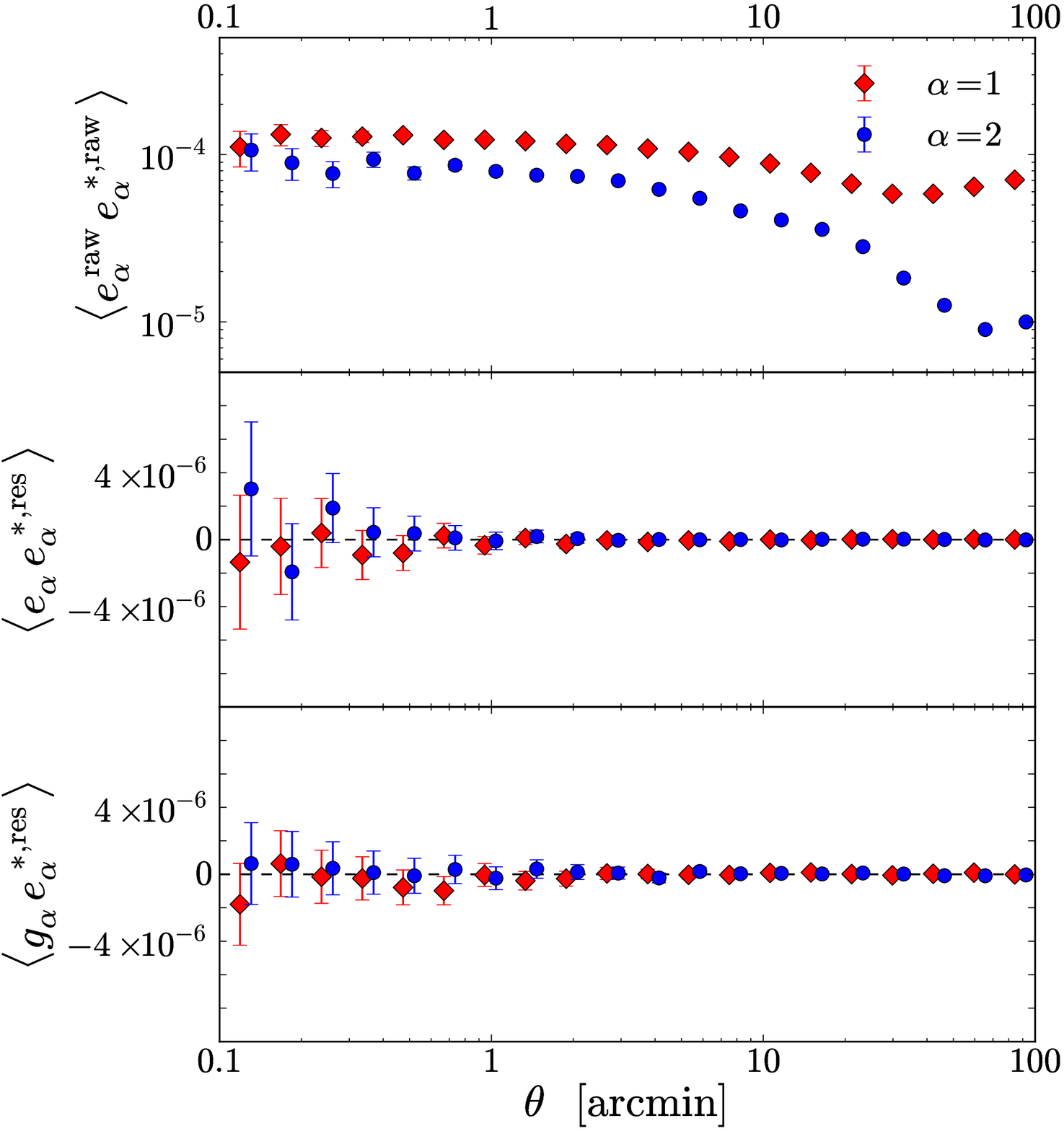}
\caption{Auto correlation function for the stellar ellipticities (left) and a cross correlation function between 
the galaxy and stellar ellipticities (right) as a function of angular separation, $\theta$.
The red diamonds and blue circles denote two components of the ellipticity ($e_\alpha$; $\alpha=1,2$), respectively.
The $x$ positions for blue circles are shifted ( multiplied by 1.1 ) from the originals.
The raw stellar ellipticities are highly correlated (left-top panel), while the residual stellar ellipticities
after the correction are consistent with zero (left-bottom panel). 
The cross correlation function between the raw galaxy and stellar ellipticities, 
$\langle e^{{\rm raw}}_\alpha e^{*,{\rm raw}}_\alpha \rangle$, shows positive values (right-top panel).
The residual cross correlation function between the galaxy and stellar ellipticities, 
$\langle e_\alpha e^{*,{\rm res}}_\alpha \rangle$, does not show any significant correlation over a wide range (right-middle panel).
The cross correlation between the residual stellar ellipticities and the reduced shear for background galaxies which we used for
lensing analysis, $\langle g_\alpha e^{*,{\rm res}}_\alpha \rangle$, does not show 
any significant feature consistent with an imperfect anisotropic PSF correction (right-bottom panel).
}
\label{fig:corr}
\end{figure}

Second, in order to more quantitatively assess the model,  we computed an auto-correlation function for stellar ellipticities and 
a cross-correlation function for the ellipticities of galaxies and stars, before and after the correction,
 respectively. These are estimated by using the average for all pairs of galaxies/stars and stars separated by the angle
$\theta$, with equal weight. A clear positive correlation between observed stellar ellipticities  is shown in the top-left panel of Figure \ref{fig:corr}. However, residual stellar ellipticities are suppressed  to zero, even 
though they are slightly anti-correlated at a separation angle $\sim 1'$.
Similarly, no correlation between the corrected galaxy and residual stellar ellipticities is found 
(right-middle panel of Figure \ref{fig:corr}). This test was conducted before applying the isotropic PSF correction, and the results support the conclusion that there is no systematic bias in the measurement of galaxy ellipticities.

Third, we investigated the cross-correlation function between 
residual stellar ellipticities and reduced shear $g_\alpha$ after the isotropic PSF correction, and found the same result.
Here, we plot the cross-correlation for our background sample after the color selection, 
estimated with a weight of $w_g$ (Equation (\ref{eq:w_g})), in the right-bottom panel of Figure \ref{fig:corr}.
This plot clearly shows the null correlation and thus indicates that the systematic bias caused by imperfect PSF correction is negligible at most.

Fourth, we computed the median for two components of residual stellar ellipticities before and after the correction.
The median for two components of residual stellar ellipticities after the anisotropic correction, 
\mbox{\boldmath $\bar{e}$}$_{\rm res}^*=(0.791\pm3.130, 1.127\pm2.331)\times10^{-5}$, improves from those of raw stellar ellipticities before the correction, 
\mbox{\boldmath $\bar{e}$}$_{\rm raw}^*=(-1.179\pm0.014, -0.520\pm0.014)\times10^{-2}$.

Fifth, we assessed the shape measurement using the same galaxies detected in all overlapping regions of different images 
(see also Section \ref{subsec:wl}).

\section{Map Making} \label{app:mapmaking}

We reconstructed the projected mass distribution following \cite{Kaiser93}.
As described in detail in \cite{Okabe08},  
we pixelize the reduced shear into a regular grid with a Gaussian smoothing  
of $G(\theta)\propto \exp[-\theta^2/\theta_g^2]$.
The resolution of the maps is defined by ${\rm FWHM}\equiv2\sqrt{\ln{2}}\theta_g$. 
The smoothed shear pattern at an angular position $\btheta$ is estimated as
\begin{equation}
\label{eq:smshear}
\bar{g}_{\alpha}(\btheta) = \frac{\sum_i G(\btheta-\btheta_i) w_{g,i} g_{\alpha,i}(\btheta_i)}{\sum_i
G(\btheta-\btheta_i) w_{g,i}} \label{eq:smshear}
\end{equation}
where $w_{g}$ and $g_{\alpha,i}$ are the statistical weight of Equation (\ref{eq:w_g}) and the reduced shear of the $i$th galaxy, 
respectively.
The error variance for the smoothed shear (\ref{eq:smshear}) is given by
\begin{equation}
\label{eq:smshearvar}
\sigma^2_{\bar{g}}(\btheta) = 
\frac{\sum_i G(\btheta-\btheta_i)^2 w_{g,i}^2 \sigma^2_{g,i}}
{ \left( \sum_i G(\btheta-\btheta_i) w_{g,i} \right)^2}.
\end{equation} 
Then, the smoothed shear field (\ref{eq:smshear}) is inverted with the kernel \citep{Kaiser93} in Fourier space 
to obtain the projected mass distribution, $\kappa(\btheta)$.
Here, we assume the weak-limit of $g_{\alpha}=\gamma_{\alpha}/(1-\kappa) \approx \gamma_{\alpha}$.
We also compute a map of the significance level for mass reconstruction, $\nu(\btheta)\equiv\kappa/\sigma_\kappa$,
with the mass reconstruction error $\sigma_\kappa(\btheta)$.
 We also make maps of luminosity ($l(\btheta)$) and number density ($n(\btheta)$) 
for member galaxies defined in Section \ref{subsec:member} and the convergence field ($\kappa_{\rm LSS}(\btheta)$) 
of LSS lensing in Section \ref{subsec:LSSlens} with a statistical weight of $w_{g,i}=1$.

\section{Model-independent Mass Measurement} \label{app:M2D}

A parameter-free estimation of subhalos is given by the aperture-densitometry, 
or the so-called  $\zeta_c$-statistics \citep{Clowe00}.
The projected mass, $M_{\rm \zeta_c}(<\theta)$, is given by
\begin{eqnarray}
M_{\rm \zeta_c}(<\theta) &=& \pi{\theta}^2\Sigma_{\rm cr}\zeta_c(\theta,\theta_{{\rm
 inn}},\theta_{{\rm out}}), \label{eq:Mzeta}\\
 \zeta_c(\theta; \theta_{\rm inn},\theta_{\rm out})
&=&\bar{\kappa}(<\theta) - 
   \bar{\kappa}(\theta_{\rm inn}< \theta <\theta_{\rm out}) \\
                     &= & 2\int^{\theta_{\rm inn}}_{\theta} d \ln \theta'
                      \langle \gamma_+ (\theta)
		      \rangle
                      +   \frac{2}{1-\theta_{\rm inn}^2/\theta_{\rm out}^2} 
                      \int^{\theta_{\rm out}}_{\theta_{\rm inn}}
		      d \ln \theta' \langle \gamma_+(\theta)\rangle  \nonumber 
\end{eqnarray}
where $\theta_{\rm inn}$ and $\theta_{\rm out}$ are the inner and outer radii of the background annulus.
The $\langle \gamma_+ \rangle$ is an azimuthal average of the
tangential component of the gravitational shear, which we take 
$\langle \gamma_+(\theta)\rangle \approx \langle g_+(\theta)\rangle$ in the weak limit. 
The uncertainty in $\zeta_c$ at $\theta_i$ is estimated as
\begin{eqnarray}
\sigma^2_i=4\sum_{j=i}^{N_{\rm inn}}\left(
\frac{\Delta\theta_j}{\theta_j} \right)^2\sigma^2_{g_+}(\theta_j)+\left(\frac{2}{1-\theta_{\rm inn}^2/\theta^2_{\rm out}}\right)^2
\sum_{i=N_{\rm inn}}^{N_{\rm out}}\left(\frac{\Delta\theta_j}{\theta_j}
\right)^2\sigma^2_{g_+}(\theta_j),
\end{eqnarray}
where $N_{\rm inn}$ and $N_{\rm out}$ are the indices for each of the discrete radial bins corresponding to 
the radii of $\theta_{\rm inn}$ and $\theta_{\rm out}$ in Equation ~(\ref{eq:Mzeta}), respectively.
An error covariance of $\zeta_c$ between each bin is given by $\sigma_{ij}=\sigma_{ji}=\sigma_j^2$ for $\theta_i<\theta_j$.
The S/N of the radial profile, which is complementary information to that of peaks in mass maps,
is computed by 
\begin{eqnarray}
{\rm S/N}=\left(\sum_{ij} M_{\zeta_c,i} V_{ij}^{-1} M_{\zeta_c,j}\right)^{1/2}
\end{eqnarray}
where $V_{ij}^{-1}$ is the inverse of the covariance matrix.

To quantify the mass of the subhalos, we estimate an average projected mass taking into account the error covariance matrix,
as follows,
\begin{eqnarray}
   M_{\rm 2D} &=& \sum_{i=N_{s1}}^{N_{s2}} \Gamma_i M_{\zeta_c,i} \\
  \sigma_{\rm M_{\rm 2D}}^2 &=& \sum_{i,j=N_{s1}}^{N_{s2}} \Gamma_i \Gamma_j V_{ij} \\
  \Gamma_i&=& \sum_{j=N_{s1}}^{N_{s2}} V_{ij}^{-1}/\sum_{i,j=N_{s1}}^{N_{s2}} V_{ij}^{-1}  \\
\end{eqnarray}
where $N_{\rm s1}$ and $N_{\rm s2}$ are the indices for each of the discrete radii where $M_{\zeta_c}$ profile is saturated.

\section{Mass Models} \label{app:massmodel}

Mass models for cluster halos and subhalos are summarized in this section.
Numerical simulations, based on the CDM model of structure formation,
predicts that dark matter halos spanning a wide mass range
can be described by a universal mass density profile \citep{NFW96,NFW97}.
In this paper, we refer to these density profiles as the NFW profile 
which is expressed in the form of
\begin{equation}
\rho_{\rm NFW}(r)=\frac{\rho_s}{(r/r_s)(1+r/r_s)^2},
\label{eq:rho_nfw}
\end{equation}
where $\rho_s$ is the central density parameter and $r_s$ is the scale radius.
The density profile has inner and outer slope values of $-1$ and $-3$, respectively.
The three-dimensional spherical masses, $M_\Delta$, enclosed by the radius, $r_\Delta$,
inside of which the mean density is $\Delta$ times the critical mass density, 
$\rho_{\rm cr}(z)$, at the redshift, $z$, is given by  
\begin{equation}
M_{\rm NFW}(<r_\Delta)=\frac{4\pi \rho_s r_\Delta^3}{c_\Delta^3}\left[
\ln(1+c_\Delta)-\frac{c_\Delta}{1+c_\Delta}\right].
\label{eq:MNFW}
\end{equation}
The NFW profile is specified by the two parameters including $M_\Delta$ 
and the halo concentration $c_\Delta=r_\Delta/r_s$.

Subhalo sizes are determined by the strong tidal field of the main cluster halo.
The mass density outside the instantaneous tidal radius of subhalos 
drastically decreases and is close to zero 
due to tidal stripping \citep[e.g.,][]{Tormen98,Hayashi03,Oguri04,Taylor04a,vandenBosch05}.
Therefore, 
the mass model for subhalos requires an additional parameter of the truncation radius, $r_t$.
We consider two models of the truncation model for NFW and SIS models described above.
We refer to these truncation models as the truncated NFW \citep[TNFW;][]{Takada03,Hamana04}.
The interior mass density profile for the TNFW model follows
the NFW model, but is zero outside the truncation radius. 
\begin{eqnarray}
  \rho_{\rm TNFW}(r)&=&\rho_{\rm NFW}~~~~~~~ (r\leq r_t), \\
                    &=&0 ~~~~~~~~~~~~~ (r> r_t). \nonumber 
\end{eqnarray}
This is an extreme case of the truncation model.
The TNFW model \citep{Okabe10a} is specified by three parameters 
including the subhalo mass ($M_t$), a truncation radius ($r_t$) and a concentration ($c_t$).
The slope of the lensing profile for the TNFW model drastically changes outside the truncation radius 
and behaves as a point source ($\propto\theta^{-2}$).
In model fitting, 
the truncation radius and the concentration are sensitive to this break and the inner profile, respectively.
The subhalo mass is also determined by the distortion signal at the truncation radius.
Since the mass densities outside the truncation radius are zero, 
the three- and two-dimensional masses within the truncation radius yield the exactly same value ($M_{\rm 3D}=M_{\rm 2D}$).

When we stack tangential distortion profiles for subhalos with different properties, 
the break in the distortion profile is smooth due to their intrinsic distribution.
In addition to the TNFW model, we compute the model taking into account a distribution function 
of the truncation radius (TNFWProb). The probability function is assumed to be Gaussian, 
$p(r_t)=\exp(-(r_t-\langle r_t\rangle)^2/2\sigma_{r_t}^2)/\sqrt{2\pi\sigma_{r_t}^2}$, where $\langle r_t\rangle$
and $\sigma_{r_t}$ are the average and the standard error for the truncation radius, respectively.
The mean lensing signal is expressed in terms of $\int p(r_t) g_+^{{\rm (TNFW)}}dr_t$.
We do not assume a distribution over the subhalo mass but instead estimate the mean subhalo mass, $\langle M_t\rangle$, 
which is sensitive to a lensing signal at $\langle r_t\rangle$.


\bibliographystyle{apj}

\bibliography{my}
\clearpage
\end{document}